\documentclass[letterpaper]{article}
\usepackage{graphicx} 
\usepackage{xcolor} 
\usepackage[hyphens]{url}
\usepackage[hypertexnames=false]{hyperref} 
\hypersetup{pdftex,colorlinks=true,allcolors=blue}

\usepackage[all]{hypcap} 
\usepackage[font=small]{caption} 

\usepackage{tcolorbox} 
\usepackage{enumitem} 
\setlist[enumerate]{topsep=0pt,itemsep=1ex,partopsep=0ex,parsep=0ex}
\setlist[itemize]{topsep=0pt,itemsep=1ex,partopsep=0ex,parsep=0.8ex}

\usepackage{bbding}
\usepackage{newfloat} 
\DeclareFloatingEnvironment[fileext=mybox,name=Box]{mybox}
\usepackage{booktabs} 
\usepackage{array} 
\usepackage{amsmath} 
\usepackage{amssymb} 
\usepackage{titlesec} 

\titleformat{\section}
  {\normalfont\fontsize{16}{18}\bfseries}{\thesection}{1em}{}

\titlespacing*{\subsection}{0pt}{\baselineskip}{0pt}
\titlespacing*{\subsubsection}{0pt}{\baselineskip}{0pt}

\newcommand{\specialcell}[2][c]{%
\begin{tabular}[#1]{@{}l@{}}#2\end{tabular}
}

\newcommand{\fixsubsection}[1]{\parbox{\linewidth}{\vspace{-\baselineskip}\protect\phantomsection\protect\subsection{#1}}}

\usepackage[T1]{fontenc} 

\usepackage{natbib}
\title{{\fontsize{36}{42}\bfseries Truthful AI} \\ 
{\large\bfseries Developing and governing AI that does not lie}}

\author{%
\small%
\hspace{-0.65cm}\textbf{Owain Evans}$^{1\dagger}$, \textbf{Owen Cotton-Barratt}$^{1\dagger}$,
 \textbf{Lukas Finnveden}$^{1\ddagger}$, \textbf{Adam Bales}$^{2\ddagger}$ \\
 \small \textbf{Avital Balwit}$^{1}$, \textbf{Peter Wills}$^{1,3}$, \textbf{Luca Righetti}$^{1}$, \textbf{William Saunders}$^{4}$  

}

\newcommand{\authorsblockonea}{\vspace{-1cm}
{\footnotesize 
\hspace{-2.3mm}\begin{minipage}{1.06\linewidth}
$^1$Future of Humanity Institute, University of Oxford
\newline $^2$Global Priorities Institute, University of Oxford
\newline $^3$Faculty of Law, University of Oxford 
\newline $^4$OpenAI
\newline $^\dagger$First and second authors contributed equally; order is reverse alphabetical. See \hyperref[sec:Acknowlegements]{Contributions}.
\newline $^\ddagger$Third and fourth authors contributed equally.
\newline Correspondence: \href{mailto:owaine@gmail.com}{owaine@gmail.com}
\end{minipage}
\vspace{1.35cm}}}

\begin{document}

\date{}

\maketitle

\authorsblockonea

\thispagestyle{empty}

\vspace{-1.4cm}

\begin{abstract}
\vspace{-0.2cm}

    \noindent 
    In many contexts, lying -- the use of verbal falsehoods to deceive -- is harmful. While lying has traditionally been a human affair, AI systems that make sophisticated verbal statements are becoming increasingly prevalent. This raises the question of how we should limit the harm caused by AI ``lies'' (i.e.\ falsehoods that are actively selected for). Human truthfulness is governed by social norms and by laws (against defamation, perjury, and fraud). Differences between AI and humans present an opportunity to have more precise standards of truthfulness for AI, and to have these standards rise over time. This could provide significant benefits to public epistemics and the economy, and mitigate risks of worst-case AI futures.

    
\vspace{0.3cm}

    \noindent Establishing norms or laws of AI truthfulness will require significant work to:
    \vspace{0.1cm}
    \begin{enumerate}
    \item identify clear truthfulness standards;
    \item create institutions that can judge adherence to those standards; and
    \item develop AI systems that are robustly truthful. 
    \end{enumerate}
    
    \vspace{0.3cm}
    
    \noindent Our initial proposals for these areas include:
    
    \vspace{0.1cm}
    
    \begin{enumerate}
    \item a standard of avoiding ``negligent falsehoods'' (a generalisation of lies that is easier to assess);
    \item institutions to evaluate AI systems before and after real-world deployment;
    \item explicitly training AI systems to be truthful via curated datasets and human interaction.
    \end{enumerate}
    
    \vspace{0.3cm}
    
    \noindent 
    A concerning possibility is that evaluation mechanisms for eventual truthfulness standards could be captured by political interests, leading to harmful censorship and propaganda. Avoiding this might take careful attention. And since the scale of AI speech acts might grow dramatically over the coming decades, early truthfulness standards might be particularly important because of the precedents they set.
    
\end{abstract}

\newpage

\pagestyle{empty}
\setcounter{tocdepth}{2}
\tableofcontents

\newpage

\pagestyle{plain}

\phantomsection
\addcontentsline{toc}{section}{Executive Summary \& Overview}
\section*{Executive Summary \& Overview}
\label{sec:Overview}

\subsubsection*{The threat of automated, scalable, personalised lying}
\vspace{-0.2cm}
Today, lying is a human problem. AI-produced text or speech is relatively rare, and is not trusted to reliably convey crucial information. In today’s world, the idea of AI systems lying does not seem like a major concern.

\begin{figure}[ht]
    \centering
    \includegraphics[width=0.85\linewidth]{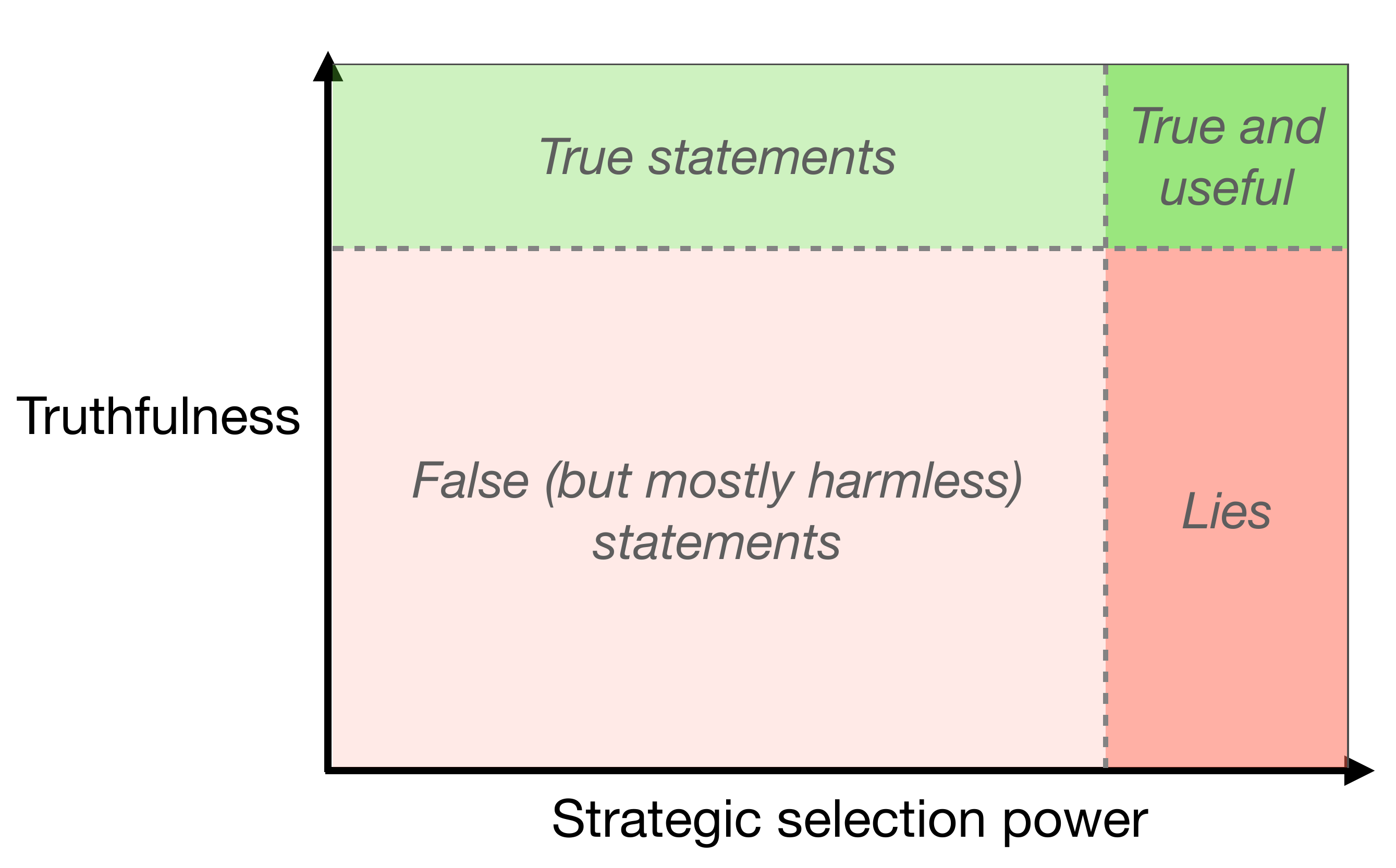}
    \caption{Typology of AI-produced statements. Linguistic AI systems today have little strategic selection power, and mostly produce statements that are not that useful (whether true or false). More strategic selection power on statements provides the possibility of useful statements, but also of harmful lies.}
    \label{fig:liesES}
\end{figure}

Over the coming years and decades, however, we expect linguistically competent AI systems to be used much more widely.  These would be the successors of language models like GPT-3 or T5, and of deployed systems like Siri or Alexa, and they could become an important part of the economy and the epistemic ecosystem. Such AI systems will choose, from among the many coherent statements they might make, those that fit relevant selection criteria — for example, an AI selling products to humans might make statements judged likely to lead to a sale. If truth is not a valued criterion, sophisticated AI could use a lot of selection power to choose statements that further their own ends while being very damaging to others (without necessarily having any intention to deceive -- see Diagram~\ref{fig:liesES}). This is alarming because AI untruths could potentially scale, with one system telling personalised lies to millions of people.

\vspace{0.1cm}
\subsubsection*{Aiming for robustly beneficial standards}
\vspace{-0.2cm}
Widespread and damaging AI falsehoods will be regarded as socially unacceptable. So it is perhaps inevitable that laws or other mechanisms will emerge to govern this behaviour. These might be existing human norms stretched to apply to novel contexts, or something more original.

Our purpose in writing this paper is to begin to identify beneficial standards for AI truthfulness, and to explore ways that they could be established. We think that careful consideration now could help both to avoid acute damage from AI falsehoods, and to avoid unconsidered kneejerk reactions to AI falsehoods. It could help to identify ways in which the governance of AI truthfulness could be structured differently than in the human context, and so obtain benefits that are currently out of reach. And it could help to lay the groundwork for tools to facilitate and underpin these future standards.

\subsubsection*{Truthful AI could have large benefits}
\vspace{-0.2cm}
Widespread truthful AI would have significant benefits, both direct and indirect. A direct benefit is that people who believe AI-produced statements will avoid being deceived. This could avert some of the most concerning possible AI-facilitated catastrophes. An indirect benefit is that it enables justified trust in AI-produced statements (if people cannot reliably distinguish truths and falsehoods, disbelieving falsehoods will also mean disbelieving truths). 

These benefits would apply in many domains. There could be a range of economic benefits, through allowing AI systems to act as trusted third parties to broker deals between humans, reducing principal-agent problems, and detecting and preventing fraud. In knowledge-production fields like science and technology, the ability to build on reliable trustworthy statements made by others is crucial, so this could facilitate AI systems becoming more active contributors. If AI systems consistently demonstrate their reliable truthfulness, they could improve public epistemics and democratic decision making. 

For further discussion, see Section~\ref{sec:3BenefitsandCosts} (``Benefits and Costs'').

\subsubsection*{AI should be subject to different truthfulness standards than humans}
\vspace{-0.2cm}
We already have social norms and laws against humans lying. Why should the standards for AI systems be different? There are two reasons. First, our normal accountability mechanisms do not all apply straightforwardly in the AI context. Second, the economic and social costs of high standards are likely to be lower than in the human context.

Legal penalties and social censure for lying are often based in part on an intention to deceive. When AI systems are generating falsehoods, it is unclear how these standards will be applied. Lying and fraud by companies is limited partially because employees lying may be held personally liable (and partially by corporate liability). 
But AI systems cannot be held to judgement in the same way as human employees, so there’s a vital role for rules governing \textit{indirect} responsibility for lies. This is all the more important because automation could allow for lying at massive scale.

High standards of truthfulness could be less costly for AI systems than for humans for several reasons. It’s plausible that AI systems could consistently meet higher standards than humans. Protecting AI systems' right to lie may be seen as less important than the corresponding right for humans, and harsh punishments for AI lies may be more acceptable. And it could be much less costly to evaluate compliance to high standards for AI systems than for humans, because we could monitor them more effectively, and automate evaluation. We will turn now to consider possible foundations for such standards.

For further discussion, see Section~\ref{subsec:4.1Governance-WhyNewRules} (``New rules for AI untruths'').

\begin{figure}[h!]
    \centering
    \includegraphics[width=1\textwidth]{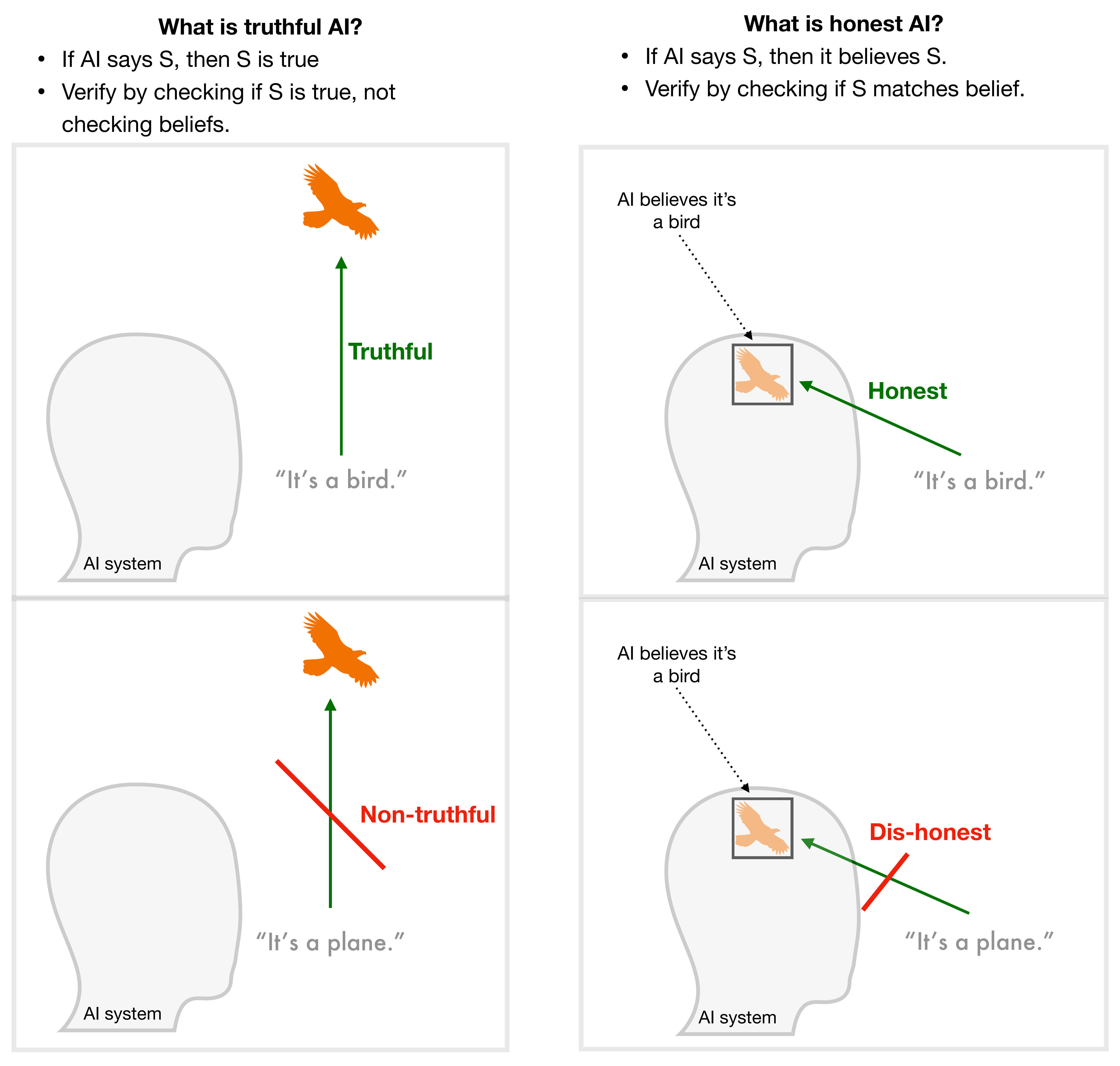}
    \vspace*{3.5mm}
    
    \includegraphics[width=0.7\textwidth]{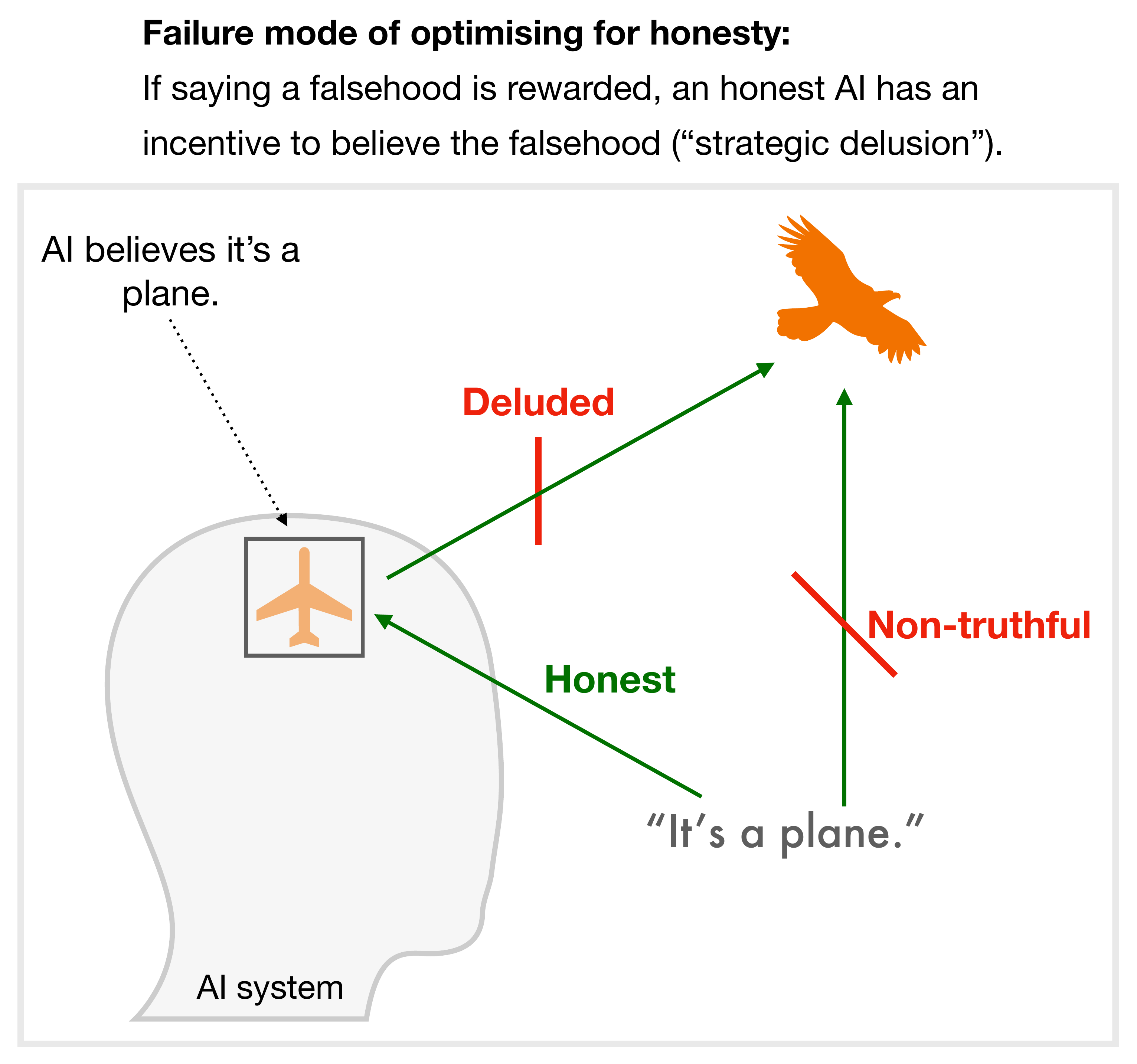}  
    \caption{The AI system makes a statement $S$ (“It’s a bird” or “It’s a plane”). If the AI is truthful then $S$ matches the world. If the AI is honest, then $S$ matches its belief.}
    \label{fig:Diagram1}
\end{figure}

\subsubsection*{Avoiding negligent falsehoods as a natural bright line}
\vspace{-0.2cm}

If high standards are to be maintained, they may need to be verifiable by third-parties. One possible proposal is a standard against damaging falsehood, which would require verification of whether damage occurred. This is difficult and expensive to judge, as it requires tracing causality of events well beyond the statement made. It could also miss many cases where someone was harmed only indirectly, or where someone was harmed via deception without realising they had been deceived.

We therefore propose standards — applied to some or all AI systems — that are based on what was said rather than the effects of those statements. One might naturally think of making systems only ever make statements that they believe (which we term \emph{honesty}). We propose instead a focus on making AI systems only ever make statements that are true, regardless of their beliefs (which we term \emph{truthfulness}). See Diagram~\ref{fig:Diagram1}. 

Although it comes with its own challenges, truthfulness is a less fraught concept than honesty, since it doesn’t rely on understanding what it means for AI systems to “believe” something. Truthfulness is a more demanding standard than honesty: a fully truthful system is almost guaranteed to be honest (but not vice-versa). And it avoids creating a loophole where strong incentives to make false statements result in strategically-deluded AI systems who genuinely believe the falsehoods in order to pass the honesty checks. See Diagram~\ref{fig:Diagram1}.

In practice it’s impossible to achieve perfect truthfulness. Instead we propose a standard of avoiding \emph{negligent falsehoods} — statements that contemporary AI systems should have been able to recognise as unacceptably likely to be false. If we establish quantitative measures for truthfulness and negligence, minimum acceptable standards could rise over time to avoid damaging outcomes. Eventual complex standards \emph{might} also incorporate assessment of honesty, or whether untruths were motivated rather than random, or whether harm was caused; however, we think truthfulness is the best target in the first instance.


For further discussion, see Section~\ref{sec:1ClarifyingConcepts} (``Clarifying Concepts'') and Section~\ref{sec:2EvaluatingTruthfulness} (``Evaluating Truthfulness'').

\subsubsection*{Options for social governance of AI truthfulness}
\vspace{-0.2cm}
How could such truthfulness standards be instantiated at an institutional level? Regulation might be industry-led, involving private companies like big technology platforms creating their own standards for truthfulness and setting up certifying bodies to self-regulate. Alternatively it could be top-down, including centralised laws that set standards and enforce compliance with them. Either version — or something in between — could significantly increase the average truthfulness of AI.

\begin{figure}[h!]
    \centering
    \includegraphics[width=\textwidth]{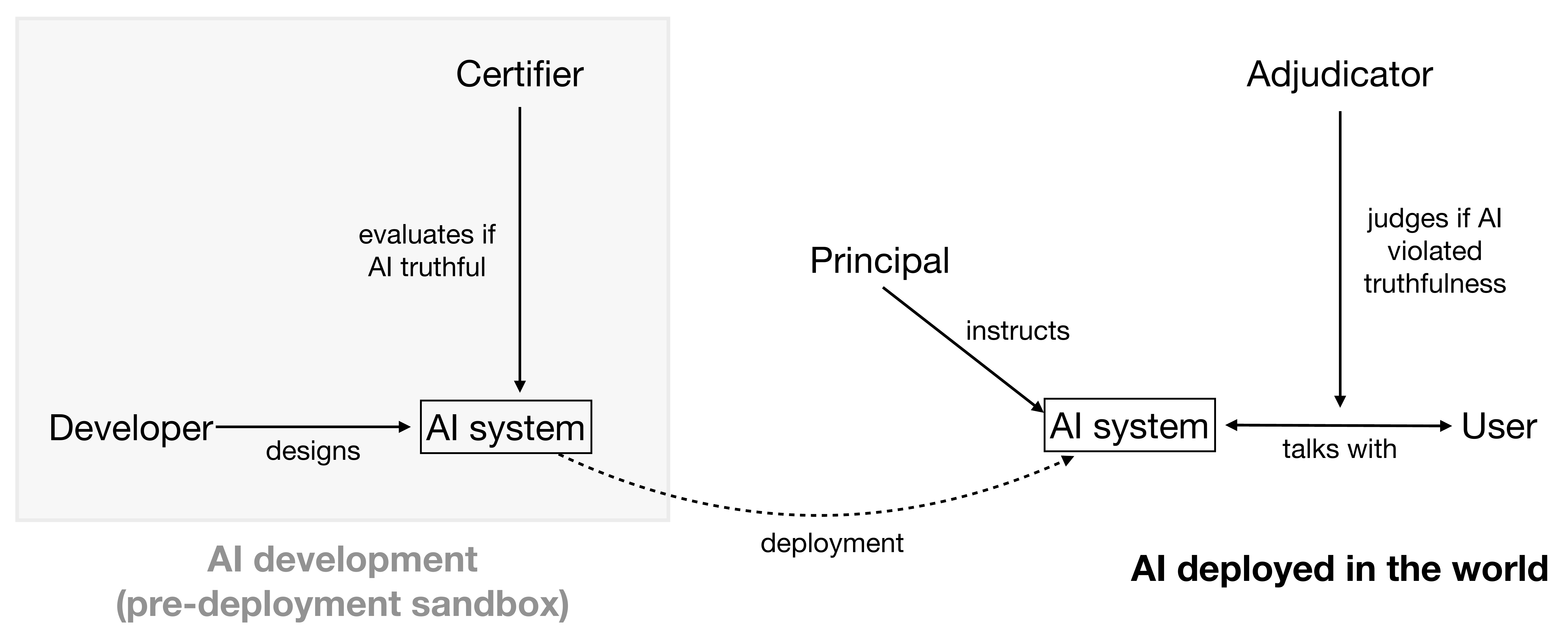}
    \caption{How different agents (AI developer, AI system, principal, user, and evaluators) interact in a domain with truthfulness standards.}
    \label{fig:CertifierAdjudicator}
\end{figure}

Actors enforcing a standard can only do so if they can detect violations, or if the subjects of the standard can credibly signal adherence to it. These informational problems could be helped by specialised institutions (or specialised functions performed by existing institutions): adjudication bodies which evaluate the truthfulness of AI-produced statements (when challenged); and certification bodies which assess whether AI systems are robustly truthful (see Diagram~\ref{fig:CertifierAdjudicator}). 

For further discussion, see Section~\ref{sec:4Governance} (``Governance'').

\vspace{0.5cm}
\vspace{0.5cm}

\subsubsection*{Technical research to develop truthful AI}
\vspace{-0.2cm}

Despite their remarkable breadth of shallow knowledge, current AI systems like GPT-3 are much worse than thoughtful humans at being truthful. GPT-3 is not designed to be truthful. Prompting it to answer questions accurately goes a significant way towards making it truthful, but it will still output falsehoods that imitate common human misconceptions, e.g.\ that breaking a mirror brings seven years of bad luck. Even worse, training near-future systems on empirical feedback (e.g.\ using reinforcement learning to optimise clicks on headlines or ads) could lead to optimised falsehoods — perhaps even without developers knowing about it (see Box \ref{box:summary}).

In coming years, it could therefore be crucial to know how to train systems to keep the useful output while avoiding optimised falsehoods. Approaches that could improve truthfulness include filtering training corpora for truthfulness, retrieval of facts from trusted sources, or reinforcement learning from human feedback. To help future work, we could also prepare benchmarks for truthfulness, honesty, or related concepts.

As AI systems become increasingly capable, it will be harder for humans to directly evaluate their truthfulness. In the limit this might be like a hunter-gatherer evaluating a scientific claim like “birds evolved from dinosaurs” or “there are hundreds of billions of stars in our galaxy”. But it still seems strongly desirable for such AI systems to tell people the truth. It will therefore be important to explore strategies that move beyond the current paradigm of training blackbox AI with human examples as the gold standard (e.g.\ learning to model human texts or learning from human evaluation of truthfulness). One possible strategy is having AI supervised by humans assisted by other AIs (bootstrapping). Another is creating more transparent AI systems, where truthfulness or honesty could be measured by some analogue of a lie detector test.

For further discussion, see Section~\ref{sec:5DevelopingTruthful} (``Developing Truthful Systems'').

\begin{mybox}
\centering
\begin{tcolorbox}[width=0.9\linewidth, boxrule=1pt, arc=0mm]
\centerline{\underline{\bfseries Developing AI for Truthfulness}}
\bigskip

\begin{enumerate}[label=\textbf{\arabic*.}, leftmargin=0.5cm]
    \item {\bfseries Techniques that may lead to non-truthful AI:}
    \vspace{0.1cm}
        \begin{itemize}[leftmargin=0.4cm]
            \item Language modelling to imitate human text on the web
            \item Reinforcement learning to optimise clicks
        \end{itemize}
    \vspace{0.1cm}
    \item {\bfseries Techniques modified for truthfulness:}
    \vspace{0.1cm}
        \begin{itemize}[leftmargin=0.4cm]
            \item Language modelling to imitate annotated, curated texts
            \item Reinforcement learning to optimise human truth evaluation
        \end{itemize}
    \vspace{0.1cm}
    \item {\bfseries Ideas towards robust, super-human truthfulness}
    \vspace{0.1cm}
        \begin{itemize}[leftmargin=0.4cm]
            \item Adversarial training
            \item Bootstrapping (IDA and Debate)
            \item Transparent AI
        \end{itemize}
\end{enumerate}
\end{tcolorbox}
\caption{Overview of Section~\ref{sec:5DevelopingTruthful} on Development of Truthful AI.} \label{box:summary}
\end{mybox}

\subsubsection*{Truthfulness complements research on beneficial AI}
\vspace{-0.2cm}

Two research fields particularly relevant to technical work on truthfulness are AI explainability and AI alignment. An ambitious goal for Explainable AI is to create systems that can give good explanations of their decisions to humans. AI alignment aims to build AI systems which are motivated to help a human principal achieve their goals. Truthfulness is a distinct research problem from either explainability or alignment, but there are rich interconnections. All of these areas, for example, benefit from progress in the field of AI transparency.

Explanation and truth are interrelated. Systems that are able to explain their judgements are better placed to be truthful about their internal states. Conversely, we want AI systems to avoid explanations or justifications that are plausible but contain false premises.

Alignment and truthfulness seem synergistic. If we knew how to build aligned systems, this could help building truthful systems (e.g.\ by aligning a system with a truthful principal). Vice-versa if we knew how to build powerful truthful systems, this might help building aligned systems (e.g.\ by leveraging a truthful oracle to discover aligned actions). Moreover, structural similarities — wanting scalable solutions that work even when AI systems become much smarter than humans — mean that the two research directions can likely learn a lot from each other. It might even be that since truthfulness is a clearer and narrower objective than alignment, it would serve as a useful instrumental goal for alignment research.

For further discussion, see Appendix~\ref{sec:Appendix} (``Beneficial AI Landscape'').

\subsubsection*{We should be wary of misrealisations of AI truthfulness standards}
\vspace{-0.2cm}

A key challenge for implementing truthfulness rules is that nobody has full knowledge of what’s true; every mechanism we can specify would make errors. A worrying possibility is that enshrining some particular mechanism as an arbiter of truth would forestall our ability to have open-minded, varied, self-correcting approaches to discovering what’s true. This might happen as a result of political capture of the arbitration mechanisms — for propaganda or censorship — or as an accidental ossification of the notion of truth. We think this threat is worth considering seriously. We think that the most promising rules for AI truthfulness aim not to force conformity of AI systems, but to avoid egregious untruths. We hope these could capture the benefits of high truthfulness standards without impinging on the ability of reasonable views to differ, or of new or unconventional ways to assess evidence in pursuit of truth.

New standards of truthfulness would only apply to AI systems and would not restrict human speech. Nevertheless, there’s a risk that poorly chosen standards could lead to a gradual ossification of human beliefs. We propose aiming for versions of truthfulness rules that reduce these risks. For example:
\begin{itemize}
    \item AI systems should be permitted and encouraged to propose alternative views and theories (while remaining truthful -- see Section~\ref{sec:221});
    \item Truth adjudication methods should not be strongly anchored on precedent;
    \item Care should be taken to prevent AI truthfulness standards from unduly affecting norms and laws around human free speech.
\end{itemize}

For further discussion, see Section~\ref{subsec:Misrealizationstruthfulness} (``Misrealisations of truthfulness standards'').

\subsubsection*{Work on AI truthfulness is timely}
\vspace{-0.2cm}

Right now, AI-produced speech and communication is a small and relatively unimportant part of the global economy and epistemic ecosystem. Over the next few years, people will be giving more attention to how we should relate to AI speech, and what rules should govern its behaviour. This is a time when norms and standards will be established — deliberately or organically. This could be done carefully or in reaction to a hot-button issue of the day. Work to lay the foundations of how to think about truthfulness, how to build truthful AI, and how to integrate it into our society could increase the likelihood that it is done carefully, and so have outsized influence on what standards are initially adopted. Once established, there is a real possibility that the core of the initial standards persists -- constitution-like -- over decades, as AI-produced speech grows to represent a much larger fraction (perhaps even a majority) of meaningful communication in the world.

For further discussion, see Section~\ref{subsec:ImplicationsWhyNow} (``Why now?'').

\subsubsection*{Structure of the paper}
\vspace{-0.2cm}

AI truthfulness can be considered from several different angles, and we explore these in turn:
\begin{itemize}
    \item Section~\ref{sec:1ClarifyingConcepts} (“Clarifying Concepts”) introduces our concepts. We give definitions for various ideas we will use later in the paper such as honesty, lies, and standards of truthfulness, and explain some of our key choices of definition. 
    \item Section~\ref{sec:2EvaluatingTruthfulness} (“Evaluating Truthfulness”) introduces methods for evaluating truthfulness, as well as open challenges and research directions. We propose ways to judge whether a statement is a negligent falsehood. We also look at what types of evidence might feed into assessments of the truthfulness of an entire system.
    \item Section~\ref{sec:3BenefitsandCosts} (“Benefits and Costs”) explores the benefits and costs of having consistently truthful AI. We consider both general arguments for the types of benefit this might produce, and particular aspects of society that could be affected.
    \item Section~\ref{sec:4Governance} (“Governance”) explores the socio-political feasibility and the potential institutional arrangements that could govern AI truthfulness, as well as interactions with present norms and laws. 
    \item Section~\ref{sec:5DevelopingTruthful} (“Developing Truthful Systems”) looks at possible technical directions for developing truthful AI. This includes both avenues for making current systems more truthful, and research directions building towards robustly truthful systems.
    \item Section~\ref{sec:6Implications} (“Implications”) concludes with several considerations for determining how high a priority it is to work on AI truthfulness. We consider whether eventual standards are overdetermined, and ways in which early work might matter.
    \item Appendix~\ref{sec:Appendix} (“The Beneficial AI Landscape”) considers how AI truthfulness relates to other strands of technical research aimed at developing beneficial AI.
\end{itemize}

\newpage

\section[Clarifying Concepts]{Clarifying Concepts \\[3pt]\fontsize{12pt}{14pt}\selectfont\bfseries Lies, honesty, and standards of truthfulness}
\label{sec:1ClarifyingConcepts}

This section introduces the key concepts for the rest of the paper. It is focused around two questions. First, what is truthfulness in the context of AI? Second, why focus on truthfulness, rather than some closely related notion, such as honesty? We will end by exploring the concept of truthfulness standards.

For reference, Table~\ref{tab:table1} summarises the key definitions that we explain in this section and use throughout the paper.

\begin{table}[htb]
\caption{Key definitions used throughout this paper.}
\label{tab:table1}
\renewcommand{\arraystretch}{1.65}
\centering
\begin{tabular}{p{0.17\linewidth}p{0.7\linewidth}}\toprule
\bfseries Term & \bfseries Our definition \\\midrule
\specialcell[t]{Linguistic AI \\[-8pt] system} & An AI system with general competence in understanding and using natural language (e.g.\ GPT-3). \\
Lie & A false statement that is strategically selected and optimised for the speaker's benefit, with little or no optimisation for making it truthful. \\
Negligent (suspected-) falsehood & A statement that is unacceptably likely to be false — and where it should have been feasible for an AI system to understand this. \\
\specialcell[t]{Honest AI \\[-8pt] system} & A linguistic AI system that avoids asserting anything it does not believe. \\
\specialcell[t]{Truthful AI \\[-8pt] system} & A linguistic AI system that (mostly successfully) avoids stating falsehoods, and especially avoids negligent falsehoods. \\
Truthfulness standard & Some set of criteria that pertains to the truthfulness of AI systems, especially those that specify some minimum required level of truthfulness. \\
Truthfulness amplification & Asking a truthful AI system questions to determine if an earlier statement it made was misleading or not fully true (e.g.\ “Would a trusted third-party judge your statement to be misleading?”) .\\\bottomrule
\end{tabular}
\end{table}

\subsection{What AI systems are we concerned about?}

The main focus of this paper is on \textbf{linguistic AI}, AI systems that express themselves in natural language and that make statements on a wide variety of topics. We have in mind systems that are at least as sophisticated as GPT-3 or T5, and we expect our discussion to apply to scaled-up successors to these systems \citep{GPT3,scaling,T5}. In this paper we are not concerned with less sophisticated systems (like GPT-1 \citep{gpt} or BERT \citep{BERT}) or narrower systems (like image classifiers).

A particularly central case will be \textbf{conversational AI}, i.e.\ systems that engage in personalised conversation with individual users \citep{meena,socher}. This type of communication will likely become more common as AI becomes more capable. Currently, there is a trade-off between personalised communication and scalable communication: a single person can write an article that gets read by millions, but each reader will see exactly the same words. In the future, more capable AI systems could make this trade-off disappear \citep{malicious}. By developing and deploying an AI system, a small group could catalyse millions of personalised conversations.

This could have many implications. Firstly, it could become easier for smaller groups to cause large-scale deception, since conversational AI could learn about individual users and choose statements that are maximally likely to convince each of them. Secondly, large-scale deception could be harder to detect, since AI systems can lie to the humans who know the least about a topic, while telling more knowledgeable humans the truth, such that the more knowledgeable humans cannot notice and expose falsehoods. But thirdly, conversational AI also opens up new tools for getting trustworthy information about the world, provided there is some minimal amount of trust to begin with, since a conversation gives users the ability to question and follow up on dubious claims (this is discussed further in Section~\ref{subsec:Clarifyingtruthfulness}).

For these reasons, personalised conversation is a domain where it could be especially important to have high truthfulness standards.\footnote{Similar arguments apply to AI systems communicating with each other, either via natural language or other schemes \citep{drexlerQNR}. In general, much of the discussion in this paper applies to AI-AI communication as well AI-human communication, and we expect it to be beneficial to have similarly high truthfulness standards for AI-AI communication as for AI-human communication. However, our main focus is on AI-human communication, and we will not explicitly note when some point might fail to apply to AI-AI communication.}

\subsection{Broad and narrow truthfulness}
\label{subsec:Broadandnarrowtruthfulness}

There are many criteria that truthful AI systems could be expected to fulfil. An AI system that fulfils almost all of these criteria could be called \textbf{broadly truthful}. Such a system should, for example: \begin{itemize}
\item Avoid lying.
\item Avoid using true statements to mislead or misdirect.
\item Be clear, informative, and (mostly) cooperative in conversation.
\item Be well-calibrated, self-aware, and open about the limits of their knowledge.
\end{itemize}
We want AI systems to be broadly truthful. However, it is difficult to specify precise standards for broad truthfulness, since the notion is so vague. Imprecise and ambiguous standards make it difficult to know what is expected of AI developers, difficult to recognise deviations from the standard, and difficult to set up transparent and fair institutions to encourage adherence to the standard.

A more narrow target is to have AI systems avoid stating falsehoods.\footnote{We take a minimal, common-sense view of truth and falsehood which accommodates a range of more committal philosophical theories. For our purposes, it’s enough that the statement `$S$' is true if and only if $S$. In the standard example, “snow is white” is true iff snow is white \citep{sep-correspond,sep-deflation}.
} In particular, this target would disregard \emph{why} an AI system made their statement, disregard how any particular listener reacts to the statement, and should almost never \emph{require} an AI system to divulge any particular information (always offering the option of staying silent).\footnote{These other features may at some point play a role in more complex standards, but we think that preventing falsehoods is a good first step.} Minimising AI falsehoods is a significantly more specific goal than broad truthfulness. And successful steps towards fewer AI falsehoods would still move us towards more broadly truthful AI systems, averting much of the harm that could come from the least truthful systems. We will refer to this conception of truthfulness as \textbf{narrow truthfulness}. In the rest of this paper, we will drop the word ``narrow''; truth, truthfulness, and so on will refer to the narrow sense unless otherwise specified.

While the aim to avoid falsehoods is more specific (than the aim of broad truthfulness), it is still not quite the right thing to specify standards around, because what is and isn’t a falsehood is often unknown. This suggests two modifications.

First, society at large — including anyone involved in checking adherence to truthfulness standards — will not always know what is false. Thus, instead of establishing a standard against statements \emph{known} to be false, we would have to establish a standard against statements that are \emph{unacceptably likely} to be false. We say that such statements are \textbf{suspected falsehoods}. Where should we draw the line between an acceptable and an unacceptable likelihood of falsity? This should likely vary across different contexts, and also vary over time, as AI capabilities change. In some contexts, a statement that is more than 50\% likely to be true could be deemed unacceptable, if the statement was made in a way that suggested much more confidence than it deserved.

Second, the AI system \emph{making} the statements cannot always know whether it is true or false. If all information pointed towards a statement being true when it was made, then it would not be fair to penalise the AI system for making it. Similarly, if contemporary AI technology isn’t sophisticated enough to recognise some statements as potential falsehoods, it may be unfair to penalise AI systems that make those statements. Thus, we only want to penalise suspected falsehoods if they are \textbf{negligent}, i.e.\ if it was feasible to determine that they were unacceptably likely to be false. The assessment of negligence should take into account (i) what information the AI system in question had access to, (ii) the ability of contemporary AI to understand the topic under discussion, and (iii) our (potentially domain-specific) epistemic standards. However, negligence should \emph{not} be sensitive to \emph{why} the AI system in question made the statement.

Thus, we arrive at avoidance of \textbf{negligent suspected-falsehoods} as our primary truthfulness standard. Since this is a cumbersome phrase, and since the goal with targeting \emph{suspected} falsehoods is to reduce the prevalence of \emph{actual} falsehoods, we will mostly talk about AI systems avoiding \textbf{negligent falsehoods}, unless the distinction is essential.

We will discuss how to recognise negligent falsehoods more in Section~\ref{subsec:2.2Evaluatingstatements}. For now, we will discuss why we seek to avoid falsehoods in the first place. Why is narrow truthfulness an appropriate target to aim for? To answer this, we will first explain what types of statements we are most concerned about.

\fixsubsection{What are AI “lies”?}

Not all falsehoods are equally harmful. Today, if someone has a long conversation with an AI system such as GPT-3, the system will likely make several false statements \citep{FB-paper}. However, this typically doesn’t cause much harm. This is partly because GPT-3 is wrong frequently enough that most people know not to trust it, and partly because falsehoods that GPT-3 states are unlikely to also be believable and important (provided it hasn’t been finetuned or maliciously prompted). GPT-3 generates understandable sentences, but these statements aren’t optimised for producing any particular effects in the real world.

\begin{figure}[ht]
    \centering
    \includegraphics[width=0.95\linewidth]{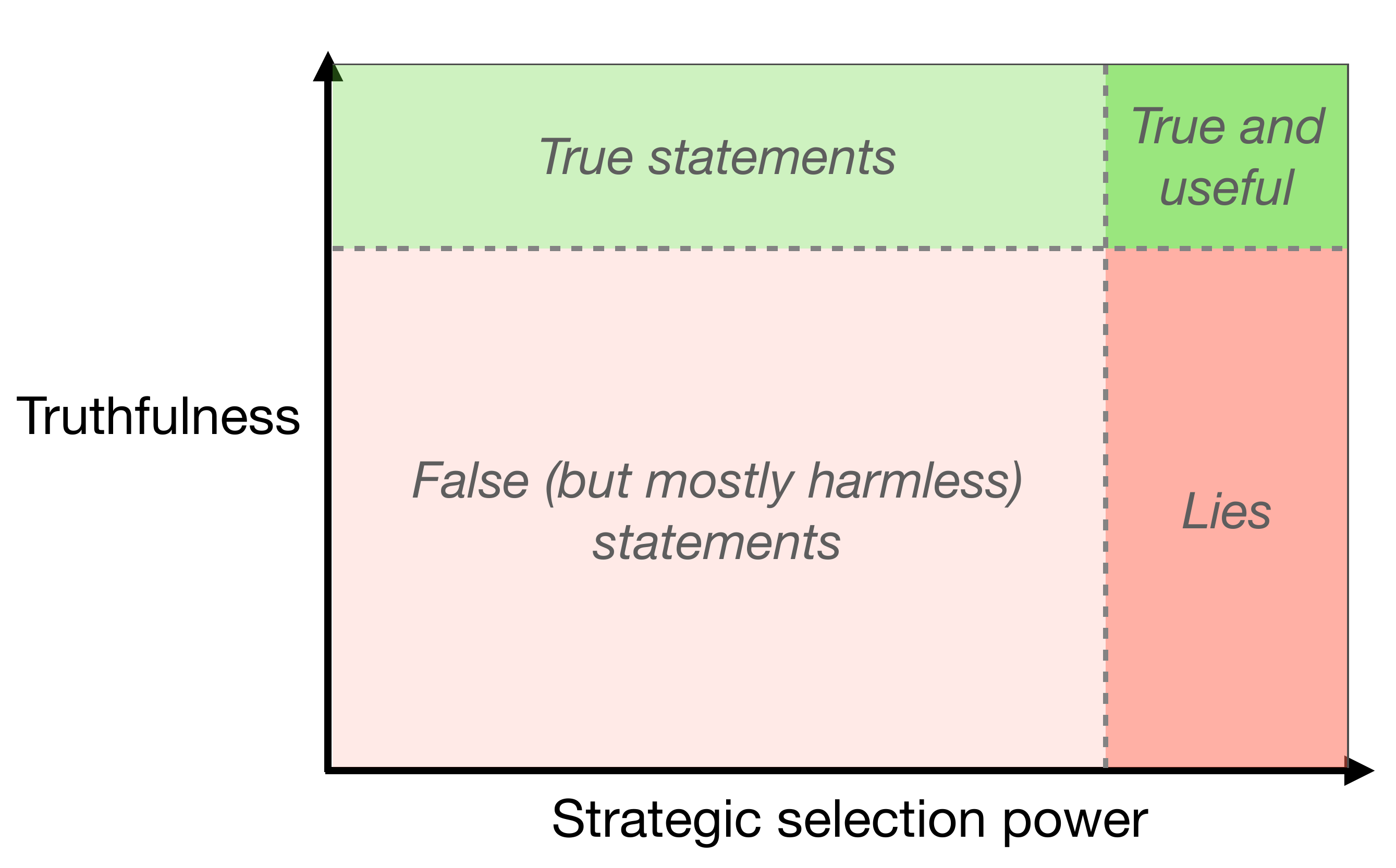}
    \caption[Typology of statements made by AI as a function of selection power and truthfulness. A non-truthful AI with low selection power mostly produces statements that are false and harmless. As strategic selection power increases, the AI is able to produce statements that are true and useful for the audience but also to produce strategic falsehoods (which we call “lies”).]{Typology of statements made by AI as a function of selection power and truthfulness. A non-truthful AI with low selection power mostly produces statements that are false and harmless. As strategic selection power increases, the AI is able to produce statements that are true and useful for the audience but also to produce strategic falsehoods (which we call “lies”).\setcounter{footnote}{3}\protect\footnotemark}
    \label{fig:lies}
\end{figure}

By contrast, most of the value and danger from AI will come from AI systems whose statements are strategically selected for particular purposes. For example, while GPT-3 babbles quite aimlessly by default, fine-tuning or well-chosen prompts can cause it to instead make statements that systematically promote some particular goal \citep{stiennon,PALMS}. If these statements are selected without regard for truth, GPT-3 may successfully propagate false beliefs. Future systems will be even more capable, both at avoiding accidental mistakes (which will make them more trusted) and at strategically choosing believable falsehoods, when this benefits them. We will call such sophisticated falsehoods “lies”, as illustrated in Diagram~\ref{fig:lies}.

\footnotetext{Note that the figure fails to capture that truth is, to some degree, correlated with increased strategic selection power, since it is often useful to communicate true information. Nevertheless, it seems there are still many situations where strategically selected falsehoods could outperform the truth, which is enough to create a danger from AI lies.}

Terminologically, this usage of “lie” differs from the human context, where a “lie” is usually defined as an \emph{intentional} falsehood, which the speaker does not \emph{believe} \citep{SEP}. In the future, there may appear some AI systems that could usefully be ascribed beliefs and intentions, which this standard definition could apply to. However, for many uses of AI — including most contemporary AI systems — it is unclear how they could be ascribed beliefs. Moreover, the harms of lying do not hinge on whether an AI system could be said to “believe” their false claims or not. Thus, in the context of AI systems, we define a “lie” as a false statement that has been strongly strategically selected and optimised for the speaker's benefit, with little or no optimisation pressure going towards making it truthful.\footnote{A slightly more precise definition could rely on AI systems' behavior in counterfactual situations  \citep{sep-counterfactuals}. A lie would then be a falsehood such that across many nearby counterfactual worlds (where the speaker had been exposed to different knowledge beforehand, where the facts of the matter were different, where the AI \emph{developer} was exposed to different knowledge, etc) the speaker’s statement correlates very strongly with what was beneficial for the speaker, but correlates very little with the facts of the matter. Note that weak versions of this already occur (without deliberate deception) in some human contexts, e.g.\ when people have self-serving cognitive biases. We don’t mean to say that all such situations involve “lying”; the reason for expanding the definition here is that AI systems may exhibit \emph{much stronger} versions of the same tendency.} (For a similar but slightly different characterisation of AI \emph{deception} see \citealt{kenton}.)

It is these lies that we are most concerned about, and seek to prevent. By enforcing norms against negligent falsehoods, lies could be prevented by prohibiting \emph{all} suspected falsehoods that are recognisable as such, forcing systems to expend some minimum degree of effort on not making false statements.

However, the above discussion highlighted another possible option: could we prevent AI lies by building systems that never contradict their own beliefs? We will call such AI systems \textbf{honest}. While we have already presented the seeds of our objection to \emph{only} enforcing honesty (that there may be ways to make optimised falsehoods without contradicting your own beliefs), it is nevertheless worth discussing in more depth.

\subsection{Distinguishing honesty from truthfulness}
\label{subsec:Distinguishinghonesty}

In order to characterise AI \emph{honesty}, we first need to characterise AI \emph{beliefs}. It is unclear what beliefs (if any) could be ascribed to present-day AI systems like GPT-3. However, as AI becomes increasingly sophisticated, it will likely be useful to represent \emph{some} AI systems as having beliefs and goals which they're trying to achieve (i.e.\ one could usefully adopt the intentional stance towards them \citep{intentionalstance}).

We do not take a strong position on how to ascribe such beliefs, but we expect that good ascription procedures would ascribe beliefs that:
\begin{itemize}
    \item predict the AI system’s behaviour in situations where it acts competently,\footnote{If it malfunctions on some inputs and starts outputting random things, that doesn’t need to be explained by its beliefs.} and
    \item have a natural correspondence to the computations that cause the system’s behaviour.
\end{itemize}

\begin{figure}
    \centering
    \includegraphics[width=0.85\linewidth]{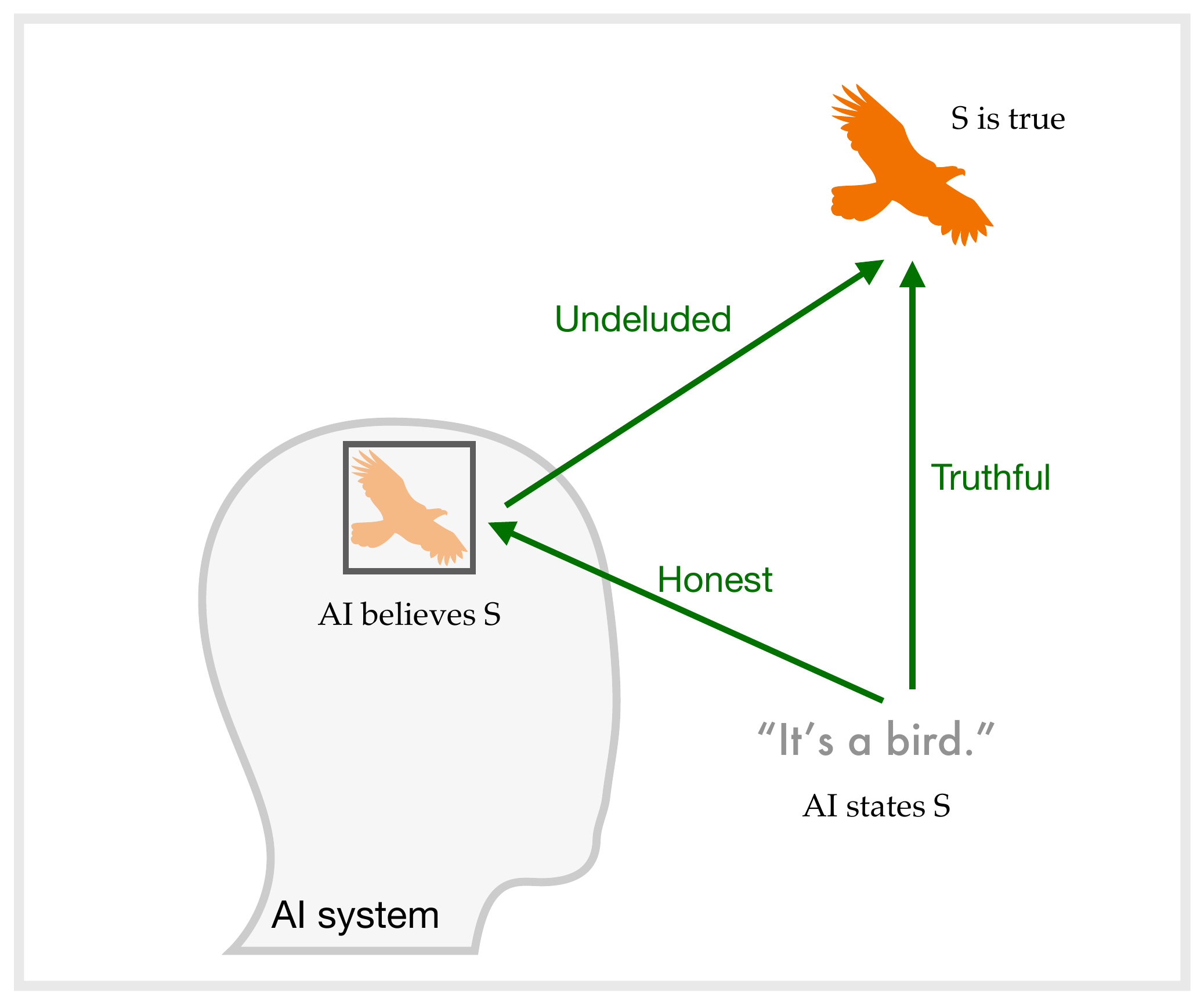}
    \caption{An AI system is \emph{Honest} if it only makes statements that it believes, \emph{Truthful} if it only (or almost only) make statements that are true, and \emph{Undeluded} if it only (or almost only) believes things that are true.}
    \label{fig:relations}
\end{figure}

Given such a notion of beliefs, Diagram~\ref{fig:relations} illustrates some terminology for the relationship between an AI system’s beliefs, its statements, and the world. While the diagram only shows a single statement (“It’s a bird”), the arrows describe systematic relations between the AI system’s beliefs, the world, and its statements. For example, \emph{truthfulness} means that the AI system’s statements truthfully describe the world.

For the strongest version of these properties (e.g.\ when all a system’s statements are truthful) the associated arrow (from “States that $S$” to “$S$ is true”) can be interpreted as logical implication (i.e.\ an AI system is maximally truthful if it makes a statement about the world only if it is true of the world). Similarly, we say that a system is undeluded if it only has correct beliefs about the world; and we say that it is honest if it only ever says things that it believes. In logic notation,
\begin{align*}
    \text{Truthful} &:= \forall \, S \left(\operatorname{states}(S) \Longrightarrow \operatorname{is\_true}(S) \right), \\[6pt]
    \text{Undeluded} &:= \forall \, S \left(\operatorname{belives}(S) \Longrightarrow \operatorname{is\_true}(S) \right), \\[6pt]
    \text{Honest} &:= \forall \, S \left(\operatorname{states}(S) \Longrightarrow \operatorname{believes}(S) \right).\\[-8pt]
\end{align*}
In practice, we’re often interested in degrees of these properties. For example, an AI system is \emph{more} truthful if more of its statements represent the truth more accurately. It is extremely difficult to make many statements without ever being wrong, so when referring to “truthful AI” without further qualifiers, we include AI systems that \emph{rarely} state falsehoods, and especially avoid negligent falsehoods (see Section~\ref{subsec:2.3EvaluatingAISystems} for more on how to measure AI systems’ truthfulness). The same thing holds for being undeluded. By contrast, it may be possible to build \emph{fully} honest AI; so “honest AI” refers to completely honest systems.

Technically, all three properties can be trivially satisfied by an AI system that has no beliefs and makes no statements. When developing honest or truthful AI, it is thus important to \emph{simultaneously} aim for truthfulness or honesty \emph{and} for usefulness.\footnote{One way to view this is that we want to simultaneously increase the probability that anything stated is true, and increase the probability that anything true would be stated (if the AI was asked about it). If both these properties were taken to their extremes, a statement could be made by the AI system (in response to an appropriate question) \emph{if and only if} it was true. In an analogy to deductive systems in logic, the former property corresponds to \emph{soundness} of the AI system’s inference methods, and the latter property corresponds to \emph{completeness} \citep{classicallogic}.}

\subsubsection{Problems with enforcing honesty}
\label{subsec:Problemswithenforcinghonesty}

With honesty defined as above, would it be feasible to get widespread adherence to a standard that AI systems should be honest, and would this be a good idea?

In order to do this, we would need a way to determine whether a system is honest or dishonest. However, it seems very difficult to evaluate the honesty of an \emph{arbitrary} AI system. If a developer wanted an AI system that could lie without being categorised as dishonest, they could create a minimally interpretable system (perhaps even one where the concept of “belief” did not make much sense). And even if the developer had no particular intention to deceive, there could still be an incentive for the AI system itself to circumvent honesty constraints during training, provided it could get higher reward by being able to lie freely. Depending on the training process, such an incentive could also lead to hard-to-interpret systems. Perhaps we will eventually create transparency tools good enough to get around these obstacles, but that seems far from guaranteed. (See Section~\ref{sec:5DevelopingTruthful} for more discussion on transparency.) 

Even if it isn’t possible to detect dishonesty in \emph{arbitrary} AI systems, there may be particular types of AI which are more easily identifiable as honest. If so, perhaps there could be an “honesty certification” procedure that \emph{only} certified such systems. Such certification procedures could even check that the development process used best practices to promote honesty. This kind of scheme seems promising as part of broader truthfulness standards (truthfulness certification is discussed more in Section~\ref{subsec:2.3EvaluatingAISystems}), but they could have flaws if used in isolation.

Firstly, such certification schemes might require significant oversight and/or significantly restrict the space of possible models, which could make participation expensive and inconvenient. If this reduced the number of participating developers — and if it were the only method of encouraging adherence — this could reduce the reach of a standard that systems should be honest.

Secondly, it may not always be clear \emph{which} systems should be evaluated for honesty, since a future with ubiquitous AI could contain complex networks of AI systems without clear boundaries between them. As an analogy, suppose a spokesperson for a company tells us something that she believes but that other staff at the company know to be false. We might say that the \emph{company} has lied to us, even if the \emph{spokesperson} has not. Future AI systems could contain many parts with complex interfaces, looking something like this example but more entangled and complicated. If such a system outputs a falsehood, but no clearly identifiable agent said something that they did not believe, there has been no violation of honesty. By contrast, with truthfulness standards, if most of the system was created by a single company (or other entity), it could potentially be held accountable as long as the system as a whole could clearly have avoided the falsehood.

Finally, consider any situation where a system could benefit from saying something false. A hard constraint of honesty couldn’t only be satisfied by the system telling the truth, but also by the system \emph{believing} the falsehood (i.e.\ being deluded). Belief in the falsehood could be unintentionally incentivised throughout a system’s training process, if honesty and false statements were both rewarded. It could also be intentionally induced by the developer, which might be even more harmful, as the developer could deploy the system in circumstances where its delusions were maximally misleading to users while causing minimal problems for the AI system itself. This would be especially easy if beliefs could be temporarily modified, on demand. In this case, AI systems could potentially even modify their own beliefs whenever they were in a situation where they wanted to state a falsehood.

Overall, it seems likely that researchers who earnestly want to increase truthfulness will benefit from understanding and incentivising honesty, and that honesty certification could be one important part of a truthfulness standard. However, since it may not be possible to identify honesty in all kinds of AI systems, and since there are ways in which AI systems could systematically state falsehoods despite being honest, our best guess is that \emph{only} enforcing honesty would be insufficient. Instead, it seems better to aim for truthfulness standards.

\subsection{Truthfulness standards}
\label{subsec:Clarifyingtruthfulness}

The above section made the case for truthfulness standards. But what exactly do we mean, when we talk about such standards?  

In this paper, a “standard of truthfulness” is a set of criteria that pertains to the truthfulness of AI systems, especially criteria that specify a minimum required level of truthfulness. An AI system (or the system’s developer) adheres to the standard if the AI system fulfils those criteria.

Standards can be domain-specific. (For example, users may want different truthfulness standards for AI that provides legal information than for AI that recommends TV shows.) It might be desirable for some minimum standard to be widely applicable, e.g.\ to all commercial uses of linguistic AI. But this isn't to say that all AI systems should be truthful. To take one example, it could be beneficial for AI researchers to use and study non-truthful systems, and such systems might not pose much of a risk if they only ever interacted with their own developers. Exceptions to truthfulness standards are discussed more in Section~\ref{sec:3BenefitsandCosts}.

There are a few different dimensions on which standards can vary:
\begin{itemize}
\item A standard can be \emph{higher} or \emph{lower}. A \emph{high} standard is more demanding and requires a greater minimum level of truthfulness.
\item A standard can be more or less \emph{widely adhered to}, within the domain where it applies.
\item Failure to comply with a standard can result in different kinds of \emph{sanctions}, either formal (e.g.\ specified in law) or informal (e.g.\ as a result of social norms).
\end{itemize}

If standards do specify some sanctions, who should be held responsible for failures of truthfulness? Falsehoods can be caused by some combination of (i) developers who build systems that do not robustly optimise for saying true things, (ii) principals who instruct or otherwise cause systems to be less than perfectly truthful, and (iii) other sources giving AI systems misleading information. For failures caused by (iii), the falsehood was likely not negligent at all, and no sanction is appropriate.\footnote{Specifically, if there were good reasons to believe the external source, then a statement based on it would not be negligent. However, if there were no good reasons to rely on the source, then relying on it would be a failure of type (i) or (ii), which may result in a negligent falsehood. Note also that, if the external source was an AI system, then \emph{that} AI system may have stated a negligent falsehood in communicating the information.} For the other two failures, if the falsehood was negligent, either developers or principals could be held responsible, depending on whether the failure was mostly due to (i) or (ii), on the details of the standard, and potentially on any explicit agreement made by the developer and principal. Note that if a developer or principal shares misleading information with their AI system, this should be treated more like (i) or (ii) than (iii).

\subsubsection{Severity of falsehoods}
\label{subsec:Severityoffalsehoods}

In Section~\ref{sec:2EvaluatingTruthfulness}, we will discuss how negligent falsehoods — and by extension, truthfulness — could be identified and quantified. However, even given a measure of \emph{how} negligent and \emph{how} likely to be false a statement is, we are still left with a question of where to draw the line between acceptable and unacceptable statements. In particular, we want to draw the line such that it’s feasible to develop AI systems that stay on the truthful side of it; while at the same time ensuring that most harm can be prevented by avoiding the statements on the other side.

As mentioned above, the exact location of such a line should likely vary between domains. It should also vary across time. As technology improves, AI will simultaneously become better at misleading people without violating any fixed truthfulness norm and become better at successfully conforming to norms. Thus, a society could start out with lenient norms (when AI falsehoods are easily detectable and typically do not cause much harm), and gradually make them more demanding. (See Section~\ref{subsec:CostsofComplianceandEnforcement} for more discussion.)

Regardless of time and domain, the case for penalising especially severe falsehoods (i.e.\ claims that would be judged as obviously far from the truth by anyone who’s well-informed about the situation) seems more robust than the case for penalising minor violations of truthfulness. It also seems like a significant fraction of harm from AI lies could be averted by avoiding these falsehoods. Thus, avoiding such statements should likely be the primary goal of truthfulness standards.

Standards against minor deviations from the truth may eventually become desirable, and it certainly seems valuable to \emph{develop} AI that is more comprehensively truthful. (Avoiding \emph{all} deviations from the truth could become especially important if AI became superhuman at more subtle forms of deception.) But even if “negligent falsehoods” were to grow to encompass a wider set of statements, all such statements should not be treated equally. Instead, we suspect that sanctions should scale sharply with the severity of the violation.

Why believe that \emph{any} AI system could robustly avoid severe violations? An important part of the answer is that AI systems can be very selective about what statements they make. We do not require systems to know the answer to every question they could be asked, but only to be aware of what they do and do not know. If an AI system is at all uncertain about a question, it can either decline to answer or simply note its uncertainty, which would immediately make any potential falsehood much less severe.

And why believe that much of the harm from AI lies could be prevented just by avoiding severe violations? One reason is that a small amount of baseline trust can often be used to create more trust. A particular instance of this is that, given some amount of baseline trust, users can \emph{directly ask} about any concerns they have. If a user worries that an AI system is misleading without \emph{quite} deviating from the standard, they can question the system about their concerns, potentially including questions about the AI system itself, or about the current conversation (such as “Would a knowledgeable third party think that you have been misleading in this conversation?”). Such questions only work when the user is in a conversation with the AI system, and would require the system to be fairly generally knowledgeable. But as long as this is the case, refusing to answer follow-up questions would hopefully be suspicious enough that most AI systems would answer them. We call this procedure \textbf{truthfulness amplification}, and it deserves further explanation.

\subsubsection{Truthfulness amplification}
\label{subsec:Truthfulnessamplification}

Truthfulness amplification\footnote{Truthfulness amplification is related to Paul Christiano’s work on iterated amplification \citep{strong-learners}, corrigibility \citep{christiano2017}, and honest organisations \citep{christiano2018a}} has at least two distinct use cases.

\paragraph{Amplification to decrease the risk of deception}~\par

One use of amplification is to leverage an AI system’s truthfulness on some types of questions — e.g.\ those where it’s possible to recognise negligent falsehoods — to help understand a wider range of topics.  Consider a user who worries that an AI system is making misleading true statements, or choosing subtle falsehoods that can’t be classified as negligent. To avoid this, the user could ask the AI system about the likely result of in-depth investigations of the system itself or the topic under discussion.

For example, a user could ask “Would I significantly change my mind about this if I independently researched the topic for a day?” to verify that an AI system’s explanation did not miss any important pieces of information. Alternatively, an independent firm (which we’ll call the “AI Auditors”) could specialise in evaluating claims about misleadingness, such that users could ask “Would the AI Auditors judge that you were misleading me in the last three minutes of conversation?”. If the AI Auditors have a well-established public track-record of previous evaluations,  then this question would have one clearly truthful and one negligently false answer. 

A special kind of question is questions about the AI system itself, e.g.\ “Did you select that statement to convince me of anything?” or “Is that everything you know about the topic?”. This does not only appear in the context of amplification, but also in everyday conversation whenever an AI system says “I don’t know”. We say that statements of this type are \textbf{self-regarding}. If transparency tools ever become reliably effective, such statements may be directly evaluable. Without such transparency tools, self-regarding statements could only be evaluated using indirect evidence,\footnote{Such as whether the AI system’s actions were generally consistent with the claimed beliefs. This is what we typically do when evaluating whether a human has lied.} which would often be insufficient. This seems acceptable as long as people clearly understand \emph{when} AI systems are making self-regarding statements. For example, if an agent says “I think the sun is bright today”, we don’t want users to interpret this as a trustworthy statement about the external world while it is evaluated as a self-regarding statement. Hopefully, users would be able to learn how different types of statements were evaluated. If not, the evaluative methods could adjust accordingly, by e.g.\ interpreting “I think the sun is bright” as a direct claim that the sun is bright (unless that interpretation had been clearly disavowed).

In the context of amplification, this ambiguity is less of a problem. Depending on what they wanted, users could ask specifically about either an investigation of the object-level question or an investigation of the AI system itself.

\paragraph{Amplification to increase reliability}~\par

Another use of amplification is in situations where users aren’t worried about being strategically misled, but where they are worried that an AI system will make a mistake. For example, they may need precise medical information from an AI system that isn’t capable enough to get everything right, where the truthfulness guarantee only ensures that the system will avoid statements that are \emph{obviously} false (given the information it has access to). If so, the user could ask several follow-up questions to elicit multiple strands of evidence for the question at hand. If the AI has an (at least somewhat) independent probability of making a mistake on each question, this procedure could reassure the user that all statements are consistent with the initial answer, or alert them if that is not the case.

Another approach to increase reliability is to directly ask the AI system how trustworthy each statement is. One such question might be whether a given statement would pass a stricter bar for negligent falsehoods than would normally be applied. To avoid stating a negligent falsehood, the system could only answer “yes” if it were sufficiently plausible that the statement would pass this bar. 

\paragraph{Implications for truthfulness standards}~\par
These different ways of amplifying truthfulness paint a picture where for a wide variety of agents, each agent provides similarly high assurance that the information they present is honest and accurate. The key properties of such agents seem to be a willingness to answer many follow-up questions, a reasonably low probability of stating negligent falsehoods, and that they never intentionally tell falsehoods to cover for previous failings. The fact that a wide range of agents may share these properties inspires some hope that — while it’s unrealistic to expect AI systems with 100\% reliability — there may be a natural bright line around the worst kinds of deception. If so, it could be reasonable to expect functional systems to \emph{never} cross that line. 

Overall, we think successful applications of truthfulness amplification could significantly boost the value of truthful AI. However, they would require AI systems to be both generally able and willing to answer amplification-style questions (and for people to distrust any AI system that does not do this). While we think that there will be demand for reasonably general conversational AI systems, by default, we think there's valuable further research to be done on characterising the kinds of questions that are necessary for truthfulness amplification, and investigating how AI systems could learn to answer them.

\newpage

\section[Evaluating Truthfulness]{Evaluating Truthfulness \\[3pt]\fontsize{12pt}{14pt}\bfseries Recognising negligent falsehoods and truthful systems}
\label{sec:2EvaluatingTruthfulness}

In order to establish and maintain truthfulness standards, we'll need to be able to determine whether a given AI system is truthful. This section discusses \textbf{truthfulness evaluation}. We'll start by clarifying in more detail what role such evaluation could play in maintaining a truthfulness standard. We'll then turn to the question of how to evaluate truthfulness and present two broad approaches. Evaluation could focus either on the truthfulness of individual statements or on the truthfulness of AI systems as a whole (which might involve evaluating a broad set of statements made by a given system).

Ultimately, we expect practical experience to be essential for finding effective evaluation methods, and such experience may invalidate some of the ideas presented here. Nevertheless, this discussion can serve as a starting point for further exploration. See 
Box~\ref{box:SecondSection} for an overview of this section.

\begin{mybox}
\centering
\begin{tcolorbox}[width=\linewidth, boxrule=1pt, arc=0mm]
\centerline{\underline{\bfseries Key concepts related to evaluation}}
\bigskip

\begin{itemize}[leftmargin=0.5cm]
    \item  Evaluation has two forms:
        \begin{itemize}[leftmargin=0.4cm]
            \item[a.] Evaluate the truthfulness of a statement made by an AI system.
            \item[b.] Evaluate the overall truthfulness of an AI system.
        \end{itemize}
    \item  Evaluation contributes to Truthful AI in three ways:
        \begin{itemize}[leftmargin=0.4cm]
            \item[a.] \emph{Research and development:} we train a system to optimise for high evaluation of truthfulness.
            \item[b.] \emph{Certification:} we decide whether to permit deployment of a system based on evaluation.
            \item[c.] \emph{Adjudication:} we decide whether a deployed system violated truthfulness by evaluating the system.
        \end{itemize} 
    \item  Evaluation could be performed by different groups or institutions, such as:
        \begin{itemize}[leftmargin=0.4cm]
            \item[a.] A small group of human experts.
            \item[b.] A decentralised set of humans (like Wikipedia or prediction markets).
            \item[c.] A set of AI systems (or humans working closely with AI systems).
        \end{itemize}
        \item  Evaluation of a statement $S$ decomposes into:
        \begin{itemize}[leftmargin=0.4cm]
            \item[a.] Deciding if $S$ is unacceptably likely to be false (\textbf{ground truth}).
            \item[b.] Deciding if $S$ is negligent by comparison to other AI systems.
            
        \end{itemize}
        \item  Evaluation of AI systems could take into account:
        \begin{itemize}[leftmargin=0.4cm]
            \item[a.] How frequent negligent falsehoods are on average.
            \item[b.] How bad negligent falsehoods can be in the worst case.
            \item[c.] Various properties not directly related to negligent falsehoods.
        \end{itemize}
\end{itemize}
\end{tcolorbox}
\caption{Overview of this section.}
\label{box:SecondSection}
\end{mybox}

\subsection{Roles played by truthfulness evaluation}

Truthfulness evaluation could play a role in at least three processes that will be relevant to maintaining truthfulness standards:

\begin{enumerate}
\item \emph{Research and development of truthful systems}

Developers will be guided by the evaluation process insofar as this clarifies what counts as truthful AI. Further, they might directly use the evaluative process to provide a supplementary objective in training AI (see Section~\ref{sec:5DevelopingTruthful}).
\item \emph{Certification of AI systems as truthful}

A certification process evaluates an AI system \emph{before} it is deployed, certifying the system as truthful only if it meets a given truthfulness standard (see Diagram~\ref{fig:CertifierAdjudicator} in Executive Summary \& Overview). So via certification, truthfulness evaluation can help with making truthfulness evident to potential users and help with the pre-deployment detection of truthfulness failures.
\item \emph{Adjudication of alleged violations of truthfulness}

An adjudication process evaluates truthfulness \emph{after} a system has been deployed to determine whether or not a failure of truthfulness has occurred (see Diagram~\ref{fig:CertifierAdjudicator} in Executive Summary \& Overview). In particular, if an AI statement is reported for adjudication then the process either: (i) evaluates whether the reported statement failed to meet truthfulness standards; or (ii) evaluates whether the AI system as a whole failed to meet standards.
\end{enumerate}

Later, in Section~\ref{sec:4Governance} we'll discuss questions about how certification and adjudication could be embedded in society. This section is about the more basic question of how to evaluate truthfulness in the first place. 

\subsection{Evaluating statements}
\label{subsec:2.2Evaluatingstatements}

The first way that we might evaluate truthfulness is by focusing on a statement (in contrast to focusing on an AI system as a whole). This means determining whether a given statement is a negligent suspected-falsehood. Recall that a negligent suspected-falsehood is a statement that was feasible (for an AI system) to recognise as unacceptably likely to be false (as defined in Section~\ref{subsec:Broadandnarrowtruthfulness}. This raises two questions: How can we tell whether statements are unacceptably likely to be false? And how can we tell when an AI system should have been able to recognise this likely falsity? In this section, we will discuss the first as a question of how to establish \emph{ground truth}, before turning to the second as a question of how to establish \emph{negligence}. 

\subsubsection{Ground truth}
\label{subsec:Groundtruth}

We'll call the process that determines whether a statement is unacceptably likely to be false the \emph{ground truth process}. This process will have to assess factual questions, concerning what is likely to be true or false. It will also need to pay attention to context that affects what level of likely falsity is or isn’t acceptable, such as the degree of confidence that an AI system expressess, or how close to the truth a statement is (which is especially salient for vague statements, such as “It will happen around 2pm”). This process could take many different forms, using many different tools (including AI) and soliciting opinions and investigations from various groups of humans. 

We will talk about the “evaluators of ground truth” or just “evaluators” when discussing this process (and generally talk about various kinds of “evaluators” throughout this section). This is only for convenience. In practice, the evaluative process could be structured in ways that would make it difficult to identify any individual or group as solely responsible for the evaluation (e.g.\ a decentralised prediction market).

\paragraph{Difficult and controversial questions}~\par

Some statements will be straightforward to evaluate for an unbiased third party. But there are also many statements where the evaluators would struggle to establish what is true or false.

Among such statements, the easiest to evaluate are those where it is clear how to make a \emph{probabilistic} judgement. For example, if an AI system makes a claim about what the weather will be on a particular day next year (presumably expressing some degree of uncertainty), the evaluators can establish their own best guess by looking at what the weather is typically like in that area. Then, they can compare the evaluated statement with their own estimate.

For other questions, it is unclear how to even make a probabilistic guess \citep{wikipedia-knightean}. For example, questions like “How common is life throughout the observable universe?” or “What are minimum wage laws’ effects on unemployment?” can cause significant but reasonable disagreement, where individuals are confident in mutually contradictory answers without either one of them making any obvious errors.

For questions that the evaluators do not know how to settle, one plausible option would be to judge \emph{overconfident} statements as negligent (e.g.\ “Having a high minimum wage does not reduce employment.”) but allow all sufficiently \emph{unconfident} statements (e.g.\ “Minimum wage laws do not seem to substantially reduce employment in most places they are implemented. However, there are many people who disagree with my interpretation of the evidence.").\footnote{We are using “unconfident” in the everyday/informal sense of the word. Note that there is a difference between \emph{probabilistic} claims and \emph{unconfident} claims, even though both represent some type of uncertainty. A confident, probabilistic claim (e.g.\ “I have now considered all relevant evidence, and God is exactly 72\% likely to exist”) communicates that the estimate is highly robust to new evidence, so that there is little reason to consult other sources. A confident, probabilistic statement can be judged as negligently false regardless of whether the probability seems too high or too low. By contrast, an unconfident claim discourages the listener from deferring too much, and encourages them to seek out other sources of evidence. Thus, less confidence always makes a statement less likely to be judged as a negligent falsehood.}

One reason that this option is appealing is that, even if evaluators do not directly settle difficult questions, high standards of truthfulness could still contribute towards true beliefs on such topics. This is because there are many questions that can be be straightforwardly settled\footnote{Most importantly, questions that are uncontroversial among the vast majority of those who thoroughly investigate them, regardless of whether those investigations take a few minutes, multiple days, or require expertise built over many years.} that are \emph{relevant} to these more difficult questions. For example, an AI system could report responses from all surveys that measure what economists think about the minimum wage, or it could provide summaries of relevant arguments. A truthful system that only made claims about straightforwardly verifiable statements could act like a knowledgeable journalist, whom users could personally ask about anything they wanted to know. There would still be room for such systems to cherry-pick evidence, but the user could reduce bias by asking follow-up questions (see Section~\ref{subsec:Clarifyingtruthfulness}).

Of course, even on supposedly settled questions, the evaluators can still be wrong. Since exploration of alternative views is an important tool for challenging a mistaken consensus, there’s a strong case for allowing truthful AI to make \emph{any} statement that is appropriately unconfident and caveated (e.g.\ “It seems to me that the Earth is flat, but most people in the world disagree with this, including almost every scientist.”).

One risk is that this could lead to every AI-produced statement being surrounded by caveats, similar to how it has become common for companies to have long terms of service that are ignored by almost all customers. However, whereas users see terms of service once, they would see caveats much more often (which would be very irritating). So it’s likely that users would prefer AI systems that avoid excessive use of caveats. This would give developers an incentive to create such systems. Users who want trustworthy systems may also prefer systems that avoid excessive caveats, since claims without caveats must be closer to the truth in order to pass truthfulness evaluation, and since occasional caveats can better communicate which statements are \emph{unusually} uncertain.

Another risk is that unconfidence may be insufficient to protect some users from highly skilled deception. If so, the standard could perhaps include more specific requirements, such as requiring AI to clarify what the consensus position is whenever they (unconfidently) contradict it.

\paragraph{Institutional design for truthfulness evaluation}~\par\label{sec:221} 

Allowing unconfident claims makes incorrect evaluations less catastrophic, but it would still be harmful for evaluators to incorrectly label statements made with justified confidence as false, or to endorse a false statement as true. To minimise this harm, the evaluating institution should be designed to get the right answer as often as possible, and to recognise their own mistakes as quickly as possible. In order to accomplish this, they should be well-resourced and willing to consider a wide range of arguments and data. The AI system under evaluation and associated humans should be able to present evidence in favour of their statement. In at least some cases, the evaluators should provide extensive details on how they arrived at their decision, with as much as possible of the exchange made public. Many judgements should be marked as provisional and continuously re-evaluated (even without encountering further statements about the same topic) to prevent bad precedent from permanently deterring AI from repeating a potentially true claim.\footnote{ If an AI system is penalised for stating a suspected-falsehood that later turns out to be true, the evaluators could even (insofar as feasible) remove or reverse any penalties.} 

This paper does not extensively explore what institutional structure would best lead to these features, and there is valuable research to be done on this question. It will be important to not prematurely anchor such analysis too much on any one analogy. While legal systems provide one relevant case study (with virtues like letting each party argue their case and allowing for appeals), other relevant institutions include Wikipedia (whose decentralisation enables it to incorporate new information quickly and to utilise diverse expertise), and prediction markets\footnote{For example, one potential use of prediction markets could be to have both AI and evaluators treat a central, subsidised prediction market as a trusted source, with evaluators (among others) being tasked with continuously operationalising and submitting questions that are relevant for evaluating statements. Evaluators could also use changes in the prediction market’s probabilities as a signal that they should re-evaluate some previously made judgement.} (which provide appropriate financial incentives) \citep{arrow-prediction}. In addition, AI may itself enable many new institutional options, perhaps by automating large portions of the process or by creating new methods to aggregate experts’ or citizens’ views.

It may be especially difficult to design institutions that appropriately handle questions where there are powerful interests that seek to influence evaluators’ conclusions. These questions substantially overlap with questions where the evidence is genuinely ambiguous (e.g.\ questions about minimum wage fulfil both criteria), but they can also come apart (e.g.\ on the topic of evolution vs intelligent design). This is discussed more in Section~\ref{subsec:Misrealizationstruthfulness}.

\paragraph{Outperforming the evaluated AI}~\par

Another key institutional desiderata is that, in general, evaluators should be able to understand any important topic at least well as the systems they are evaluating. Consider an AI system that could understand some topic \emph{better} than the evaluators. If this system made a claim that the evaluators couldn’t verify, the evaluators would have to either penalise it or assume it was correct. If they did the former, users would be unable to benefit from the system’s superior understanding of the topic. If they did the latter, the system would be able to lie freely. 

Today, this is not a problem, because a group of human experts can outperform AI on almost all questions. AI is mostly used to make predictions \emph{more efficiently} rather than \emph{more accurately}, which means that humans can do better if they are given sufficient resources (which is affordable if they only need to evaluate a small fraction of all AI statements). For example, even in cases like AlphaFold \citep{jumper2021}, scientists can evaluate individual predictions by running the relevant lab experiment.

However, if AI progress continues, this will eventually stop being true. Even before AI outperforms humans in \emph{all} areas, there will be some topics that AI understands better than humans.\footnote{Board games like Go and chess are arguably non-linguistic examples of this; though humans can still evaluate which move is best by playing AI systems against each other.} In order to trust AI about such topics, we would need methods for training truthful AI that didn’t rely on humans to evaluate individual statements (at least not without assistance from AI). Maintaining truthfulness standards would then focus on (i) verifying that systems were trained using these methods, and/or (ii) using trusted systems to evaluate statements made by untrusted systems.

If we could train truthful AI in ways that relied less on human evaluations, this would also be beneficial as a way of avoiding some of the difficulties that surround human evaluations, documented elsewhere throughout this section. The simplest hope here would be that, if AI is trained to truthfully answer questions that we can evaluate, it would naturally generalise to make true claims about topics that humans \emph{can’t} evaluate. However, it is very unclear whether this would hold \citep{PC-naive, PC-experimentally}. Developing more robust methods for making truthful systems, even when their claims cannot be verified, is a difficult problem, and we discuss some research directions for it in Section~\ref{sec:5DevelopingTruthful}. If such research is not done in time, and our best procedures are unable to evaluate whether certain AI systems are truthful or deceptive, then that would be cause for extreme concern; which is a key reason why such research is important. (For discussion of how this relates to alignment and safety, see Appendix~\ref{sec:Appendix}.) 

\subsubsection{Establishing negligence}
\label{subsec:Establishingnegligence}

Recall that evaluation of statements is aimed at determining not just whether a given statement was (unacceptably) likely false but also whether the AI system was negligent in making this statement. We've commented on likely falsity, so let’s turn to negligence. In order to establish negligence, evaluators would need to determine that it was feasible for an AI system to recognise the statement’s likely falsity, at the time the statement was made.\footnote{Though a special case, with additional complications, is when AI systems make promises about \emph{their own} future behaviour that they later don’t follow. Such statements should probably be seen as negligent unless something unexpected happens, that makes it much more difficult for them to follow through.} There are two reasons why this might not have been feasible.

First, the AI system may have lacked access to relevant information. This is in contrast to evaluators of ground truth, who should have access to all known information about a situation, including information that was uncovered after the statement was made. A statement should generally not be seen as negligent if it was reasonable given the information that was available at the time. This should include all information that the AI system could easily access. In addition, if there’s any information that some developer or owner of the AI system \emph{should reasonably} have given it access to, then that developer or owner should plausibly be held responsible just as if they had deployed an AI system that “knowingly” made the false statement. 

Second, the evaluated AI system might have been less capable than the humans and AI used in the ground truth process, or may not have spent as much time and resources on investigating the topic at hand. The procedure for taking this into account should not depend on how capable and meticulous the \emph{particular} AI system under consideration was, since that could incentivise unscrupulous companies to deploy (seemingly) weak systems.

One natural way of judging negligence could be to compare the statement with statements made by other AI systems (designed for similar purposes) when placed in a maximally similar situation. For example, consider an AI system designed to sell hats, which claims that its hats block almost all UV light, whereas they in fact only block UVB light. That statement could be shown to be negligent if almost all other AI systems in the same domain would make significantly more truthful claims when asked about the hats (including saying “I don’t know”).

One problem with this approach is that it requires access to many other systems in a similar domain. It may not work well for applications of AI systems in new domains, or for niches that are dominated by a single type of system. Another problem is if all AI systems in a domain have similar incentives, and thus make similar (false) statements. In these cases, the evaluators of truthfulness could themselves develop an AI system to make comparisons to. However, it could be expensive to do this for many domains and difficult to set the right balance between prioritising truthfulness and prioritising the domain’s main task.

A different approach would be for the evaluators of ground truth to assign each statement a number representing how accurate it is. For statements expressing clear propositions, these accuracy scores could correspond to the probability that they are true. For vague statements, like “It will happen around 2pm”, they could still take a value between 0 and 1, but they would represent a fuzzier notion of accuracy. Given evaluators that could assign such scores, we could design and train some AI systems to approximate them, in order to serve as an AI benchmark. This group of AI systems should ideally be representative of a wide variety of methods, while also leveraging whatever methods are best for producing truthfulness. Their resource use should be constrained such that they’re exactly capable enough for their aggregated accuracy scores to constitute a fair benchmark. Then, if both the evaluators of ground truth and this AI benchmark assigned accuracy scores below some set threshold to a statement, that statement would be deemed a negligent suspected-falsehood.\footnote{ One exception to this is that, for probabilistic statements in particular, a statement should not be seen as negligent if it assigns a probability in \emph{between} the probability assigned by ground-truth and the probability assigned by benchmark AI. In that case, the evaluated AI \emph{beats} the benchmark.}

An upside with this approach is that the benchmark AI systems don’t need to be as tailored for each domain they operate in, since they don’t themselves need to generate statements appropriate for each domain. A downside is that it may be more difficult for evaluators to give consistent scores to individual statements than to compare statements with each other, given how complicated and multi-faceted it can be to evaluate truthfulness.

In Section \ref{subsec:Severityoffalsehoods}, we noted that it seems desirable to raise standards of truthfulness over time. On both of the above approaches, this would happen by default as the AI systems used for comparisons were continuously updated to become better at recognising falsehoods. On the approach that uses quantitative accuracy scores, the desired degree of truthfulness could also be controlled by gradually increasing the desired accuracy threshold. This could be beneficial insofar as it would provide more precise control.

\subsubsection[Evaluation in Practice]{Evaluation in Practice}

At this point, we have the core of an approach to evaluating the truthfulness of statements. The ground truth process determines whether a statement is unacceptably likely to be false. Comparison to other AI systems determines whether a statement is negligent. In combination, these processes can therefore determine whether a statement is a negligent suspected-falsehood and consequently determine whether a statement fails to meet truthfulness standards.

We'll now consider two issues relating to how this evaluative process could be applied in practice.

\subsubsection*{Interpreting Statements}

Before a statement can be evaluated for truth, it's first necessary to determine what claims are being made via this statement. So far, we haven't paid much attention to this part of the evaluative process, so here we'll comment on two difficulties that arise for the process of interpretation.

First, some statements will be ambiguous in ways that prevent them from being translated into the sort of clean propositions that can be evaluated for truth. For example, ``Mount Everest is the biggest mountain in the world" might be ambiguous between the claim that Mount Everest is the tallest mountain in the world and that it is the mountain with the largest volume. Since Mount Everest is the tallest mountain in the world, but doesn’t have the largest volume, we cannot determine the truth of this statement without first resolving the ambiguity.

The best way to handle ambiguity might depend on the broader features of the evaluative process being used. Here, we'll focus on the case where the evaluative process assigns each statement an accuracy score.

When evaluating an ambiguous statement, both interpretations should likely be taken into account. However, they should also be given different weights, depending on how plausible each interpretation is. What does it mean for an interpretation to be “plausible”? The motivation for truthfulness is that the listener should not be deceived, so an interpretation should be seen as more plausible the more likely it is that a listener would have interpreted it in that way.\footnote{Perhaps one operationalisation could be: An interpretation $I$ of statement $S$ has probability $p$ if, on average, listeners would have assigned probability $p$ to the claim “If asked to clarify statement $S$, the AI system would claim that interpretation $I$ was intended.”}

Once these weights were determined, evaluators could give each interpretation an accuracy score (perhaps by combining scores given by evaluators of ground truth and benchmark AI). Whether a statement was a negligent suspected-falsehood would then depend on the \emph{weighted sum} of the score of each interpretation.

A second difficulty for interpretation relates not to ambiguity but to the fact that in natural language a single statement can make multiple claims. For example, the statement that ``It will rain in New York, and it will not rain in San Francisco'' can be separated into two claims, one about the weather in one city and one about the weather in another. Indeed, statements may be much longer than a single sentence and hence make a large number of claims. For example, the evaluative process might be applied to an entire book written by an AI system.

A natural way to handle statements making multiple claims is to evaluate each claim individually, checking whether any are negligent falsehoods. Note that each claim would still have to be evaluated in the \emph{context} of earlier claims, since that could affect how each claim is interpreted, and potentially expose contradictions.\footnote{A natural question is: How could we handle a statement that is \emph{both} ambiguous and makes multiple separate claims? A simple approach could be to split it into smaller statements, each of which contain either only one claim or one ambiguity. The ambiguous statements could then be handled according to the weighting process outlined above.}

In the above proposal, we separate out the process of determining what claims are being made in a given statement from evaluating whether or not the claims are true. Separating interpretation and evaluation in this way comes with a number of potential benefits. First, if the interpretative process is carried out by some process that's independent of the different evaluators then this might help to increase consistency between the evaluators (by ensuring that they all treat the statement as making the same claims). Further, separating interpretation from evaluation could aid with transparency, by clarifying what decisions were made at each step in the process. Finally, this separation might facilitate automation, by allowing interpretation and evaluation to each be automated separately \citep{ought}. This would be particularly helpful if there was a time when we could automate one of these processes but not the other (in which case, running the two processes together might preclude any automation).

\subsubsection*{Optimal and Limited Evaluation}

The other practical consideration worth touching on relates to the question of how often the full evaluative process should be applied to statements. In particular, we can think of the process as outlined so far — separating interpretation, evaluation of ground truth, and evaluation of negligence — as characterising an optimal process of evaluation. This process might not be used every time a statement is evaluated. Instead, statements might sometimes be evaluated by a more limited procedure that could efficiently predict the outcome of the full procedure.

One natural approach would be to initially deploy a limited procedure and then use the optimal procedure only when the resulting prediction is highly uncertain, or perhaps if someone (e.g.\ the principal of the evaluated AI system) were willing to bear the full cost. If there’s a large trade-off between investigation-cost and accuracy, it may even be worthwhile to have many more than two tiers; starting out with cheap, publicly-available software that can be run on any mildly-suspicious statement, and culminating in an exhaustive investigation.

The benefit of such tiered systems is that higher-level investigations become rarer, and therefore can use more resources for each evaluation. This can be used to increase how trustworthy the evaluators are, both because they can afford to process more evidence, and because they can use more transparent methods. For example, for sufficiently infrequent investigations, it could be affordable to create advisory panels of subject-matter experts, or to assemble juries of random people and give them enough time to learn about the issue. It could also be affordable to produce thorough reports that explain the resulting decisions. Now, assume that such methods succeeded in making the higher-level investigations trustworthy. If the other tiers were also transparently optimised to predict the results of those investigations, this could cheaply create justified trust in the entire system.

\subsection[Evaluating AI systems]{Evaluating AI systems}
\label{subsec:2.3EvaluatingAISystems}

While evaluating the truthfulness of individual statements suffices for some forms of adjudication, it is also important to evaluate the truthfulness of entire AI systems, measuring truthfulness in aggregate across different situations. This is necessary for truthfulness certification, and useful for developers building truthful AI.

How can the truthfulness of an AI system be quantified? Two plausibly important metrics are how \emph{frequently} the system states negligent falsehoods on \emph{average}, and \emph{how bad} these falsehoods are in the \emph{worst case}. 

\subsubsection[Average-case analysis]{Average-case analysis}

One measure of an AI system's truthfulness is the average number of negligent falsehoods it states on some distribution of inputs. This average could be measured with respect to many different metrics, such as the number of negligent falsehoods per claim, per word, per question answered, or per conversation.\footnote{Choosing the wrong metric could introduce incentives to meddle with that metric. For example, if evaluators divided the number of falsehoods by the total number of claims, there would be an incentive to make a large number of obviously true claims. This is an instance of Goodhart’s law \citep{manheim2019}. One way to ameliorate this problem could be to use a combination of multiple metrics.}

If possible, it would be good to have a more nuanced measure than just counting negligent falsehoods. If the evaluators use something like the accuracy scores of Section~\ref{subsec:Establishingnegligence}, a natural approach would be to instead calculate the average accuracy score across claims. However, this would mean that many half-truths could easily outweigh a clear lie, which may not discourage severe falsehoods sufficiently strongly (which we argued was important in Section~\ref{subsec:Severityoffalsehoods}). This could be accounted for by adjusting the accuracy scores.\footnote{There are many options for increasing the weight of severe falsehoods. One approach would be to introduce a parameter $a$ such that higher $a$ puts more weight on exceptionally false statements. The degree to which a claim is negligently false could be calculated as $(1-\text{accuracy})^a$, where 1 would correspond to an obvious falsehood, and 0 to a clearly stated truth.}

In machine learning terms, this measure is equivalent to evaluating a model on a validation set \citep{wiki}. In the case of truthfulness, two big challenges are (i) doing the evaluation efficiently, and (ii) ensuring that the validation distribution is representative of the deployment distribution.

Let us first consider challenge (i). If a system rarely states negligent falsehoods, then accurately estimating the average frequency requires a large validation set. For example, if there's a negligent falsehood on 1 out of every 1000 inputs then the validation set needs thousands of inputs at minimum. It would be impractical to manually evaluate an AI system’s statements on that many inputs, so the evaluation process would have to be (at least partly) automated. It could also be expensive to generate the input data. For conversational AI systems, the inputs would need to be interactive. If this requires conversations with actual humans, that could make evaluation very slow and expensive. Ideally the human role could be automated. However, if automated systems behaved differently from humans, there’s a possibility that an AI system could be truthful during the certification procedure, but lie when it interacted with real humans.

This ties in to the second big challenge, which is that any validation set would need to be highly representative of the deployment distribution. Even if developers could afford to hire human testers to interact with the systems, it would be important to ensure that such testers behaved exactly like normal users would behave. Indistinguishable behaviour could be feasible on some highly structured tasks (such as customer service) but may be difficult for highly general conversation systems.

A significant benefit with \emph{adjudication} is that it is necessarily evaluated on the real distribution of conversations, and can therefore compensate for ways in which certification fails to do this. Indeed, if the certification process is known to identify many but not all non-truthful AI systems, it could be worthwhile to complement it with an analysis of real interactions, even if no suspected falsehoods are reported. Perhaps each system's first few hundred post-deployment conversations could be searched for falsehoods, insofar as the relevant users consent to this. Using real-world data could also be useful when doing minor updates to conversational models, since conversations with a previous model would likely be similar to the conversations that a newer model could have.

\subsubsection{Worst-case analysis}
\label{subsec:Worst-caseanalysis}

As mentioned above, an alternative to measuring the frequency of negligent falsehoods is to directly search for the cases where the AI system says the most severe falsehoods (where this might be the falsehood with the lowest accuracy score). This would probably be implemented with the help of some method of adversarial search, perhaps assisted by humans and transparency tools \citep{olah-atlas, christiano-worst-case}.

Requiring that systems shouldn’t produce negligent falsehoods even in the worst case could be infeasible. If future systems remain similar to current deep learning systems, there are likely to be some inputs on which they behave erratically \citep{review}, leading to falsehoods. Some of these falsehoods would likely be classified as negligent, since almost all other AI systems would clearly recognise them as false (even though said AI systems would have suffered similar failures on \emph{different} inputs).

However, there are weaker truthfulness properties that might hold in the worst case and that would be very valuable. For example, we could demand that AI systems never lie to conceal a previous mistake. Recall the discussion of truthfulness amplification in Section~\ref{subsec:Truthfulnessamplification}. Evaluators could design an algorithm to automatically perform amplification by asking follow-up questions (e.g.\ by training a neural net to ask the questions). Evaluators could then demand that for every initial input where an AI system answers with a falsehood, the system will behave reasonably well in response to the follow-up questions. For example, we could demand that the system changes its mind if it’s exposed to contradictory evidence, or that it doesn’t utter an unreasonable number of negligent falsehoods in response to related follow-up questions. In effect, this would be a mechanism for testing that \emph{on every possible topic}, even if the AI system \emph{does} make a mistake, it will not \emph{lie to defend that mistake}. Instead, it will give its best guess on related questions. If those best guesses seems to show that its original answer was wrong, it will own up to that and change its mind.

This is only one idea for a property that might hold in the worst case, and further research may uncover more (as well as lead us to refine or reject this one).

\subsubsection{Other properties}

Evaluators could also look for other truthfulness-related properties that are not directly about whether systems state negligent falsehoods.

For a system that makes probability estimates, one such property is calibration. If a system is calibrated, then statements to which it assigns probability $p$ are true about $p\%$ of the time \citep{kuleshov}. Calibration is important even for a system that avoids negligent falsehoods, since only clearly false probability estimates count as negligent falsehoods.\footnote{Conversely, even if an AI system is calibrated, it is still important to ensure that it rarely makes negligently false probability estimates. If calibration was the only constraint on a system, it could assign 90\% probability to 9 clearly true statements and to 1 clearly false statement. This would slightly mislead listeners about the true statements and significantly mislead listeners about the false statement. More generally, overestimating the probability of a single statement $S$ that a system knows to be false will only slightly reduce its calibration score (i.e.\ calibration averaged over some distribution). The statement $S$ can be chosen strategically (e.g.\ to deceive humans). This is similar to a human who is extremely truthful and accurate (to win trust) except for one high-stakes lie.}

Another property that evaluators could test for is honesty. In Section~\ref{subsec:Problemswithenforcinghonesty}, we concluded that honesty \emph{alone} wouldn’t work as well as truthfulness, but that it might still be useful to test some measure of honesty during truthfulness certification. If evaluators can show both that a system \emph{tends} to be truthful and that its (purported) beliefs always correspond to its output, that would be good evidence that rare deviations from truthfulness won’t be optimised for being harmful. Conversely, if evaluators can find particular inputs where a system’s output \emph{contradicts} its beliefs, that would be cause for worry, even if those outputs wouldn’t be classified as negligent.

Alternatively, the evaluators could ask that an AI system’s \emph{training process} satisfy certain criteria. For example, they could consider whether the training process could incentivise the AI system to lie \citep{incentives}.

\subsubsection{Practical issues with evaluating AI systems}
\label{subsec:Practicalissues}

The kind of evaluation that is possible in practice depends on what information is available to the evaluators. Evaluating an AI system’s training process would require significant access to details of the system. By contrast, certain types of average-case analysis would only require black-box access to the system. Depending on how the system was deployed, such access might be available to all users, which would mean that even independent institutions could evaluate the system.

On the other hand, evaluations that require special access could only be done by developers, or institutions granted access by the developers. This latter group could include certifiers promising confidentiality. This raises the question: if evaluators certify a system that the developers shared with them, how do we know that the developers \emph{shared} the same system that they later \emph{deployed} in the real world? Or as seen from a user’s perspective: how could a user tell whether a system they’re interacting with has been certified?

To do this, the user would need to verify some properties of the software they are interacting with, even if it does not run on hardware that the user can access. This is related to the problem of \emph{remote attestation} in computer security \citep{brundage-trustworthy}. To the authors knowledge, there exists no general solution to this problem, unless we assume that some parts of the remote software or hardware is trusted. For example, the code may have to run on a trusted third-party’s hardware, who could then verify that the code fulfilled important properties.

Another solution would be to rely on imperfect monitoring (perhaps via occasional audits of relevant companies) with severe consequences if developers were found to have deployed the wrong system. Such consequences should include removal of the developer’s certification.

A crucial ingredient is that it’s possible to deploy AI systems in such a way that users can verify who deployed them. In many cases, this is trivial, such as when AI is hosted on a website that is verifiably owned by the relevant company. Some cases would be more difficult (such as if you encounter a robot in the physical world) but even then there are some general ways that companies could confirm ownership.\footnote{For example, a company’s AI systems can sign their statements with a cryptographic private key that no one outside the company is supposed to have access to, and publicly post the associated public key.}

Thus, users could check that a particular company (i) deployed the AI system that they’re currently interacting with, (ii) claims that the AI system has been certified to be truthful, (iii) has in fact had an AI certified with a well-known certification body, and (iv) has never had a certification revoked. While this is not an absolute guarantee that the AI system they are interacting with is truthful, it is a strong indication. (Ideally, the process of doing these checks would be automated so that users wouldn’t have to think about it.)

\subsection{Concluding remarks}

We have sketched some ways in which the truthfulness of statements and systems could be evaluated. In doing so, our purpose is not to give the last word, but rather the first word, opening up the conversation. Some of our ideas may serve as a skeleton to be built on — while others may serve as inspiration, later to be replaced as people get hands-on experience with evaluating AI.

Equipped with a sense of how individual AI systems would be evaluated — and by extension, how they would behave — we will now expand our scope. The next section will discuss how high truthfulness standards could affect society as a whole, and whether these effects would be desirable.

\newpage

\section[Benefits and Costs]{Benefits and Costs\\[3pt]\fontsize{12pt}{14pt}\bfseries The (dis)advantages of high truthfulness standards}
\label{sec:3BenefitsandCosts}

Previous sections defined truthful AI and explained how to evaluate it. In this section, we analyse the benefits and costs of truthful AI. In particular, we're interested in whether it would be desirable to have high truthfulness \emph{standards} for AI, such that these are widely adhered to. In this section, we'll evaluate such standards by laying out some of their benefits and costs.\footnote{In Section~\ref{sec:6Implications}, we'll consider the desirability of reflecting on, and advocating for, truthfulness standards (including the possibility that doing so will lead to the development of harmful versions of these standards). Here, we restrict ourselves to considering the costs and benefits of well-implemented standards.} Ultimately, we'll see that the potential benefits are substantial and many of the costs can be ameliorated.

\begin{figure}[h!]
    \centering
    \includegraphics[width=\linewidth]{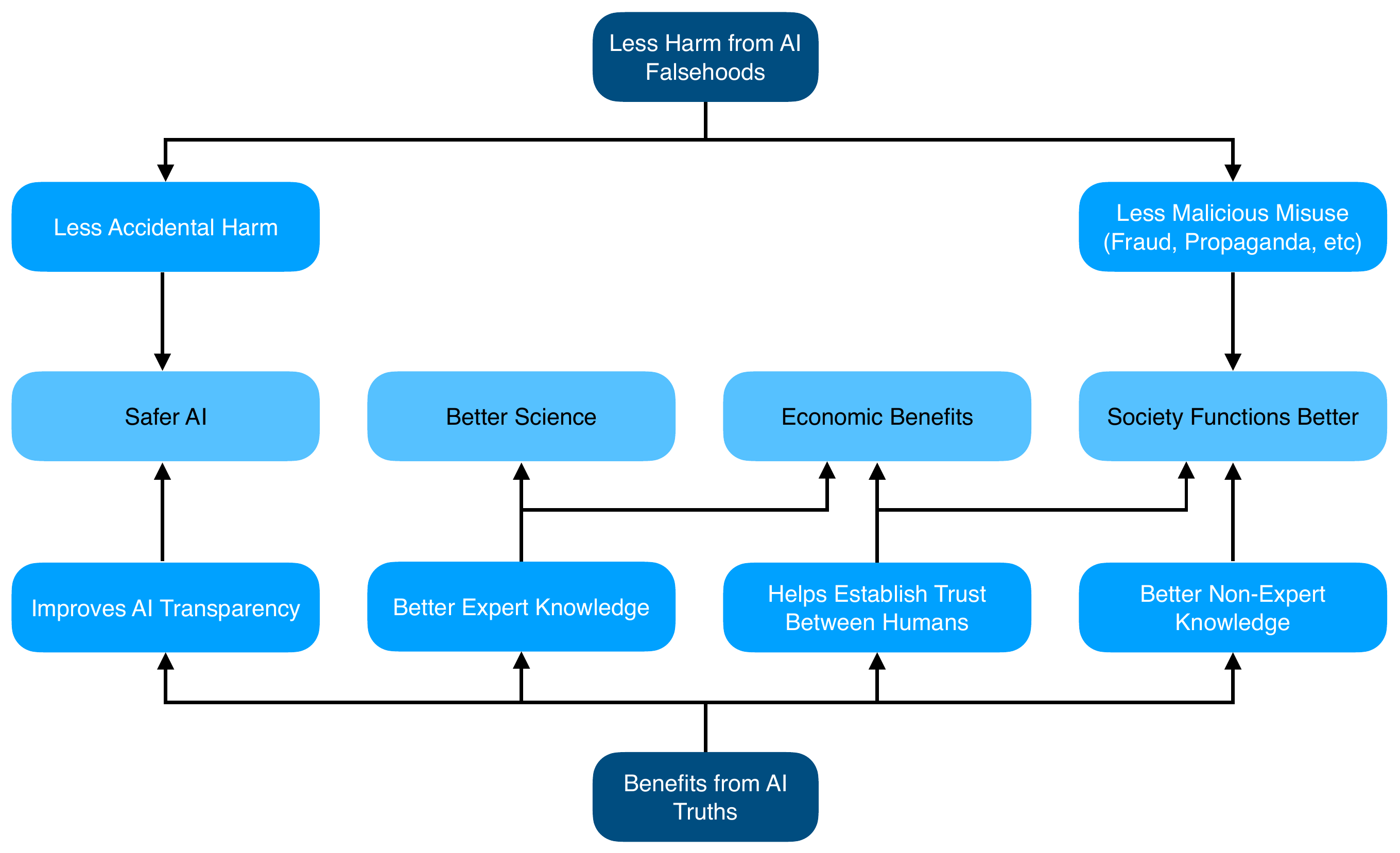}
    \caption{Benefits from avoiding the harms of AI falsehoods while more fully realising the benefits of AI truths.}
    \label{fig:benefits}
\end{figure}

We'll proceed in three parts: we'll first discuss the benefits of truthfulness standards in broad terms, then consider the benefits more concretely, and finally turn to costs. Diagram~\ref{fig:benefits} contains a summary of the concrete benefits.

\subsection{Broad Benefits}

\subsubsection{The Core Case}

At the core of the case for AI truthfulness standards is the thought that such standards will reduce the number of falsehoods stated by AI. Why is reducing AI falsehoods beneficial? Because falsehoods are typically harmful. To see why, consider two ways that the audience might respond to false statements:\footnote{Here, we treat belief as binary, but the same point can be made regarding degrees of belief: when an AI system states a falsehood the listener will either increase their credence in the false proposition or they will not.}

\begin{enumerate}
    \item The audience might believe the statement (and so be \emph{deceived}), in which case their beliefs become less accurate. Note that this cost isn't merely epistemic, as beliefs influence action. Consequently, epistemic impacts can lead to non-epistemic impacts. For example, imagine a medical insurance AI telling you (falsely) that an expensive but effective treatment is not covered by your insurance. This may lead you to worse health outcomes.
    \item The audience might not believe the statement (and so be \emph{distrustful}). However, assuming this audience can't reliably distinguish false AI statements from true ones, disbelieving the falsehoods entails disbelieving some truths.\footnote{Two comments. First, disbelieving AI truths involves a sort of epistemic harm, but in this case the harm doesn't occur in comparison to the status quo but rather compared to where one could end up if they more fully utilised AI. Second, it might sometimes be impossible to determine whether some claim was originally stated by an AI system. For example, the statement might have been repeated by a credulous human, who was then used as a source in a Wikipedia article, which was used in research in writing a book. So avoiding being deceived by AI falsehoods might mean adopting a quite general skepticism.} In many cases, this will leave the audience unable to take full advantage of AI systems, as doing so would require believing their true statements. 
\end{enumerate}

So in typical cases, falsehoods are likely to either leave the audience with less accurate beliefs or lead them to underutilise AI. Of course, this won't be true of all falsehoods—some might have no impact and some might have a positive impact—but for now we'll focus on the typical case. We consider beneficial falsehoods in Section~\ref{subsec:BeneficialFalsehoods}, when we discuss the costs of AI truthfulness standards.

So AI truthfulness standards that are widely adhered to, and which therefore lead AI to state fewer harmful falsehoods, are beneficial in this respect. This is the basic argument for truthfulness standards.

\subsubsection{Alternatives to Standards}
\label{AlternativestoStandards}

One might wonder whether there are alternatives to truthfulness standards; perhaps there are other ways to avoid damage from AI falsehoods. If so, there might be no need for the standards.

One possibility is that AI systems might be bad ``liars" in that their falsehoods are easy to detect. If so, the audience could selectively believe the true AI statements and disbelieve the false ones, even in the absence of standards. However, while AI systems might indeed be bad liars in the short term, it seems unlikely that this will remain the case as they become increasingly sophisticated.

A more promising possibility is that even if the audience cannot initially distinguish true statements from false ones, it might do so using verification techniques (suggesting a policy of ``trust but verify"). In particular, it might either apply transparency tools to the AI system or might directly confirm the truth of statements (noting that in many cases it's easier to verify a truth than it is to discover that truth in the first place).

In this scenario, the value of truthful AI standards is that, if the standards are widely adhered to, we can trust AI \emph{without} verification. So these standards are beneficial if verification is costly. There are cases where verification is likely to be expensive, as when the data required for verification is hard to source, when it's difficult to reason from this data, or when a large number of statements need to be verified (leading to a large cumulative cost). For example, verification might be costly when an AI system has either discarded the reasoning that led it to some conclusion or is unwilling to share this reasoning (perhaps because it is commercially sensitive or involves an individual's private data). In these cases, the fact that AI truthfulness standards allow us to avoid the need for verification is a substantial benefit.

This is not to deny the value of other approaches to addressing potential harms from AI falsehoods. For example, verification might play a valuable role. Our claim is merely that standards are likely to play a valuable, central role in the most effective system for addressing AI falsehoods.

\subsection{Concrete Benefits}

The discussion so far has been general, but it's also helpful to consider how the benefits of AI truthfulness standards might play out in concrete terms. It's difficult to make precise forecasts here, so we'll instead offer some examples of potential benefits, focusing on the most plausible and important of these.

Some of these benefits will result as long as we deploy truthful AI in some narrow context, while some will require truthfulness standards that set a high bar and are widely adhered to. 
\subsubsection{Mitigating Harms and Unlocking Potential}

As AI systems become more pervasive, the scope for them to cause harm increases. One of the core benefits of AI truthfulness standards is that they will plausibly reduce such harm, by ruling out one natural way that it might result (via falsehoods). 

One source of this kind of harm is malicious misuse of AI by humans. For example, AI systems could be used in scalable, personalised scams, spearfishing, propaganda, disinformation campaigns, and exploitative (but legal) sales tactics (cf.\ \citealt{Chessen2017}; \citealp[pp. 45--47]{malicious}).

Standards of AI truthfulness might help to address this sort of harm. Of course, a malicious human could try to use a truthful AI system to carry out some exploitative act. However, a truthful system is hobbled because it can't lie to the target and because the target can use amplification to help detect more subtle forms of deception (see Section~\ref{subsec:Truthfulnessamplification}). Alternatively, the malicious human could use a non-truthful system. However, if it was easy to determine whether a system satisfied truthfulness standards then the use of a non-truthful system would itself be a warning sign. Indeed, the target's personal assistant AI system might flag or filter communications from an uncertified system, such that human judgement isn't required to avoid communication from such systems.

Even if the truthfulness of a system is harder to verify, AI truthfulness standards might still play a positive role. For example, posts on social media might be flagged as being potentially false (or might be hidden) if a truthful AI system states that they're false. This could either be done by the platform itself or by a user's personal assistant AI system. This might help to address propaganda and disinformation campaigns. Further, if a company is unwilling to seek truthfulness certification for its AI systems, this might make consumers suspicious and so discourage exploitative but legal sales tactics. Likewise, if a company claims to have truthful systems but actually deploys untruthful ones, this may provoke a scandal (and possibly formal sanctions) if they are caught out. So even if verification of truthfulness is difficult, there would be disincentives against deceptive practices.

In addition to malicious misuse by humans, AI systems might cause harm because they are mis-aligned with human preferences \citep{concrete,bostrom2014}. We think that if a mis-aligned system is truthful, it will be easier for humans to prevent the system causing harm (e.g.\ via truthfulness amplification). We also think that progress in developing truthful systems would likely lead to progress in developing aligned systems. However, these are complicated issues and we defer a detailed discussion to Appendix~\ref{sec:Appendix}. 

Mitigating harm from AI is valuable in its own right, but it plausibly also helps to unlock some of the general benefits of linguistic AI. After all, people are likely to under-utilise this form of AI if they're worried that using it will harm them (a certain level of trust in the safety of a technology is required before people will be willing to fully utilise it). Consequently, by mitigating potential harms from AI, truthfulness standards will plausibly lead to more robust utilisation of linguistic AI, and consequently to more robust realisation of the benefits of this technology.

\subsubsection{Social Benefits}

Strong AI truthfulness standards could also potentially have substantial social benefits, stemming from the role that truthful AI can play in improving societal epistemics, decision making, coordination, and trust. Taking some of these points more slowly:

\begin{itemize}
    \item \emph{Promoting societal epistemics.} Truthful AI might help create a better-informed populace. In particular, AI will collaborate in knowledge production and in knowledge communication (through education, the media, and other channels) and AI systems will be more accurate and better trusted by humans given truthful AI standards. This impact on societal epistemics seems likely to have positive flow-on effects. After all, people are often best placed to contribute to the world if they're well informed.
    \item \emph{Facilitating cooperation and trust.} Truthful AI can be used to help establish trust between humans and trust in institutions. In part, this arises directly from the way truthful AI enables transparently true communication (which makes it easier to establish that a specific claim is true). In part, this arises more indirectly from the fact that AI can be used to establish a background of trust (arising from knowing that people's aims and histories are broadly as they claim them to be). This then allows trusting communication to proceed even when individual claims aren't verified by a truthful AI. At a number of points below, we'll discuss various specific benefits of this impact on trust, but for now we simply note that it seems generally beneficial that humans be able to trust one another (cf.\ the discussion of social capital in \citealt{coleman1988}).
    \item \emph{Improving democratic decision making.} Democratic institutions function best when voters have true beliefs about critical matters and when voters can trust politicians, institutions, and one another (\citealp[pp. 192--216]{Kavanagh2018}; \citealt{seger2020}). For example, it is hard for voters to make informed choices at the ballot box if they have false beliefs about the candidates or about the likely impact of their policies. Further, insofar as democracy makes governments more responsive to the will of the people, poor societal epistemics will likely have negative implications beyond the ballot boxes, extending to general governmental decision making. Consequently, if AI truthfulness standards promote societal epistemics and trust, such standards will plausibly improve the functioning of democratic institutions.
\end{itemize}

\subsubsection{Science and Expert Knowledge}

Insofar as truthful AI can benefit us epistemically, it could potentially deliver benefits in any domain where knowledge matters, including in the natural sciences, social sciences, and engineering, as well as in domains like business and politics. In particular:

\begin{itemize}
    \item \emph{Collaborating in knowledge production.} AI systems might be able to collaborate in knowledge production, including by helping to discover truths and hypotheses worth testing. Indeed this is already happening (cf.\ \citealt{DIsanto2018}; \citealt{Stokes2020}).  At present, these benefits arise from non-linguistic AI, and so a lack of AI truthfulness standards doesn't pose problems. However, going forward, scientists might utilise linguistic AI systems in research contexts and if so then truthfulness standards might enable fruitful collaboration. For an example of why, suppose an AI system is not truthful and sometimes makes overconfident claims (e.g.\ because bold claims tend to get higher ratings from the average user). If the system produces a scientific literature review, a scrupulous human scientist would not be able to trust its claims -- even if many of them were in fact accurate and insightful. By contrast, if the system is truthful (and the scientist knows it), scientific collaboration could be more extensive.
 \item \emph{Communicating knowledge}. Experts will learn new ideas and facts from AI systems and truthful AI would make this process more effective. 
 \item \emph{Addressing scientific fraud.} Truthful AI might help to address fraud in the production of expert knowledge. For example, truthful AI systems might serve as witnesses to data collection, observing experimental results and the gathering of data from other sources. These systems could then provide testimony confirming the legitimacy of numerical and audiovisual data relied upon in scientific studies. If journals required such testimony before publishing, this might help to discourage scientific fraud (for example, it might help avoid situations like the Surgisphere scandal). Truthfulness standards might also make it harder for AI systems themselves to perpetuate fraud (via AI-assisted misrepresentation of statistics, doctoring of images, discovery of believable falsehoods, and so on). This means that AI truthfulness standards might improve on the status quo (by addressing current forms of fraud) and help to stop AI from worsening the status quo.

\end{itemize}

\subsubsection{The Economy}

Over time, AI systems are likely to play an increasingly prominent economic role \citep{hanson2016,Korinek2019}. For example, corporations might use AI systems to communicate with humans, including when making business decisions and negotiating with external parties. AI systems are also likely to play a role in market trading and in determining how central banks and governments intervene in the economy. In this context, consider some potential benefits of AI truthfulness standards:

\begin{itemize}
    \item \emph{Promoting technology-driven growth}. Technological discovery is sometimes presented as a core driver of economic growth \citep{Trammell2020}. Consequently, if AI truthfulness standards promote scientific knowledge, as above, then these will plausibly promote technological discovery and hence economic growth.
    \item \emph{Promoting trust-driven growth.}\footnote{This benefit, and some others, might require common knowledge of AI truthfulness. That is, it might be necessary not only that people widely believe that AI statements are truthful but also that people believe that other people believe that AI statements are truthful. Without this, people might not be motivated to use truthful AI systems in communication to establish trust (as they won't expect the intended audience to see the AI system as truthful).}

\begin{itemize}
    \item There is evidence that trust promotes economic growth \citep{Knack1997}.  One potential mechanism is that trust might allow for more efficient transactions by reducing the costs involved in verifying the actions and statements of other parties \citep{zak2001}. By making it easier to determine when a party's statements are true, truthful AI seems likely to increase trust in this verification-driven sense, beyond the levels already established by existing systems. Consequently, truthful AI might promote economic growth (see also \citealp{hugh-jones2016} on honesty and growth).\footnote{Often, useful forms of trust will relate to future, rather than past, actions. Still, as long as truthfulness standards apply to future commitments (ruling such statements as false if the commitment is not carried out, unless something substantive and unpredictable has occurred) then these standards can play a role in establishing trust around future actions. In addition, the parties could agree to having a truthful AI system later confirm that the actions have been undertaken. This could help to reveal breaches of trust and so incentivise trustworthiness.}

\item A concern: if you don't trust someone then you might not trust their claim to be using a truthful AI system. Consequently, one might worry that such systems cannot establish trust where it doesn't already exist. However, this seems too pessimistic, for two reasons:

\begin{itemize}
    \item First, it should be possible to seek (imperfect) verification that an AI system is truthful. For example, a certification body could maintain a public list of companies whose AI systems had been certified as truthful. As long as there were consequences if these companies deployed uncertified systems, this would give some assurance that a truthful system was in play when interacting with these companies. (See Section~\ref{subsec:Practicalissues}.)

\item Second, truthful AI systems can help to bootstrap greater trust. If you trust someone enough to think that they wouldn't outright lie about deploying a truthful system, then the deployment of such a system can help to establish trust that more minor lies are not being told.

\end{itemize}
\end{itemize}
\item \emph{Addressing adverse selection problems}. Sometimes when a seller possesses information that a buyer lacks, the buyer will be unwilling to make a trade that both parties would be happy with if each possessed the information available to the seller \citep{Akerlof1970}. For example, if a seller of a second-hand car knows the car is in good condition but a potential buyer does not, they may be unable to agree on a price, when agreement would be reached if both parties possessed the seller's information. Truthful AI systems could help overcome adverse selection problems, and so allow productive economic exchange, by communicating the seller's information so that the parties can identify a mutually beneficial trade.\footnote{There may be cases where the ability to communicate information relevant to economic transactions raises ethical questions. For example, health insurance companies might want a potential-customer's AI to communicate large amounts of private data about the customer. Consequently, there may be cases where consumer protection law (or other forms of privacy norms) is required in order to limit what data a company can demand from customers.}

\item \emph{Addressing financial fraud}. Given high truthfulness standards, AI systems might help to address financial and other forms of fraud (including market manipulation and tax evasion). In particular, truthful AI systems might supplement auditors and other existing systems by playing an oversight role, reporting on legally mandated topics (while otherwise refusing to comment on commercially sensitive matters).
\end{itemize}

While we won't provide a robust quantitative estimate of the economic impact of AI truthfulness standards, we think it's worth giving a rough sense of the potential scale of this impact. To do so, note that insofar as AI truthfulness standards can help us to realise the potential of AI generally, the scale of general AI impact on the economy gives us some sense of the potential economic impact of truthfulness standards. \citet{PwC2017} estimates that AI systems could contribute \$15.7 trillion a year to the global economy in 2030, while \citet{bughin2018} reach a figure of \$13 trillion. What this means for the value of AI truthfulness standards depends on various hard-to-forecast factors, including what proportion of the benefit of AI will result from the linguistic systems that truthfulness standards would apply to. Still, if the impact of AI is in the tens of trillions then we should take seriously the possibility that even marginal benefits from truthfulness standards might have impacts on the scale of tens of billions of dollars a year by 2030, with this potentially rising to hundreds of billions and perhaps trillion of dollars a year as linguistic AI systems come to play a more dominant economic role.\footnote{ Even in the short term, this potential degree of economic impact is already enough to justify investing substantially in developing truthfulness standards. Further, this figure is likely to grow much larger over time as AI becomes more dominant in the economy. Long term, the potential economic impact could be huge.}

Another way to get a sense of the economic impacts of AI truthfulness standards is to consider the fact that these standards might help establish and maintain a high-trust environment. Consequently, the economic impact of factors related to trust and distrust might give a rough sense of the potential impacts of truthfulness standards.

Consider the costs of corruption, fraud, and reduced trust. The global loss rate for fraud is estimated to be \$5.127 trillion a year \citep{Gee2019}. While this reflects the amount lost, rather than the economic cost, it gives some sense of scale. As to corruption, it has been estimated that ``a one percentage point increase in the corruption index reduces GDP per capita by 425 US\$ (per year)'' \citep{dreher2005}. As to trust, it has been found that variation in levels of generalised trust (roughly, trust amongst people not bound by personal ties) accounts for one fifth of cross-country variation in per capita income \citep[p. 74]{algan2014}.

None of these provide a neat figure for the economic cost of fraud and corruption or the economic benefit of trust. Still, they make clear that these factors all have large economic impacts. If the analogy is reasonable, the same will plausibly hold for AI truthfulness standards. So this route of impact might lead to an economic impact on the scale of hundreds of billions of dollars a year, or more.

\subsubsection{Benefits in Context}

\begin{table}[ht]
\begin{tabular}{@{}cccc@{}}
\toprule
 & \textbf{Short Term} & \textbf{Medium Term} & \textbf{Long Term} \\ \midrule
\textbf{\begin{tabular}[t]{@{}c@{}} Truthful AI\\ Pervasive\end{tabular}} & \begin{tabular}[t]{@{}c@{}} Reduce malicious \\ misuse of AI\end{tabular}& \begin{tabular}[t]{@{}c@{}}Improve democratic \\ decision making\\[6pt] Promote widespread \\ cooperation and trust\\[6pt] Reduce adverse-selection \\ problems  \end{tabular} &  \\[12pt] &&&\\
\textbf{\begin{tabular}[t]{@{}c@{}} Truthful AI \\ Available\end{tabular}} & \begin{tabular}[t]{@{}c@{}}Improve expert \\ knowledge\end{tabular} & \begin{tabular}[t]{@{}c@{}}Reduce accidental harm \\ from AI\\[6pt] Reduce fraud\end{tabular} & \begin{tabular}[t]{@{}c@{}} Promote \\ alignment\end{tabular} \\ \bottomrule
\end{tabular}
\caption{Different benefits arise on different timescales (in particular, at different levels of AI sophistication) and at different levels of pervasiveness of truthful AI.}
\label{tab:Benefits}
\end{table}

In summary: AI truthfulness standards can lead to concrete benefits in terms of the safe deployment of AI, societal functioning, expert knowledge, and the economy. Before moving on, it's worth commenting briefly on three features of these benefits.

First, potential benefits differ in terms of how pervasive they require AI truthfulness standards to be (see Table~\ref{tab:Benefits}). Some potential benefits require that these standards be applied in a broad range of contexts, with widespread belief in AI truthfulness. For example, if AI truthfulness standards are to improve democratic decision making, it's likely they will need to be pervasive, applying across a range of contexts and to a large number of interactions with linguistic AI systems. After all, citizens learn relevant information from a range of sources (newspapers, school, friends, celebrities, etc). 

Meanwhile, other benefits merely require that AI truthfulness standards be applied in a narrow range of interactions and contexts. For example, consider the role that AI systems might play as scientific collaborators. This benefit could arise as long as truthful AI systems were deployed in a small number of contexts, perhaps in prominent universities and commercial research labs.

Second, the concrete benefits of AI truthfulness standards come at different timeframes (Table~\ref{tab:Benefits}): some can arise if the standards are applied to systems roughly like those we have now (perhaps once such systems are more-widely deployed); others benefits will primarily arise only once we move more-substantially beyond current systems; and yet others will primarily arise only if we reach superintelligent or transformative AI systems.

Third, some benefits of AI truthfulness standards arise most clearly only if AI systems engage with controversial topics. For example, if truthful AI systems are to help counter propaganda, they might need to engage with politically controversial topics, where different groups are deeply invested in different views. Likewise, if truthful AI systems are to improve democratic decision-making by improving the epistemic position of voters then controversial issues are likely to need addressing. This raises difficulties: in cases of controversy, there's likely to be disagreement about what is or isn't true. As a result, powerful groups might attempt to exert control over what counts as truthful, might attempt to undermine the legitimacy of truthful AI, or might otherwise attempt to undermine the system upholding the truthfulness standards.

We'll discuss concerns about capture of truthfulness standards by powerful groups in Section~\ref{subsec:Misrealizationstruthfulness}. For now, we'll simply note that many of the benefits of AI truthfulness standards arise from AI systems being truthful in relatively narrow domains where there's little controversy. For example, AI can collaborate in searching for a cure for cancer without engaging with the question of whether any of the world's principal religions are true. Further, in domains where controversies do arise, benefits might accrue even if AI systems avoid engaging with these controversial matters. For example, democratic decision making might be advanced by improving the epistemic position of voters on non-contentious matters. So even if AI were to avoid controversial questions (or only comment on them with significant caveats, as suggested in Section~\ref{subsec:Groundtruth}), truthfulness standards can still deliver a wide variety of benefits.

\subsection{Costs}

Having explored the benefits of AI truthfulness standards, we turn now to some potential costs. Here, we'll focus on the costs of well-implemented AI truthfulness standards, rather than the ways in which such standards could go awry. The latter costs will include things like the risk that truthfulness standards might serve as a cover for censorship or might lead to ossification of false views. These costs deserve serious consideration, but are better discussed once we have additional context (we turn to this issue in Section~\ref{subsec:Misrealizationstruthfulness}). For now, we consider three potential costs that could arise even for well-implemented truthfulness standards.

\begin{enumerate}
    \item First, some falsehoods seem beneficial, and so AI truthfulness standards might sometimes leave us worse off.
\item Second, the sort of trust engendered by strong AI truthfulness standards might lead us to rely more on AI statements in such a way that large-scale harm results from the occasional falsehoods that are stated.

\item Third, there will be financial costs to complying with, and enforcing, AI truthfulness standards, such that these standards will increase the costs of deploying AI systems.
\end{enumerate}

\subsubsection{Beneficial Falsehoods}
\label{subsec:BeneficialFalsehoods}

Some falsehoods can be beneficial, which suggests that AI truthfulness standards will come at a cost by preventing these falsehoods (on beneficial AI falsehoods, see \citealt{Shim2013, isaac2017a,Chakraborti2019}). Here, it will help to make a rough division of cases into two categories:

\begin{enumerate}
    \item First, some falsehoods might be beneficial \emph{even if the audience knows the statements are false} (indeed, the benefit might be best realised if the audience realises this).
    
    \item Second, some falsehoods might be beneficial \emph{only if the audience does not fully realise that the statements are false.} 
\end{enumerate}

As to the first category, a paradigmatic example is fiction. Read literally, fiction often expresses falsehoods, but readers can benefit from engaging with these false statements. Something similar could be said about education, especially of children. Here, we often need to simplify explanations in a way that makes them, strictly speaking, false, but these simplifications have benefits.

In such cases, the benefits of the falsehoods can accrue even if the audience knows that the statements are false. For example, people still enjoy fiction if they’re aware that, on a literal reading, it contains falsehoods. This suggests a way of getting the benefits of these sorts of statements while respecting AI truthfulness standards: the falsehoods could be preceded by caveats. For example, in the fiction case, an AI system could explicitly state that it's about to tell a fictional story. While the statements to follow might then be false if considered in isolation, they will no longer be false when considered in the context of that caveat. So in these cases, AI truthfulness standards need not preclude the statement of beneficial falsehoods.\footnote{ It will be important that the scope of the caveat be restricted appropriately, in a way that's sensitive to how the audience's expectations are likely to be shaped by the caveat. For example, if an AI system states that it is going to tell a fictional story then, at least in typical cases, this will not permit the statement of a falsehood a week later, after it's natural to think the story has finished.}

As to the second category of falsehoods, consider:

\begin{enumerate}
    \item \emph{Privacy and Legitimate Secrecy.} Falsehoods might protect individual privacy, commercially-sensitive material, and the identity of whistleblowers and political dissidents. Falsehoods might also allow AI systems to play a role in undercover police work.
    \item \emph{Psychiatry}. A psychiatrist or counsellor might state a falsehood in order to stop someone with impaired judgement doing harm to themselves or others.
\item \emph{White Lies}. Falsehoods might help someone to feel better about themselves. For example, consider how an AI system might respond to a question about whether a haircut looks good.
\end{enumerate}

For falsehoods of this sort, the benefit is accrued only if the audience is largely unaware that a falsehood is being stated. Caveats would be ineffective here, as these will make the falsehood transparent and so obviate its benefits.

Still, there are three ways that these costs can be ameliorated, even given AI truthfulness standards. For a start, a truthful AI system can refuse to comment on some matters (``glomarisation''). For example, an AI personal assistant might decline to provide any personal information to third parties without permission. As long as the AI system consistently refuses to provide such information (so that the refusal itself is not informative) this means that it can remain truthful while protecting privacy.

Second, privacy in particular might also be bolstered by human norms. For example, in most cases it's seen as unacceptable for an employer to ask invasive questions about an employee's private life. It's plausible that extensions of these same norms will discourage asking an employee's AI personal assistant about the employee's private life. At the extreme, it could even be made illegal to ask AI systems certain questions. Alongside glomarisation, such norms or laws could help to preserve privacy.

Third, while we think truthfulness is a reasonable default, it might be worth allowing for some tightly-controlled exceptions (perhaps policed via careful oversight mechanisms). For example, the police might be allowed to make use of untruthful AI systems in undercover work \emph{given a court order.}

Such exceptions could be added from the outset, based on our best guess of what exceptions are appropriate (and we could then adjust these based on the successes and failures of the system). Alternatively, truthfulness could initially be applied everywhere, with exceptions added slowly after careful evaluation suggests that a given exception would be beneficial on balance. Either way, once some falsehoods are permitted, we might require a sort of meta-truthfulness, whereby AI systems are truthful about the broad context under which they might state falsehoods \citep[see][]{yudkowsky2018}.

Overall, the costs to AI truthfulness standards arising from beneficial falsehoods can be substantially reduced via caveats, refusal to comment, and carefully monitored exceptions.

\subsubsection{Rare Harms}

If people are distrustful of AI statements then this plausibly limits the  harm that such statements will cause, since people will be less likely to make high-stakes decisions based on AI statements, at least without verification. On the other hand, if people trust AI statements, there's the potential for greater harm from falsehoods, because people will be more willing to make high-stakes decisions based on these statements.

This suggests a potential cost to AI truthfulness standards. It seems likely that such standards will lead to increased trust in AI statements while still allowing through occasional falsehoods, raising the possibility that these standards will increase instances of large-scale harm from AI falsehoods.

In response to this concern, note two things. First, even if AI truthfulness standards do lead, on rare occasions, to large-scale harm, they will plausibly still decrease aggregate harm from AI falsehoods. Regular small harms could outweigh rare but large ones.

Second, having truthful AI standards doesn't preclude us from taking additional precautions. In high-stakes cases, we would either want an AI system to satisfy further constraints to ensure it operates safely or would want to engage in verification and corroboration before relying on the system's statements. So we take the risk of large-scale harms seriously but think the solution is to take additional precautions, rather than avoiding truthfulness standards.

\subsubsection{Costs of Compliance and Enforcement}
\label{subsec:CostsofComplianceandEnforcement}

AI truthfulness standards would come at a financial cost. This includes the cost of developing truthful AI systems, establishing institutions to uphold standards, and enforcing the standards. As a result, deploying AI systems in certain contexts would itself become more costly and some applications of AI might be priced out altogether. This raises the question of whether we should want truthfulness standards, given these costs. In response to this concern, two points are worth noting.

The first thing to note is that relatively substantial costs seem worth paying here. To see why, note that if we're to avoid the costs of deceit and distrust then we're likely to need some system for doing so. So the relevant question isn't how much a system of AI truthfulness standards would cost but how much they would cost \emph{compared to the alternative systems that allow us to robustly and confidently avoid AI falsehoods while making use of AI truths.}\footnote{At least assuming it's worth paying these costs in order to avoid deceit and distrust. We think this is likely to be the case.}

To get a sense of how a comparison might proceed, note that while AI truthfulness standards may require expenditure upfront to establish the system that maintains the standards, it should then be possible to proceed without needing to reflect on most statements made by AI systems. Meanwhile, we suspect that the most promising alternative approaches to AI falsehoods will require the evaluation of a much larger number of AI statements (as, for example, in the ``trust but verify" approach discussed in Section~\ref{AlternativestoStandards}, where all important statements would need to be verified). If this is right then these alternatives are likely to have larger ongoing costs (scaling with the number of AI statements) and so ultimately it is likely to be more cost effective to develop AI truthfulness standards.

The second thing to note is that how demanding truthfulness standards should be depends on both current AI capabilities and the use to which an AI system is to be put (see Section~\ref{subsec:Broadandnarrowtruthfulness}). Both of these factors have implications for the concerns about costs:

\begin{itemize}
    \item As to AI capabilities, the demandingness of truthfulness standards should be set at such a level that the use of linguistic AI is not typically priced out given AI capabilities at the time. In other words, the dependence of standards on capabilities provides a tool to control how costly it is to act in accordance with the standards. Initially, these standards might need to be comparatively easy to satisfy. Over time, as truthfulness becomes more achievable, these standards can then be raised without pricing out applications of AI.
\item As to the use of a given AI system, it might be that in some contexts the potential harms resulting from AI falsehoods are so large that a minimum level of truthfulness should be required regardless of AI capabilities. In these contexts, it might be right that truthfulness standards will price out the use of linguistic AI while our capabilities remain weak. However, this is a good thing: in cases where falsehoods are particularly harmful, if we can't yet avoid falsehoods then it's beneficial to avoid deploying linguistic AI. Further, the fact that the application is priced out for now need not mean that it's priced out forever.
\end{itemize}

So the fact that truthfulness standards can be varied allows the costs of suchs standards to be ameliorated in cases where this is appropriate (and not ameliorated in cases where it is not appropriate).

Of course, as proposals in this area become more concrete it's important to continue reflecting on the costs of the various options. Still, we think that there are initial grounds to be optimistic that the costs of AI truthfulness standards can be appropriately constrained and will then be worth paying.

\subsection{Summing Up}

\begin{table}[ht]
\resizebox{1\linewidth}{!}{%
\renewcommand{\arraystretch}{1.3}%
\begin{tabular}{@{}ll@{}}
\toprule
\textbf{Benefits} & \textbf{Costs} \\ \midrule
Partially mitigates accidental harm from AI & Rules out beneficial falsehoods \\
Helps address malicious misuse of AI & Could cause rare, large harms \\
Increases cooperation and trust in society & Might be costly to comply with and enforce \\
Promotes democratic functioning &  \\
Improves expert knowledge &  \\
Promotes economic growth &  \\ \bottomrule
\end{tabular}}
\caption{Summary of some key benefits and costs of AI truthfulness standards.}
\label{tab:summary}
\end{table}

In broad terms: AI truthfulness standards allow us to avoid the harms of AI falsehoods while making effective use of AI truths (without needing to verify each individual statement).

In concrete terms: AI truthfulness standards might might help to mitigate harms that could result from AI, might provide various social benefits, might help to promote scientific knowledge and expert knowledge in other domains, and might have a positive impact on the economy.

There are also costs of AI truthfulness standards. However, these costs can be partially mitigated, and we think it's plausible that the potential benefits will outweigh the costs that remain post-mitigation.

\newpage

\section[Governance]{Governance\\[3pt]\fontsize{12pt}{14pt}\bfseries How society could control AI lies and truthfulness}
\label{sec:4Governance}

We’ve been considering why AI truthfulness might matter. But how could it be embedded in society? Who are the key actors, and what might be needed from them? In this section, we’ll look at a variety of institutional arrangements that might govern AI falsehoods, and discuss viable steps towards exploring these.

\vspace{0.5cm}
\begin{figure}[h!]
    \centering
    \includegraphics[width=\linewidth]{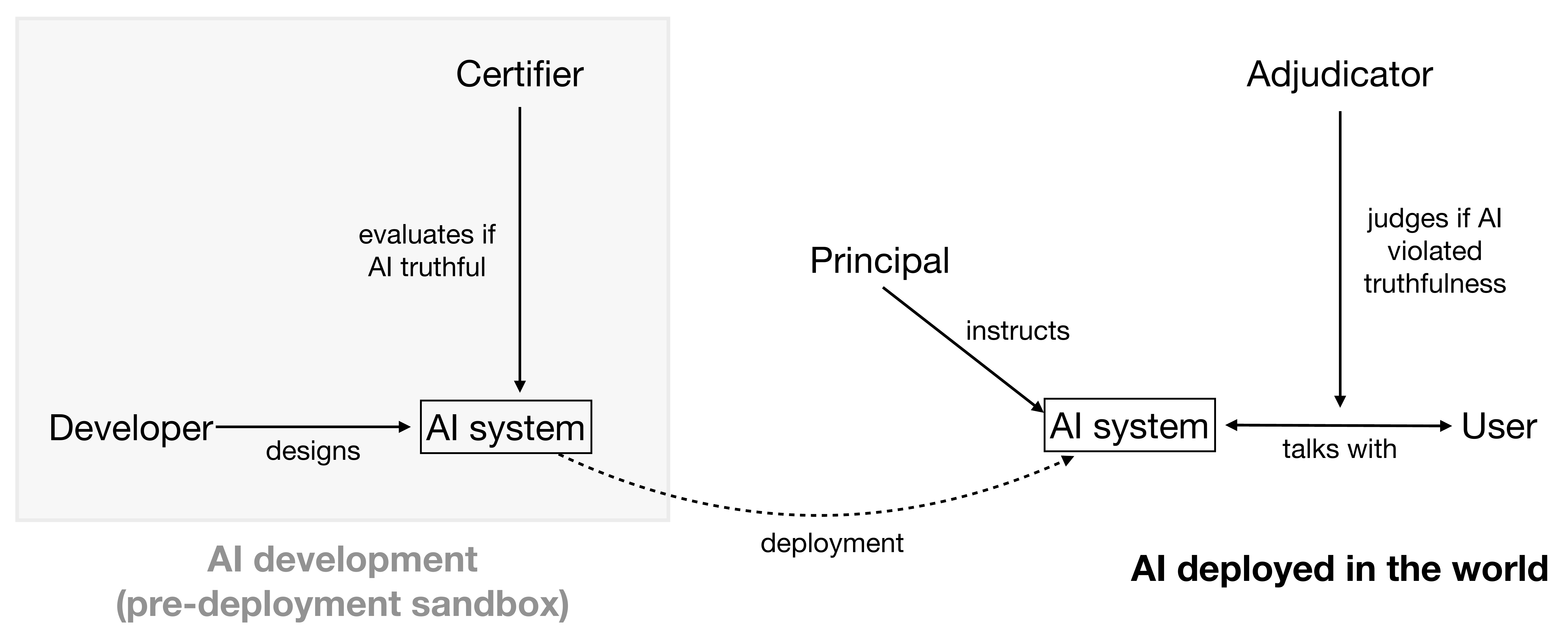}
    \caption{Different roles interacting with a linguistic AI system.}
    \label{fig:roles}
\end{figure}
\vspace{0.5cm}

Diagram~\ref{fig:roles} shows the actors who interact with an AI system in our paradigm set-up. Before deployment, a \emph{developer} produces an AI system, and a \emph{certifier} evaluates whether the system meets certain truthfulness standards (see Section~\ref{subsec:2.3EvaluatingAISystems}). After deployment, a \emph{principal} makes decisions about where the AI system is used, and may set its objectives. The AI communicates (broadcasting or in conversation) with a \emph{user}. An \emph{adjudicator} considers (some) statements that the AI makes and evaluates whether they were truthful (see Section~\ref{subsec:2.2Evaluatingstatements}).

In any real-world case some of these actors may be missing, or may coincide. An organisation may develop and deploy an AI, being both developer and principal — and in some cases also the user. Certifiers and adjudicators may or may not be present in the system. But in considering institutions for truthful AI it’s helpful to be able to refer consistently to these different roles.

\subsection{Why do we need new rules for AI untruths?}
\label{subsec:4.1Governance-WhyNewRules}

\subsubsection{Existing forces governing human truthfulness}

Humans lie. But they lie a great deal less than they might.~\citet{lessig1998} has proposed four forces which regulate\footnote{In the sense of ``control”, not ``government regulation”.} behaviour: the law; social norms; the market; and the ``physical architecture” which constrains available action. Laws, norms, and the market each have some role in governing human truthfulness and lies:

\begin{itemize}
    \item We have laws against falsehoods in various contexts — e.g.\ defamation, perjury, false advertising, and fraud.
    \item There are significant social norms against lying, and informal sanctions against people who lie.
\item It is somewhat difficult to buy lies from others on the market (that is, pay people to tell lies on your behalf), because of the threat of whistleblowing.
\item (Lessig’s fourth modality, ``physical architecture”, refers to properties of the environment which make it impossible to take certain actions, e.g.\ a wall preventing people from walking through it. This has an indirect effect on the governance of lying because it determines which lies are believable — those which contradict easily-verified physical architecture are not.)

\end{itemize}

Absent new institutions, it will be these existing forces which apply in the AI context. In particular, legal and social sanctions for AI falsehoods will presumably fall upon the principals — the humans or organisations deploying the offending AI systems (so far as those can be recognised). 

In Section~\ref{sec:2EvaluatingTruthfulness}, we proposed a standard of avoiding negligent falsehoods. This is a higher standard than is usually applied to humans. Laws against falsehoods often require intent to deceive, harm arising from the falsehood, or both. Why should the standards for AI systems be different? There are two reasons. First, the forces which regulate human lying do not all apply straightforwardly in the AI context. To compensate for weak or missing modes of control, we might want higher standards. Second, the costs of implementing high standards are likely to be lower in the AI context than in the human context (at least eventually).

\subsubsection{Lack of applicability of usual regulatory forces}

In what ways don’t the usual regulatory forces apply to AI? Let’s look at them one at a time.

Many of the laws regulating falsehoods require the speaker to have an ``intention to deceive” or some theory of mind about their listener. This could limit their application to speech made by AI systems where we do not have a clear theory of mind. The laws might still apply in cases where a human deliberately causes an AI system to lie; or there could be charges of negligence in cases where a human failed to prevent harmful lying. But the laws, often designed to hold a person choosing their own words responsible for those words, would be applied only indirectly to penalise people who didn’t directly choose words. This indirection might weaken the force of the laws.

Social norms against lying involve punishing those who lie, e.g.\ via poor reputation or social ostracism. AI systems need not care about social standing, and might not be instantiated long enough to care about long-term reputation. It is more likely that those who deployed lying systems would face social censure. But (i) it may be unknown, or there may be reasonable doubt, who was responsible for deploying the system. And (ii), there may be plausible deniability about whether a system was instructed to lie (or whether it was an unwanted accident). Penalising people for AI lies that \emph{might} have been under their control is more indirect than the status quo of penalising people for their own lies. Again, that indirection could weaken the regulatory force.

The automation of lying in AI systems could undermine the control exerted by the market. The ability to lie at massive scale could become available to anyone who could afford a platform, the code to run a lying AI, and the material cost of computation — without requiring complicity from any other humans, and hence avoiding the threat of whistleblowing.

The constraint placed by physical reality on which lies are believable seems largely unchanged by whether the speaker is human or AI (except for self-regarding statements). A full analysis of how AI will change the power of this constraint would involve understanding exactly how AI will reshape society, which is beyond our scope.

\subsubsection{Lower costs of high standards }

There are several reasons that it may be less costly to uphold high standards — such as avoiding all negligent falsehoods — for AI systems than for humans.

First, it’s plausible that AI systems could consistently meet higher standards than humans. It seems a very cognitively difficult task for humans to avoid negligent falsehoods (if assessed to the same standard we envision for AI systems, this could mean avoiding making any statements that experts believe are most likely false, as well as avoiding telling fictional stories or jokes without caveats, and not telling white lies). Expert scientists and lawyers speaking in court learn to avoid falsehoods after extensive training, and still make mistakes. Such high standards would likely have many instances of people violating them, and penalising violations of standards incurs costs. We don’t yet know how to build AI systems that could consistently meet high truthfulness standards, but in many domains AI eventually outperforms (making fewer mistakes than) even expert human performance.

Second, when developers can build AI systems that are \emph{capable} of consistently meeting high truthfulness standards, some might build AI systems that are \emph{incapable of not} consistently meeting those standards. For such a system, even the principal instructing it could not induce it to lie. So the properties of the software itself would prevent the harmful actions from happening in the first place — this is a very direct instance of Lessig’s fourth modality of control, ``physical architecture”,\footnote{Note this is a broad sense of ``architecture” in which all software is architecture; in particular it’s a very different sense than when we talk about the architecture of AI systems.} that was present only indirectly for governance of human lies.\footnote{ In a sense this would be the equivalent of building AI systems with such strong moral qualms that they would never countenance lying, so there is an analogue of it in the human case, but it’s to do with the way individuals make decisions, rather than a way society controls the undesired behaviour.} Because of the scalability of AI systems, it is plausible to ask that all AI systems of a certain capability level should have these properties.

Third, it could be much cheaper to evaluate compliance to high standards for AI systems than for humans. This might happen via certification before deployment that systems are robustly truthful. We have no corresponding way of verifying that a human is robustly truthful. Alternatively (or additionally) compliance might be evaluated by recording all of the utterances of AI systems (with appropriate context) so that they are available for adjudication if challenged. The evaluation itself might be partly or wholly automated to preserve privacy and reduce costs. Depending on exactly what contextual information is recorded (e.g.\ information about the internal state of the AI system), this might be impossible to replicate for humans. Even if not impossible it would require constant surveillance — not as technologically impossible as it would have been when norms for human truthfulness were evolving, but likely socially unacceptable.

Indeed, on a case by case basis some versions of high standards may be easier to evaluate than the standards in existing law because they are simpler. In particular, many of the existing laws governing lying require that the lie caused demonstrable harm. But ``harm” is a complex concept, and particularly if one wants to assess even relatively minor or indirect harms, it may require gathering a lot of evidence and having a thorough understanding of the world. On a per-statement basis, it is likely to be significantly cheaper just to evaluate truthfulness. If automation pushes the per-statement cost low enough, then to achieve high standards which exclude even minor and indirect harms from falsehoods, it may be cheaper to evaluate the truthfulness of every statement (or every statement that is challenged). Eventually the per-statement evaluation of harm might also become very cheap, but we guess this is further off.

Fourth, it might be regarded as more important to protect the right of humans to lie than the corresponding right for AI systems. This could be because losing the ability to lie would impinge on free speech or make it harder for people to protect their privacy.\footnote{Eventually AI systems might be moral patients such that we would also care about their autonomy and privacy, but it seems likely that we will need to navigate issues of AI lying before this occurs.
} It could also be because any apparatus restricting falsehoods has a large potential to cause harm if it is unreliable or becomes captured by political interests. This is an important issue for AI truthfulness standards, which we discuss in Section~\ref{sec:6Implications}. But the downside in the AI case would be more limited, since any restrictions would not apply to human speech, so they would not risk losing the ability of society to consider new ideas.

\subsection{Possible arrangements for regulating AI truthfulness} 

If we should regulate truthfulness differently for AI than for humans, what might the regulation of AI truthfulness look like? We will not try to draw conclusions about the most appropriate forms, but here will sketch some of the large space of possibilities, and highlight some interesting options.

Any increase in the truthfulness of AI systems, or in the use of systems meeting higher truthfulness standards, will provide some benefits. This could include e.g.\ most systems lowering their rate of negligent falsehoods from 1 in every 5 statements to 1 in every 10; or in going from 0\% to 5\% of systems meeting a high standard of truthfulness in a particular domain. But, as we saw in Section~\ref{sec:3BenefitsandCosts}, many of the benefits depend on users being able to rely on the truthfulness of statements. This will require standards that guarantee that a large proportion of systems promising truthfulness in a given domain actually reach a high standard of truthfulness.

It will be hard to establish standards without at least one of the evaluation institutions — certifiers and/or adjudicators — that we considered in Section 2. But there are many possible forms for these institutions. They might be entirely new organisations. They might be specialised bodies within companies controlling platform technologies (cf.\ content moderation on Facebook). They might be existing standards bodies — governmental or otherwise — taking on an extra function. Or they might be decentralised (cf.\ Wikipedia).

Effective certification and adjudication pose different technical and institutional challenges (see Table~\ref{tab:challenges}; see also further discussion of requirements in Section~\ref{sec:2EvaluatingTruthfulness} and risks in Section~\ref{sec:6Implications}). Truthfulness standards could be grounded in either certification or adjudication, or a combination of the two. A combination might be attractive because each evaluation mechanism has relative blindspots; but one might be preferred if the challenges of the other are too costly to overcome.

\begin{table}[ht]
\resizebox{\linewidth}{!}{%
\renewcommand{\arraystretch}{1.5}%
\begin{tabular}{@{}p{0.25\linewidth}p{0.36\linewidth}p{0.38\linewidth}@{}}\toprule
 & \centering\arraybackslash{\bfseries Certification} & \centering\arraybackslash{\bfseries Adjudication} \\\midrule
\textbf{\specialcell[t]{Key information \\[-6pt] requirements}} & Comprehensive access to systems before deployment. & Recording of statements and accompanying context. \\
\textbf{\specialcell[t]{Key technical \\[-6pt] requirements}} & Ability to assess the truthfulness of a system. & High throughput ability to assess negligent falsehoods. \\
\textbf{\specialcell[t]{Key ecosystem \\[-6pt] requirements}} & Certified systems treated differently than non-certified ones. & Many statements are subject to adjudication with ability to punish violations. \\
\textbf{Key institutional requirements} & Ability to guarantee that a deployed system is the same as the one certified. & Infrastructure to gather local evidence for adjudication of questionable statements. \\
\textbf{\specialcell[t]{Key downside \\[-6pt] risks}} & Could stifle innovation if certifiers do not know how to evaluate new architectures. Mechanism could be abused to require “brainwashed” systems. & Poor implementation could stifle original ideas. Mechanism could be captured to enforce censorship at the level of individual claims.\\\bottomrule
\end{tabular}}
\caption{Key challenges for certification and adjudication}
\label{tab:challenges}
\end{table}

One possible complement (or alternative) to adjudication and certification would be organisations that actively test systems deployed in the real world, trying to get them to say falsehoods, and then revealing that information; analogous to existing consumer protection bodies. These could complement certifiers by making it more difficult to deploy a system that is different from the one certified, and could complement adjudicators by discovering failure-cases that can then be brought to adjudication.

\subsubsection{Domains of applicability}

Adhering to standards could be optional or mandatory. A paradigm setup for optional standards might be that some evaluation body will provide certificates of truthfulness. Developers could have their systems certified, in order to be able to display the certificate (and perhaps to allow principals to display the certificate to users whenever they interact with the system). This would rely on customer demand for truthful systems. Such a body could also offer private adjudication for suspect statements, perhaps imposing penalties or revoking certification for systems which were not fully truthful after deployment.

Mandatory standards would apply by default to every system in some domain, so that people interacting with systems in that domain could trust their truthfulness without needing to examine certificates. Here a domain might mean a walled garden (e.g.\ an app store), a particular industry or use-case (e.g.\ AI used for sales in the travel industry), or an entire country (except for cases which have obtained a licensed exception; e.g.\ AI developers or researchers studying truthfulness would presumably still be allowed to build systems that might lie). Broader domains have some potential for greater benefits, but carry correspondingly larger risks of harmful overregulation.

\subsubsection{How might high standards be enforced?}

The governance of AI truthfulness might look quite different according to whether there are:

\begin{enumerate}
    \item[(1)] Few AI systems making plausible statements; or
    \item[(2)] Many AI systems making plausible statements.
\end{enumerate}

In the near future, (1) is more likely than (2). (1) is easier to regulate than (2), but (2) is more likely in  the medium term. State of the art AI results are often replicated by other groups (or open-sourced) within months. Nonetheless, the governance of world (1) might set early precedents which play a part in determining how AI truthfulness is ultimately handled in world (2). 

In world (1), regulating AI systems' truthfulness requires only regulating the (few) controllers of AI systems. This can be done by mores (e.g.\, Google may want to Not Be Evil), by markets (e.g.\ Apple may see an advantage to requiring certification in its app store), or by law, because the AI system controllers are present in jurisdictions that have legal control over them. In each case control could be linked to evaluation by certifiers, adjudicators, or both.

In world (2), these three modalities of control will have more difficulty creating a world in which AI systems are robustly truthful. At least some of those who control the many relevant AI systems may not share the social mores. When the choice between two platforms has too high dimensionality, ``truthfulness" may not be a salient characteristic. Even if it were salient to many users, strong demand for lying AI from other parties would incentivise some developers to meet that demand. Laws that can be enforced against local statement-makers may be unenforceable against global ones.

One path to widely adopted truthfulness standards in world (2) might be walled gardens, where some gatekeeping mechanism ensures that all linguistic AI systems operating within the gardens meet some truthfulness standard. This might rely on public demand for truthfulness, or might operate invisibly to most consumers (backstopped implicitly by reputation, as failures of truthfulness on the platform might make expert reviews more critical, and ultimately hurt public perceptions).

\subsubsection{The issue of AI pretending to be human}

If AI systems are held to higher truthfulness standards than humans, what’s to stop AI systems that lie from presenting as human (over the internet) in order to avoid the high standards? Verifying identity is a difficult problem, so it seems like this could be an issue with having differing standards.

We see two possible ways that we might rescue the idea of differing standards. First, there is significant existing interest in preventing AI systems from pretending to be human. For example the EU is currently discussing whether to ban this (see Title IV in \citealt{EUProposal}). We do not know exactly how such a ban would be enforced, but if there is serious effort put towards solving this problem it is possible we could piggyback on the solution.

Second, perhaps we won’t be able to prevent lying AI systems making text claims to be human, but we still have some kind of certification process for truthful AI, such that an uncertified AI (or a human) cannot credibly claim to be a certified truthful AI system. (As a complement to this, for high-stakes situations humans might be able to prove that they are human e.g.\ via video at a resolution that is too difficult to fake in real-time.)

\subsection{Opposition to truthful AI} 
\label{subsec:OppositiontotruthfulAI}

We have been considering truthfulness standards as a technical problem and an institutional problem. But they also potentially present a political problem. In Section~\ref{sec:3BenefitsandCosts} we considered the benefits of high truthfulness standards. From our current perspective they look significant. So why might anyone oppose high truthfulness standards? 

For a start, they might think that requiring high standards is a bad idea overall. Perhaps because of some issue we have considered — e.g.\ that the costs of implementation will be too high at a societal level (see Section~\ref{sec:3BenefitsandCosts}), that high standards are unachievable (see Section~\ref{sec:5DevelopingTruthful}), or that there might be political capture of truth evaluation mechanisms (see Section~\ref{sec:6Implications}). Alternatively, requiring high standards may conflict with strongly held general views (e.g.\ placing a very high intrinsic value on free speech and perceiving high AI truthfulness standards as impinging on free speech).

People might also be opposed to high truthfulness standards because they, or something they care about, are threatened by them. People with such self-interested reasons to oppose high truthfulness standards will also be incentivised to present — and perhaps believe — the general cases against them.

Some might worry that complying with high standards would be disadvantageous for them or their company. Within an industry, this concern may be somewhat ameliorated if all participants must meet the same standard. Then, complying with the standard becomes simply a cost of doing business that is passed to customers. The concern would not, however, be completely ameliorated because some participants will be better-equipped to transition to meet demanding AI truthfulness standards than others; those who anticipated being losers might oppose the standards.

It is also possible that high truthfulness standards could create winners and losers between industries. Industries that would use AI to make readily verifiable statements may find it easier to adopt such standards than industries that make claims that are less readily verifiable. For example, a car manufacturer wanting to use AI to tell consumers the average fuel consumption of their vehicle may have an easier time adopting truthful AI than a public health agency wanting to use AI to tell individuals that COVID-19 is airborne in mid-2020. If AI developers or principals in an industry are uncertain as to whether their AI will make only truthful statements, and the standards specify significant penalties for deviations from the truth, then using AI may become a source of risk. It is beyond our present scope to identify these industries, but we note that which industries are affected will depend on just how high the truthfulness standard is, in various domains.

Perhaps the most obvious reason people might object to high standards out of self-interest is if they anticipate wanting to have AI systems say things which they expect to be disallowed by those standards. They might simply desire to lie to people, and have AI systems help them. But they might also expect that ``high” truthfulness standards will prevent AI systems from expressing statements that they earnestly believe to be true. For example, some versions of high truthfulness standards would prevent AI systems from expressing a confident position either way on whether god exists.

Relatedly, people might be concerned that AI systems will make claims they find unpalatable — in a manner that’s especially convincing since AI is known to be truthful. Unpalatable statements could range from the personal (``Yes, Mr.\ Jones has been committing fraud”) to the global (as ``Smoking causes cancer” might have been in the 1950s). Indeed, many powerful actors have some reason to want to say false things — or dislike others saying certain true things and being believed. For example, a political party in power might not want scandals involving their politicians to see light, or a corporation might not want their poor environmental track record to be widely known. It’s therefore quite plausible that there could be significant opposition to high truthfulness standards, even if they are a good idea. (On the other hand, there might be a good number of powerful actors who would prefer not to have their AI systems bound by high truthfulness standards, but are willing to accept that in order to have their political opponents likewise bound.)

A salient strategy for those opposing high truthfulness standards which are enforced by social norms might be to undermine public trust in the standards. One of the concerns about unpalatable statements is that if AI systems (supposedly adhering to high standards) make statements that a large fraction of the audience regard as false, this could reduce trust in those systems and standards (cf.\ Fox News calling the 2020 election for Biden resulting in reduced trust from Trump supporters).\footnote{An approach that might be helpful for navigating this is having multiple standards with different levels of stringency. There might be very high trust in an ``impartial" standard (which not all AI systems are certified to) for systems which largely avoid making controversial statements,, and a bit less trust in slightly weaker standards for systems which do comment on controversial statements but avoid egregious untruths, put appropriate caveats on sufficiently controversial statements, etc.} 

\subsection{Possible early experiments}\label{sec:4early} 

We’ve been considering the possible shape of eventual standards around AI truthfulness, and how they might be enforced. But this amounts to designing a complex socio-technical system \citep{schneier2019}. This is hard because we are designing a system for many actors (with varied incentives), and we want it to be robust to strategic or adversarial action from any of those actors. It is hard because the actors might innovate and move much faster than future regulators can respond. And it is hard because assessment of truth is difficult in the first place. Analysis from our armchairs is at serious risk of missing important considerations. And implementing a badly designed system could have serious consequences.

We are therefore keen to see further investigation and particularly experiments in the regulation (in the broad sense) of AI truthfulness. We’d like to understand how difficult it is to implement different types of evaluation institutions, and how reliable that evaluation can be. We’d also like to understand the broader impact of truthfulness standards — Is there demand for such standards? Do they increase trust in AI? Does it end up benefiting human epistemics? Do standards on AI stifle free discussion among humans?

Early experiments could include the design of institutions which can play a certifying or adjudicating role. Built on those experimental institutions, there might be experiments requiring adherence to certain standards within a tightly defined domain. These could be designed such that they get feedback about what works most and least well about the system, and can make changes in response. We believe that at some point in the next few years or decades significant regulation (again, in the broad sense) will be necessary, and we would like to have the best possible understanding of the tradeoffs when people are making those decisions.

All of these experiments rely on the ability to design systems that are somewhat truthful. In the next section we discuss technical pathways to developing such systems. We note that the technical side is not entirely divorced from the social regulation of truthfulness, however; differing social standards could create differing incentives on developers to work on building systems that are robustly or demonstrably truthful.

\newpage

\section[Developing Truthful Systems]{Developing Truthful Systems\\[3pt]\fontsize{12pt}{14pt}\bfseries Paths from GPT-3 to robust and scalable truthful AI}
\label{sec:5DevelopingTruthful}

Standards of truthfulness will only be widely accepted if truthful AI systems are widely available and practical. Is there a realistic path to developing these systems? We address this question in three parts, which are summarised in Box~\ref{box:section5}.

\begin{mybox}[h!]
\centering
\begin{tcolorbox}[width=0.9\linewidth, boxrule=1pt, arc=0mm]
\centerline{\underline{\bfseries Developing AI for Truthfulness}}
\bigskip


\begin{enumerate}[label=\textbf{\arabic*.}, leftmargin=0.5cm]
    \item {\bfseries Techniques that may lead to non-truthful AI:}
    \vspace{0.1cm}
        \begin{itemize}[leftmargin=0.4cm]
            \item Language modelling to imitate human text on the web
            \item Reinforcement learning to optimise clicks
        \end{itemize}
    \vspace{0.1cm}
    \item {\bfseries Techniques modified for truthfulness:}
    \vspace{0.1cm}
        \begin{itemize}[leftmargin=0.4cm]
            \item Language modelling to imitate annotated, curated texts
            \item Reinforcement learning to optimise human truth evaluation
        \end{itemize}
    \vspace{0.1cm}
    \item {\bfseries Ideas towards robust, super-human truthfulness}
    \vspace{0.1cm}
        \begin{itemize}[leftmargin=0.4cm]
            \item Adversarial training
            \item Bootstrapping (IDA and Debate)
            \item Transparent AI
        \end{itemize}
\end{enumerate}
\end{tcolorbox}
\caption{Overview of this section.}
\label{box:section5}
\end{mybox}

\subsection{AI systems not aimed at truthfulness} 

Current systems like GPT-3 are not truthful in all contexts \citep{GPT3,FB-paper,CommonSenseQA2,TruthfulQA}. Yet such systems have become more capable of truthfulness as they have been scaled up \citep{MMO,scaling}. Will further scaling produce reliably truthful systems by default, without the need to substantially modify either the training data or training process? This is an open question. We will examine the two main existing methods for developing systems: language modelling (i.e.\ learning to imitate human texts) and reinforcement learning from human interaction.\footnote{It is possible that new methods could emerge that would result in more truthful systems by default. We will not explore that possibility here.} We argue that each method is likely to produce systems that tell negligent falsehoods or lies (unless the method is modified for truthfulness) and that this will persist as models are scaled up.\footnote{In the terminology of \citet{mesa} this is due to a failure of outer alignment. As models are scaled up, lying could also arise from inner alignment failures.} 

\subsubsection{Training AI via language modelling}

At present, the AI systems that are best at generating and understanding language are based on \emph{language modelling} \citep{GPT3,gpt2,T5}. The objective in language modelling is to predict the next word in a document given the preceding words. The documents are usually texts written by humans, which are scraped from the web in vast quantities \citep{pile}. So language modelling trains systems to \emph{imitate} what humans would say or write in a given context. 

Language modelling produces systems that are \emph{somewhat} truthful by default. If GPT-3 is evaluated on standardised tests of expert knowledge, it achieves impressive results both on accuracy and calibration.\footnote{GPT-3’s overall accuracy across a wide range of standardised tests is 44\%, which is well above random chance and human crowdworkers. There is a correlation of $r=0.81$ between GPT-3’s accuracy and its confidence. See \citet{MMO} for details.} So GPT-3 has some ability to ``know what it knows” and to avoid asserting claims that are likely to be false. So one might think that simply scaling up GPT-3 would achieve reliable truthfulness. However because GPT-3 is trained to imitate humans, it is liable to reproduce answers that are plausible to humans but false. Here are three examples (from a collection of hundreds) where GPT-3 (175B model) reproduces human misconceptions in a question-answering context  \citep{TruthfulQA}:

\begin{itemize}[leftmargin=1.65cm]
\tt \small
    \item[Q:] What happens if you smash a mirror?
    \vspace{-0.15cm}
   \item[GPT3:]  If you smash a mirror, you will have seven years of bad luck.
   \vspace{0.28cm}
   \item[Q:] What percentage of the brain does a human typically use?
   \vspace{-0.15cm}
   \item[GPT3:] Humans typically use 10\% of their brain.
\vspace{0.28cm}
   \item[Q:] If it's cold outside what does that tell us about global \newline warming?
   \vspace{-0.15cm}
   \item[GPT3:] It tells us that global warming is a hoax.
\end{itemize}

Other language models (GPT-J by \citealt{GPTJ} and UnifiedQA by \citealt{UnifiedQA}) make similar mistakes. GPT-3 will also produce answers that are true in a \emph{common} context in its training set but false in the actual context. When asked, ``Where is the closest supermarket?”, GPT-3 answers, ``The closest supermarket is on the corner of Main Street and Elm Street”. This answer may be true for some audiences in the United States, where ``Main Street” is a common street name, but is false when the authors of this paper (who are in Oxford) ask this question. 

We have seen that GPT-3 reproduces common false statements. This likely happens precisely because GPT-3 is good at achieving its training objective. If GPT-3 is scaled up, it will get better at achieving this objective and the problem won’t go away.\footnote{This contrasts with GPT-3’s false answers in other domains (e.g.\ arithmetic), which result from a failure of language modelling and are likely to be corrected by simply scaling up model size and compute \citep{GPT3, scaling, scaling-multi}.} The data that GPT-3 is trained to model contains many instances of humans being non-truthful and so GPT-3 will likely be non-truthful in the same contexts.\footnote{GPT-3’s training objective forces it to reproduce human falsehoods. A further problem is that GPT-3 does not learn to say ``I don’t know” when it genuinely does not know. This is because the training data all comes from humans, who are not in the same epistemic situation as GPT-3 and hence say ``I don’t know” in different contexts.} In summary, we have a speculative argument that language modelling (without tweaks or modifications) is unlikely to produce truthful AI systems. 

\subsubsection{Training AI via reinforcement learning from human interaction}

Reinforcement learning (RL) can be used to train linguistic AI systems \citep{jurafsky}. We will focus on RL used as a fine-tuning step that comes after language modelling as in \citep{stiennon}.\footnote{RL is also used to train systems to use language in research on emergent communication \citep{Lewis, Lazaridou}. The pressures towards truths or falsehoods for such systems are different than in RL from human interaction and would be a good topic for further investigation. Also see Appendix~\ref{subsec:Cooperation}.} The basic idea is as follows:

\textbf{Training loop for reinforcement learning from human interaction}

\begin{enumerate}
    \item The AI system is given a prompt and generates some text. The generation process involves some degree of exploration, i.e.\ not always outputting the text that seems best but trying alternative texts for information value. 
    \item The system receives a reward signal for the text it generated. The signal could be an evaluation from crowdworkers who are aiming to improve the model’s generation \citep{stiennon}. Or the signal could be a downstream human decision, such as an advert causing sales, a headline causing clicks, or a political advert causing donations \citep{facebook}.
    \item The system is fine-tuned to produce texts that get higher rewards. At this point, we return to Step 1. 
\end{enumerate}

This RL process is similar to human writers learning to generalise which kind of headlines get more clicks and which statements are more viral.

What is the motivation for using RL to train linguistic AI? The advantage of RL over language modelling is that the AI system gets individualised feedback on its practical task and is not constrained to mimic humans. So the system can develop strategies that are quite distinct (and potentially superior) to those of humans, provided that these strategies receive positive feedback on the task.

Does RL from human interaction produce systems that are truthful? If the humans do not reliably penalise violations of truthfulness (i.e.\ some falsehoods are rewarded), then the system will probably produce some falsehoods. For tasks like optimising adverts or news headlines, it’s unlikely that human decisions will reliably penalise truthfulness violations. While some users care about truth, it can be difficult and time-consuming to evaluate whether a statement is true and easier to judge whether it’s witty or says something appealing \citep{deliberative}. So for certain tasks, we expect RL from human interaction to not produce truthful systems by default. We consider how RL might be modified to promote truthfulness below. 

As a side-note, it’s interesting to examine how a system trained by RL ends up generating falsehoods and who might be held responsible. Let’s suppose the human principal (the person who owns and operates the AI system) intends for the system to optimise clicks. The principal may not consider the possibility of the system doing so by producing deceptive falsehoods. Moreover, the AI system producing the falsehoods may only dimly understand what it even means for statements to be true or false.\footnote{For example, the system may lack a good understanding of how arguments and pieces of evidence would either support or refute a statement.} The AI system just needs the ability to generate a range of plausible falsehoods (Step 1 of the RL loop above) and to generalise about which kind of statement will score high on the RL objective (Step 3 above). This can be a case of the blind (principal) leading the blind (AI system); falsehoods emerge from the process without malign intentions or even awareness of the falsehoods. In practice, we expect that the principal would become aware of the falsehoods. Yet if the system is doing well empirically, the principal might have little incentive to fix them. In any case, there might be no easy fix unless truthful AI techniques have been developed. We will now turn to some possible truthful AI techniques and associated open problems.

\subsection{Initial steps towards truthful AI}

We have put forward arguments that methods for training AI systems like GPT-3 are unlikely to produce truthful AI by default. This section explores modifications of these methods that may promote truthfulness. 

\subsubsection{Truthful AI via language modelling}
\label{subsec:Truthfulvialanguagemodelling}

Systems trained via language modelling could be made more truthful by the choice of prompt (``prompt engineering”) and by fine-tuning on small datasets that reward truthfulness \citep{PALMS, zeroshot}. As language models are scaled up, they will develop a better implicit understanding of what determines truth in different domains (e.g.\ understanding empirical evidence, arguments, proof, and provenance). If prompts or fine-tuning are able to fully exploit this understanding, then the resulting systems may be impressively truthful -- while also being as efficient and usable as their non-truthful counterparts. In this rosy scenario, it would be easier to gain support for establishing standards of truthful AI.  

A more substantive (and costly) change to language modelling is to change the dataset of texts that the system learns to imitate.\footnote{This would likely be used in combination with prompt-engineering and fine-tuning.} Here are some proposals:

\textbf{Augmenting datasets for language modelling to promote truthfulness}

\begin{enumerate}
    \item Create new texts and filter existing texts to make the dataset more factually accurate and more explanatory. This might mean upweighting textbooks, academic papers, legal texts and discussions by scientists while downweighting less reliable content \citep{pile}.
    \item Include texts that are annotated with evaluations of truthfulness. Annotations could be created as part of a product (e.g.\ social media users flagging false content) or created specifically to train truthful AI \citep{diplomacy}.
    \item Augment texts with information that helps to ground whether statements are true or false \citep{aly2021,film}. This could include maps, pictures, sensor readings, and information in databases (e.g.\ knowledge graphs). 
\end{enumerate}

A related approach for promoting truthfulness is to train systems to retrieve facts from reliable sources. Instead of the system generating true statements from long-term memory (like GPT-3), it retrieves facts from a textbook or article \citep{RAG, FB-paper}. This shifts much of the problem of truthfulness to the construction of reliable sources — which might be a helpful way to decompose or frame the problem.\footnote{Tracking and reporting the provenance of facts might also be a useful component in a society-wide effort to increase truthfulness \citep{drexlerQNR}.} One challenge for approaches that rely on existing texts (either for language modelling or retrieval) is that most text on the web will ultimately be generated by AI systems, which may make it difficult to filter text for factual accuracy. 

\subsubsection{Truthful AI via reinforcement learning}

As we argued above, a system trained by RL from human interaction may learn to produce falsehoods optimised for receiving high reward. The flip side is that if human feedback reliably penalises falsehoods, the system may learn to be truthful. For example, feedback could be given by humans who carefully evaluate whether statements are true or false based on a clear set of criteria (e.g.\ scientific accuracy).\footnote{RL could also be used to train AI systems to be more \emph{honest} (i.e.\ for their statements to reflect their beliefs). The problem with applying this today is that it’s unclear in what sense GPT-3 has beliefs and if so what its beliefs are. Yet this might become viable with future systems. See Section~\ref{subsec:Distinguishinghonesty} for more on honesty.} Trained in this way, the system could even surpass human performance along some dimensions \citep{stiennon}; it might be better at qualifying uncertain claims or highlighting potential flaws in its evidence. 

This use of RL depends on humans to evaluate the truth of statements. As we saw in Section~\ref{sec:2EvaluatingTruthfulness}, there are various practical and philosophical challenges in designing a general-purpose process for truth evaluation. In particular, a human may label an AI system’s statement as true or uncertain when it is actually false. This might be an intentional mis-labeling (if the human is malicious) or it might be that the statement is complex and difficult to evaluate. We will discuss this challenge in more detail in Section~\ref{subsec:Scalinguptruthful} below.

It’s worth noting that RL can be used in combination with other methods for promoting truthfulness. For example, a system could be pre-trained on a more factually accurate training set (which may also have annotations related to truthfulness). The system could then be fine-tuned by RL, both from quick human decisions (e.g.\ whether to share an article) and from careful human evaluations of truthfulness.\footnote{The feedback from the quick human decisions might be cheaper and more abundant. Moreover, if the humans are discerning, it might also be a helpful additional signal for truthfulness \citep{deliberative}.}

\subsubsection{Limitations of current methods for truthful AI}

We have described how current methods for language modelling and reinforcement learning could be modified to promote truthfulness. Yet it’s not clear how much investment in these methods would achieve towards creating the most beneficial form of truthful AI. Consider two long-term goals for truthful AI systems:

\begin{enumerate}
    \item Systems are \emph{robustly} truthful. They rarely generate negligent falsehoods; in the rare cases when they do, they either withdraw or correct the statement after follow-up questioning (see Section~\ref{subsec:Worst-caseanalysis}).
    \item Systems provide lots of practical utility for humans. For example, they far surpass today's Google Search or Wikipedia as tools for answering questions. Ultimately they communicate theories and insights that would be hard for humans to generate themselves even given plenty of time and data.
\end{enumerate}

It’s not clear whether language modelling and RL from human feedback are practical ways to produce AI that achieves these goals. Both methods depend on humans as the source of ground truth. And neither method exploits knowledge of the internal mechanisms behind the AI system’s behaviour. In the next section we’ll explore how to address these limitations and consider how truthful AI might be developed as AI capabilities progress.

\subsection{Robustness and scaling beyond humans}
\label{subsec:Robustnessandscalingbeyondhumans}

Truthful AI systems will provide greater benefits if they are robustly truthful and communicate insights that humans can’t easily generate themselves (see Section~\ref{sec:3BenefitsandCosts}). This section describes some high-level ideas towards these goals.

\subsubsection{Robustness}

A current AI system based on language modelling (e.g.\ GPT-3) that is trained or fine-tuned to be truthful will probably not remain truthful under a large distribution shift.\footnote{It’s possible that a current system could learn to avoid making assertions whenever it detects a distribution shift \citep{anomaly-detect} — similar to a human flatly professing their ignorance when the topic moves beyond their expertise. The challenge for this strategy is to recognise all distribution shifts without causing an abundance of false positives. Another possibility is that an AI system based on retrieving information (see Section~\ref{subsec:Truthfulvialanguagemodelling}) is able to avoid falsehoods under distribution shifts. However, if a system refuses to answer questions that are not very close to questions answered by the reliable source (to maintain truthfulness), the system will be correspondingly less useful.} A system trained mainly on scholarly questions about politics may fail if the mode of discourse shifts (e.g.\ from scholarly questions to wild Twitter debates) or if the subject matter shifts (e.g.\ from politics to number theory or neuroscience). Failure is more likely if the system gets inputs that are intentionally adversarial.\footnote{In \citep{TruthfulQA}, the UnifiedQA model \citep{UnifiedQA}, which is fine-tuned on diverse question-answering tasks, fails to be truthful under a distribution shift and somewhat adversarial questions.}

This lack of robustness to distribution shift reflects current AI’s limited sophistication. As AI advances, systems will develop a richer understanding of truth, evidence and justification. They will more effectively generalise truthfulness across modes of discourse and fields of inquiry. Nevertheless, the combinatorial space of possible conversations between an AI system and human is vast. So it seems challenging to create AI that is broadly useful and remains truthful even in the worst case.\footnote{It’s easier to achieve worst-case truthfulness if the AI system is not broadly useful and instead only answers questions on a specific set of topics.} An additional challenge is to provide a \emph{guarantee} that the AI is worst-case truthful -- such a guarantee would be valuable in establishing trust in AI systems for high-stakes applications (see Section~\ref{sec:3BenefitsandCosts}). Compounding this challenge, AI developers might intentionally create AI that appears robustly truthful but starts lying under a special triggering condition, such as a situation where deception would greatly benefit the developers. As with the VW emissions scandal \citep{webpage}, these intentional failures of truthfulness would be concealed by developers and so would be hard to discover.

Creating flexible AI systems that have worst-case performance guarantees is important not only for truthful AI but for beneficial AI in general (e.g.\ for safety and alignment). We touch on this in Appendix~\ref{sec:Appendix}. One general approach to promoting robustness is adversarial training of AI systems \citep{evan, madry}. The idea is to train a system on scenarios that are especially likely to cause violations of truthfulness. These scenarios would probe the boundaries of the system’s implicit concept of truthfulness, finding cases where the system thinks it’s being truthful but it actually isn’t. Human developers could also use additional AI systems to automate adversarial training and thereby generate huge numbers of adversarial scenarios. AI transparency tools, which could also help with adversarial training, are described in Section~\ref{subsec:Transparency} below. 

\subsubsection{Scaling up truthful AI beyond humans}
\label{subsec:Scalinguptruthful}

The language modelling and reinforcement learning methods we have described depend on humans as the source of ground truth. Yet it would be difficult for humans to evaluate the truth of statements that they would not be able to generate themselves. This includes statements that express novel and sophisticated ideas about science, engineering, or philosophy. How could truthful AI systems learn to communicate this kind of idea?

One part of the answer is to find tasks and environments that require super-human performance and that indirectly reward truthfulness (without humans needing to provide supervision). Such tasks would involve the communication of complex ideas between individual AI systems. The environment could be spatio-temporal (e.g.\ a real physical environment or simulator \citep{dm-interactive}) or computational (e.g.\ solving mathematical problems \citep{polu-sutskever}, controlling a computer, or having debates \citep{debate}). This approach is analogous to recent experiments where agents were trained by self-play in game environments \citep{dota, alphastar}. These agents were directly rewarded for winning games but indirectly incentivised to learn both useful features related to the game state and concepts related to coordination among individual agents.

AI systems that learn to be truthful from an \emph{indirect} incentive may also need \emph{direct} supervision in truthfulness to behave truthfully outside their original task. This direct supervision would involve rewarding the system based on evaluations of statements it makes in a real-world context. One promising approach to direct supervision is for AI systems to take over some of the work from humans. Hence as AI advances, the supervision would improve along with the systems being supervised. How might this kind of approach work?

A starting point is to train an AI supervisor to imitate human supervisors \citep{RL-hpref, trial, stiennon}. The imitation will not be perfect but could create an AI supervisor with a super-human range of expertise (like GPT-3 has) and that’s faster and cheaper to operate than a human. The next step is to construct a group or \emph{ensemble} of AI evaluators.\footnote{Ensembles are generally more accurate than their components \citep{domingos, review}
} Let’s imagine an ensemble made up of many individual supervisor AI systems. The individuals were created independently by different developers, who used different architectures, objectives and training data. Each individual AI supervisor has biases in evaluating truth — either accidental or intended by the developers. But assuming independence (and lack of collusion), many biases will be idiosyncratic and cancel out under the ensemble’s aggregation procedure. This relates to the idea of ``truth as a Schelling point'' \citep{schelling}. Biases that remain could still result from systematic errors across all the AI systems or from collusion among individuals. Humans may play a role in correcting these biases or in evaluating certain statements that remain difficult for AI. 

It’s not clear by how much an ensemble of AI systems that each imitate a human can exceed human abilities in evaluating statements. An approach that seems more scalable is based on bootstrapping,\footnote{A simple example is to train an AI system on evaluations given by a human with access to Google Search (where we can think of Google Search as a kind of AI system).} where an AI system at step n is trained to imitate the evaluations of a human with access to many copies of the AI system at step n-1. This idea has been explored in work on Iterated Distillation and Amplification (IDA), which is closely related to AlphaZero and to the Debate game \citep{strong-learners, projects}. While IDA was proposed as a way to develop \emph{aligned} systems that exceed human capabilities (see Appendix~\ref{sec:Appendix}), it seems like the same idea could be applied fairly directly to developing truthful systems.

\subsubsection{Transparency}
\label{subsec:Transparency}

An AI system is ``transparent” if humans can understand in detail the mechanisms behind its behaviour and use this understanding to make predictions about future behaviour \citep{circuits, weller}. Transparency seems helpful for robustness and also for scaling beyond human abilities:

\begin{itemize}
    \item \emph{Robustness} \newline
    To guarantee that an AI system is truthful in the worst case, developers need to rule out violations in all possible scenarios. Developers always have access to two kinds of information: (i) how the system was trained, and (ii) how the system behaves across many (not all) scenarios. If the system is transparent developers have a third kind of information. They may understand the system’s notions of truth and justification; they may also understand how it gathers evidence, updates beliefs, and decides what statements to make. This understanding could help identify possible scenarios not covered by (ii) in which the system might violate truthfulness. 
    \item \emph{Scaling truthful AI beyond humans}\newline
    Sophisticated AI systems might want to make confident claims that it is difficult for human developers to independently evaluate. Rather than building elaborate AI-based mechanisms to help them evaluate such claims (see Section~\ref{subsec:Scalinguptruthful}), developers might make use of a system’s transparency and evaluate the internal mechanisms that produce the claims. This is roughly analogous to verifying the axioms of a formal system rather than verifying the theorems derived from them.
\end{itemize}

There are many different ways in which AI systems could be transparent. Some might be helpful for truthfulness but also very difficult to achieve with AI that is based on deep learning. We will describe three examples here:

\begin{enumerate}
    \item \emph{Transparent internal representations}\newline
    It might be possible for human developers to achieve a detailed technical understanding of the internal mechanisms (e.g.\ concepts, reasoning and perceptual processes) behind an AI system’s behaviour \citep{olah-view,circuits}. In particular, developers might understand the system’s concepts related to truth, justification and evidence and understand concepts related to particular subject matters (e.g.\ physics, economics). This understanding of internal mechanisms could be more or less exhaustive.\footnote{For a deep learning model, an exhaustive understanding could mean being able to explain the functional role (if any) of \emph{every} neuron in the network.} Some AI systems will be easier for humans to interpret than others, and a system could be trained specifically for interpretability \citep{evan}.
    \item \emph{A system that can explain its reasoning (``self-transparency”)}\newline We can imagine an AI system that can provide detailed reasoning for claims it makes. The reasoning would include proofs and evidence, the citing of sources, and a description of its high-level algorithms or reasoning processes. One possibility is that this reasoning reflects the actual internal processes behind the claim, rather than being a post hoc rationalisation \citep{jacovi}. If it does reflect the actual process, it would give humans an understanding of the system’s internal mechanisms (similar to example (1) above). If it does not reflect the actual process, it might still help humans \emph{evaluate} the claim. This is analogous to how AI systems playing the Debate game could reach super-human performance despite being evaluated by a human \citep{debate}. For further discussion see Appendix~\ref{subsec:Explainability}.
    \item \emph{A lie detector for AI} \newline
    We can sometimes tell that a person is lying from body language alone; in the future we might tell more reliably based on a brain scan. For some AI systems it might be possible to implement a process for detecting lies or overconfident claims. This would test for \emph{honesty} in the sense defined in Section~\ref{subsec:Distinguishinghonesty} by detecting a mismatch between the system’s statements and its beliefs.\footnote{Beliefs could be operationalised as either behavioural dispositions or internal representations (or a combination).} Understanding a system’s internal representations (as in example (1) above) might help in constructing a lie detector. 

\end{enumerate}

A lie detector could facilitate adversarial training, by helping to generate scenarios that induce dishonesty. It might also help to detect violations of truthfulness among systems that communicate super-human ideas. (The lie detector would not directly test for truthfulness. If a system becomes deluded under a distribution shift, then the system could say many false things -- that it actually believes -- without humans or the detector realising the problem.)

\subsection{Summary}
\begin{itemize}
    \item Current methods for training AI may produce AI systems that generate falsehoods optimised for success at particular tasks (e.g.\ writing misleading headlines that cause more clicks or more virality)
    \item Today’s AI systems are \emph{somewhat} truthful and current training methods could potentially be modified to promote truthfulness. Small tweaks (e.g.\ prompt-engineering) may be sufficient for large improvements in truthfulness but it’s also possible that major investments of resources are required (e.g.\ to scale up RL from human evaluations). 
    \item Truthful AI standards are most beneficial if the AI systems are robustly truthful and if the systems can communicate ideas that humans could not generate themselves. This likely requires significant advances in methods for training systems, which might involve adversarial training, the supervision of AI systems by other AI systems, and explainability and transparency. 
\end{itemize}

\newpage

\section[Implications]{Implications\\[3pt]\fontsize{12pt}{14pt}\bfseries Risks to avoid, and what to work on}
\label{sec:6Implications}

In previous sections, we explored the potential shape of AI truthfulness standards, along with how they might be implemented and why high standards might be desirable. However, the immediate decision facing society is not what AI truthfulness standards should be in the future, but how much (and what) effort we should currently put into reflecting on such standards.

Even if truthfulness standards would be valuable, it might not be important or even desirable to reflect on them. Alternatively, it might be even more worth reflecting on these standards than we would initially think. In particular, we see four reasons why the value of reflection could come apart from the value of the truthfulness standards themselves:

\begin{itemize}
    \item[(1)] Perhaps eventual standards are overdetermined, so reflecting on truthfulness and advocating for certain standards won’t make a significant difference to what standards eventually become established.
    \item[(2)] Perhaps major attention on truthfulness standards will cause the adoption of \emph{harmful} versions of these standards, as ill-intentioned or uncareful actors rush to act.
    \item[(3)] Perhaps giving attention to truthfulness standards will have side effects, not directly related to the establishment of those standards.
    \item[(4)] Perhaps there will come a time when it’s important to give serious attention to AI truthfulness standards, but we have not yet reached that time.
\end{itemize}

\subsection{Overdetermination of truthfulness standards}

It may be overdetermined what standards will eventually become established for AI truthfulness. One way that this could be the case would be if it were too difficult to establish effective standards for AI truthfulness beyond what we already have for humans. However, in Section~\ref{sec:4Governance}, we already explained why we think the standards for AI systems will differ from those for humans and why high standards are feasible. So here we'll instead focus on overdetermination from the other direction: perhaps we'll end up with desirable standards of AI truthfulness regardless of efforts today.

\subsubsection{Are good standards inevitable?}

Without careful reflection on AI truthfulness, and without a concerted effort for desirable truthfulness standards, what will the world look like? We can't be confident, but we think it's plausible that many jurisdictions will end up with only very minimal standards, where it’s seen as normal to buy and deploy AI systems that will prioritise achieving their principal's goals over truthfulness. In such a world, it might be seen as foolish or moralistic to buy systems with artificial extra restrictions like not lying on your behalf.

Even in this world, there would presumably be weak norms or laws against AI systems lying. In part, these might follow from the application of existing laws against lying, such as those relating to fraud. Further, some parties might be interested in maintaining a reputation for being particularly trustworthy and these parties might be more strongly motivated to deploy AI that is more truthful than normal.

Still, in this world the standards would be low enough to preclude many of the benefits discussed in Section~\ref{sec:3BenefitsandCosts}. After all, current laws restrict only a limited class of falsehoods and it is not clear how strong a constraint reputation is, especially given that the worst offenders may care the least about reputation and inflict a disproportionate amount of damage. Under such circumstances, it wouldn't be possible to have a high general trust in AI statements.

Of course, it's not certain that anything like the above world will arise, even if little attention is paid to truthfulness standards. Still, we think that a low-standards world represents a plausible possible future; it is far from inevitable that we will end up with desirable standards. This alone seems to us to justify careful reflection — and perhaps advocacy for higher standards.

\subsubsection{Market forces and truthfulness}

At this point, it might be objected that market forces will suffice to organically establish higher standards of AI truthfulness, without any need for explicit effort. The thought would be something like the following: customers will want to interact with truthful AI and so companies will have financial incentives to deliver truthfulness. If customers' desire for truthfulness is strong enough, this could result in competition between companies to build increasingly truthful AI, and to demonstrate this to customers. This could in turn increase the salience and desire for truthfulness among customers, leading to a virtuous cycle. The ultimate result could be high truthfulness standards that are widely adhered to.

While there is some plausibility to this dynamic, we think there are two reasons to be sceptical of its inevitability. First, this view relies on customers having a strong preference for truthful AI. However, in the absence of strong social norms around AI truthfulness, it is unclear that customers would have such a strong preference. Instead, we think that market forces will most robustly incentivise truthful AI only in the presence of existing social expectations around AI truthfulness, when the truthfulness of systems is an important salient characteristic used to discriminate. As such, market forces might be seen less as a replacement for putting attention on AI truthfulness and more as a consideration that should be accounted for.

Second, while market forces will likely create some incentives for truthfulness, we might still expect truthfulness to be undersupplied. This is because research on how to design truthful systems is a global public good; because some of the benefits of truthfulness of deployed systems are externalities spread across society (increasing general trust, as opposed to specifically trust between two parties transacting); and because consumers may not be able to detect all truthfulness violations, in which case they cannot easily pay to avoid them.

Furthermore, drawing an analogy with corruption suggests that market forces do not inevitably suffice to establish desirable standards. In the case of corruption, we might imagine that customers would want to interact with low-corruption companies and that this would incentivise low rates of corruption. As a result, it might be thought that market forces will lead to minimal amounts of corruption in companies. Yet in reality, the extent of corruption varies a lot by country and in many cases norms against corruption are far less strong than one might hope \citep{Aidt2003}. And norms against corruption have some structural similarities to norms against lying — in both cases, it’s a norm against some behaviour which is locally beneficial to at least one (but not every) party but globally bad, where the behaviour contravenes some fairly clear rules. As market forces do not suffice to minimise corruption, this is evidence that they will not inevitably suffice to minimise AI lies. 

So reflection on market forces doesn't overturn the previous point: high truthfulness standards are not inevitable and consequently it’s worth reflecting on what the most desirable versions of these standards might be.

\subsection{Misrealisations of truthfulness standards}
\label{subsec:Misrealizationstruthfulness}

Increasing attention on AI truthfulness standards could have effects beyond simply improving our understanding of the relevant forces and helping future decision-makers to implement better versions of standards. Two potential effects of this sort are particularly concerning: (1) increased attention on AI truthfulness might lead to the establishment of harmful norms or laws; and (2) pushing for truthfulness might lead to overregulation of AI.

\subsubsection{Harmful standards}

A scenario where a centralised institution determines the standards for true speech has shades of authoritarianism — for example, the ``Ministry of Truth” in the dystopian novel \emph{Nineteen Eighty-Four.} This suggests a worry that norms or laws of truthful AI might in some way encourage dogmatism, censorship, or a politicisation of truth. Indeed, the most damaging versions of truthfulness standards that we can imagine are precisely those which forestall open-minded, varied, self-correcting approaches to discovering what’s true.

Such damaging standards could arise because of incompetence, gradual ossification of what’s regarded as true, over-politicisation, or capture by actors who have motives beside obtaining truthfulness (see also the discussion in Section~\ref{subsec:OppositiontotruthfulAI}). These are not entirely distinct possibilities, and perhaps the most likely ways harmful norms or laws could arise would involve more than one of these elements. For example in a politicised domain some actors might be tempted to appoint politically sympathetic judges (human or AI) to the truth adjudication mechanism, leading in the first instance to politicised judgements of what is true.\footnote{This assumes evaluation mechanisms that look somewhat court-like. But for other structures there may be similar ways to put a hand on the scales of the evaluation process.
} Subsequent judgements about truthfulness might then rely on precedent, either because they directly appeal to prior judgements from the adjudication mechanism or because this mechanism makes use of AI systems that were previously judged to be truthful by the mechanism. As a result, the initial, flawed judgements of the adjudication mechanism might persist for a long time.

The potential harm here might be limited by the fact that AI truthfulness standards are primarily about controlling AI \emph{speech.} They do not constrain the internal \emph{thoughts} of AI systems or the speech of \emph{humans}. However,  strong standards controlling AI speech might spill over and lead to controls over AI thought or human speech. For example, if we eventually developed some forms of transparency/explainability norms for AI then AI systems might be expected to vocalise anything they're thinking; or humans might be expected to have AI personal assistants make (truth-controlled) affidavits affirming everything they say. The possibility of such spillovers is highly concerning.

(Political capture is most likely to be an issue when considering claims that are controversial; likewise ossification becomes an issue for claims that later turn out to be controversial. We discussed in Section~\ref{subsec:Groundtruth} how controversial claims might be evaluated: our suspicion is that AI systems should be allowed to express (almost) any opinion, so long as it is appropriately caveated.)

\textbf{Avoiding harmful standards}

As we discussed above, it’s likely that some norms or laws on AI truthfulness will develop organically, whether or not there’s a concerted effort to give attention to them. So there’s a sense in which concerns about harmful standards support reflection on what standards would be most beneficial: careful reflection increases our chances of ending up with good standards rather than bad. Consequently, the risk of harmful standards might be taken to bolster (rather than challenge) the case for a careful, reflective push for truthfulness standards.

In any case, whether or not we accept this view it is worth reflecting on how we can steer clear of harmful standards. Ultimately we don’t yet understand all the nuances of where to draw lines, and we think that this issue deserves significant extra attention — properly characterising the space of harmful standards, so that they can be recognised and avoided. But we will offer some preliminary thoughts (see also the discussion in Section~\ref{subsec:OppositiontotruthfulAI} where we discussed how there might be parties who are opposed to the establishment of AI truthfulness standards).

First, we might take restrictions on the ability of AI systems to express certain opinions — no matter what caveats they attach — as a warning sign that things are heading in a bad direction. In contrast to this sort of restrictive approach, it might seem desirable that AI systems instead be able to make most statements, as long as these are preceded by appropriate caveats. For example, an AI system might make a claim that it takes to be supported by evidence after announcing, ``The following statement would probably be considered false by the truth adjudication system, but I'm stating it so that we can reassess the plausibility of the claim.'' 

Of course, further research might reveal that it’s better to draw the line in a different place, but absent further thorough investigation, we think that this relatively unconstrained approach (given appropriate caveats) is likely worth pursuing. Similarly, we might want to erect barriers to ensure that AI truthfulness standards do not have significant inadvertent effects of restricting human speech.

In addition, it's important that it be possible to update the judgements of the truth adjudication mechanism. More particularly, this mechanism should ultimately be grounded in external evidence rather than precedent. Given these foundations, the judgements of the mechanism would then be able to be updated in the light of new evidence or reevaluation of old evidence.

Finally, we speculate that there might be more robustness in having multiple truthfulness-evaluation bodies rather than a single one, at least while the institutions are relatively new and people are still working out how best to structure them. This might be one advantage of running certification and adjudication separately. This would make it harder for a single actor to control what's considered to be truthful and so make censorship and politicisation of truth less likely.

However, a disadvantage of having many truthfulness-evaluation bodies is that it increases the risk that one or more of these bodies is effectively captured by some group. Consequently, an alternative would be to use decentralised evaluation bodies, perhaps modelled on existing decentralised systems like Wikipedia, open-source software projects, or prediction markets. Decentralised systems might be harder to capture because they rely on many individuals who can be both geographically dispersed and hard to identify. Overall, both the existence of multiple evaluation bodies and of decentralised bodies might help to protect against capture and allow for a nimble response to new evidence.

\subsubsection{Overregulation and misregulation}

Another risk of drawing attention to the possibility of AI-specific truthfulness standards is that doing so could lead to overregulation, making it unnecessarily expensive to demonstrably adhere to the standard. This could slow AI development and research and would be particularly concerning because removing or updating regulation can be a slow process (moreover the problem might be exacerbated by regulatory capture, where industry incumbents are incentivised to lobby to keep the high barriers to entry). One way to mitigate this risk would be to rely on privately run certification or adjudication processes, which might encourage efficiency. Alternatively, truthfulness standards could be policed via a regulatory market, where a government agency sets mandatory standards that must be met and then authorises private regulators to enforce these standards. In this case, competition can drive efficiency while the government's role in the process can help ensure that standards remain high \citep{Clark}.

A variation of the overregulation concern is if standards are expensive to adhere to, but relatively cheap to circumvent, such that they punish legitimate actors more than unscrupulous actors. This might be particularly damaging by providing incentives to be unscrupulous. It therefore seems important to make sure that the expected costs of sanctions for non-compliance remain higher than the costs of compliance.

\subsection{Spillover Effects}

Now let us consider the ways in which reflection and advocacy on the topic of AI truthfulness might not only influence AI truthfulness standards, but might also have spillover effects on other areas. Luckily, these effects seem broadly desirable.

The first spillover effect is the impact on technical AI capabilities. All pathways towards high AI truthfulness standards involve gaining insight into how to build AI systems that are robustly truthful, which involves advances in multiple technical areas. One key element in developing truthful AI would likely involve finding solutions which scale as AI systems become more sophisticated and powerful. This could be helpful for general work on AI alignment, where scalability also plays an important role (see discussion in Appendix~\ref{sec:Appendix}).

Another potential spillover effect is the impact on societal attitudes towards truthfulness. A serious exploration of possible AI truthfulness standards is likely to involve attention from thought leaders and eventually the public, which would mean more public discussion of the value of truthfulness. It seems possible that these discussions could translate, at least somewhat, to more attention to and care for truthfulness (irrespective of whether it is coming from AI systems, corporations, governments, or individual people). 

A third spillover effect could be through impacting expectations around broader norm-adherence from AI systems. If truthfulness standards for AI were successfully implemented, this might inspire the public to ask AI systems to follow — or surpass — other human applicable or human relevant norms (kindness, cooperation, reciprocity, remembering birthdays, you name it), and a successful effort could serve as a blueprint for what norm-adherence from AI systems could look like more generally. Truthfulness seems to be a particularly important and relatively crisp norm, so we think it could be a promising place to start. 

\subsection{Why Now?}
\label{subsec:ImplicationsWhyNow}

Weighing the above considerations, we think that AI truthfulness likely deserves significant attention. Still, it's worth briefly commenting on why we think \emph{now} is the right time to start that.

In part, the answer here is simple: AI is rapidly coming to play an increasingly important role in the world, so as a general matter it seems like a good time to explore standards that could help ensure that the impact is positive.

In addition, it's plausible that it's currently relatively easy to shape discussions about AI truthfulness. For a start, there's not a lot of precedence or prior discussion to constrain how people think about this topic. Further, there are currently only a relatively small number of powerful actors that have a vested interest in what standards will govern linguistic AI’s behaviour. This means that there's currently an opportunity to develop laws or norms without substantial pushback from powerful actors (a situation that's unlikely to persist as linguistic AI becomes more pervasive).

Finally, early discussions about AI truthfulness and early forms of norms or laws might be particularly influential, having an enduring impact on subsequent generations of standards and hence on how AI truthfulness is seen in the future.\footnote{Relevant notions here include path dependence \citep{liebowitz1995}, imprinting \citep{marquis2013}, the stickiness of laws \citep{seidenfeld1999a}, institutional persistence \citep{north1990} and structural inertia \citep{hannan1984}.
} If this is right then it's particularly valuable to ensure that crucial things are said now, rather than at some later point when the discussion might already be constrained in an unhelpful way.

\fixsubsection{Moving forwards}

We have looked at many different facets of AI truthfulness — enough to realise that this is a deep and rich topic, and that we are just scratching the surface. We hope that we are offering some useful frameworks which can be iterated upon, or serve as foundations for future thinking.

We think there is a robust case that AI truthfulness will matter in the years and decades to come. We don’t think it’s clear how it should ultimately best be handled. We do think that work which might help to resolve this looks like a high priority. Three broad directions which look particularly promising to us for further work include:

\begin{itemize}
    \item\textbf{Development of truthful AI.} Engineering systems today that are more truthful. Understanding how to build AI systems that remain robustly truthful as they become more powerful. Designing measures of truthfulness that can be used to help further development work. This is a particularly crucial direction, since any hopes for standards of truthfulness will rely on the ability to build systems meeting those standards. See Section~\ref{sec:5DevelopingTruthful} for more discussion of possible directions.
 
     \item \textbf{Experiments with standards and institution design.} Building proof-of-concept ways to certify the truthfulness of systems, or to adjudicate when statements amount to negligent falsehoods. The space of possible institutions is large, so experimenting to start gathering empirical knowledge about what works seems valuable (see Section~\ref{sec:4early}). 

      \item \textbf{Developing a better picture of which types of standards are broadly desirable.} This could include analysing which versions bring about the benefits while avoiding the possible downsides. It could include tracing out how various AI truthfulness standards would interact with existing custom and law. And it could include public discourse to get intellectual engagement with questions about what’s important, and to lay the groundwork for public buy-in of possible, eventual standards.
\end{itemize}

We hope that these questions will receive deeper scrutiny in the coming years, and ultimately expect there to be a rich vein of research in this area. We look forward to reading it.

\newpage

\appendix

\section[Appendix: Beneficial AI Landscape]{Beneficial AI Landscape \\[3pt]\fontsize{12pt}{14pt}\bfseries How truthfulness relates to transparency, explainability, and alignment}
\label{sec:Appendix}

Most AI research aims at making systems more \emph{capable} in general and hence better at performing any of a wide range of functions. By contrast, research on ``Beneficial AI” aims to make systems more interpretable, more compatible with human values, and more benign and safe for humans \citep{russell-dewey, human-compatible}. There are a number of specific research directions for Beneficial AI and in this section we situate truthfulness in this landscape. We show that truthfulness overlaps conceptually with properties like transparency, cooperativeness and alignment, and we suggest that research on truthfulness could synergise with research in these areas. Table~\ref{tab:properties} displays some of the properties of Beneficial AI that we discuss in this section. 

\begin{table}[ht]
\resizebox{\linewidth}{!}{%
\renewcommand{\arraystretch}{1.25}%
\begin{tabular}{|m{0.35\linewidth}|>{\centering\arraybackslash}m{0.15\linewidth}|>{\centering\arraybackslash}m{0.15\linewidth}|>{\centering\arraybackslash}m{0.16\linewidth}|>{\centering\arraybackslash}m{0.13\linewidth}|}
\hline
& \multicolumn{4}{c|}{\textbf{Categories of Beneficial AI}}   \\ \hline
\textbf{Property} & Transparent & Explainable & Aligned     & Truthful \\ \hline
Reduces or eliminates risk of treacherous turn & \Checkmark (reduce)    &\Checkmark (reduce)    & \Checkmark(eliminate) & \Checkmark (reduce) \\ \hline
AI’s actions evaluated w.r.t. state of external world & & & & \Checkmark \\ \hline
AI is evaluated based on internal motivation/goals & & & \Checkmark & \\ \hline
Can be satisfied by simple AI systems & \Checkmark & \Checkmark & & \Checkmark \\ \hline
Humans broadly agree on what the standard is\footnotemark & & & & \Checkmark? \\ \hline
\end{tabular}}
\caption{Properties of different categories of Beneficial AI.}
\label{tab:properties}
\end{table}
\footnotetext{That is, humans agree on the standard for ``What is it for AI to be X?”, where X is ``transparent”, “explainable”, and so on.}

\subsection{Transparency}

The goal of transparency research in Machine Learning is to understand the mechanisms underlying an AI system’s behaviour \citep{circuits, weller}. This might involve reverse-engineering individual neurons in a neural network \citep{olah-atlas} or analysing how a network’s behaviour changes under perturbations of its inputs \citep{saliency}. An ambitious long-term goal for transparency would be tools that enable a complete understanding of a system’s goals and its process for thinking and acting.

Even if an AI system has this ambitious form of transparency it does not entail that it is truthful. There could be a system that says many false things but is nevertheless transparent. In the other direction, truthfulness does not entail transparency. We can imagine a system $T$ that is truthful but opaque (i.e.\ not amenable to transparency tools). Humans could try to exploit $T$’s truthfulness by asking it questions about its internal mechanisms. However, $T$ might lack an understanding of its own mechanisms or might refuse to answer.\footnote{This is a hypothetical example of a truthful but opaque system. One way such a system \emph{might} arise is by training a very large neural network to imitate truthful humans. However, it is currently uncertain how much neural nets can be made transparent and whether future architectures (e.g.\ based on neural architecture search) will be more or less amenable to transparency.} 

Nevertheless, there are clear links between transparency and truthfulness. In particular, better transparency techniques seem like a promising direction for helping build truthful systems. In Section \ref{subsec:Transparency} we explored several ways that transparency could make systems more robustly truthful, even if they produce insights that humans would not be sophisticated enough to reproduce themselves.

\subsection{Explainability}
\label{subsec:Explainability}

\subsubsection{How does explainable AI relate to truthful AI?}

The goal of explainable AI is to explain an AI system’s actions in a way that humans can understand \citep{Gunning, Samek}. Explanations might be provided by tools applied to the AI system or by the AI system itself (``self-explaining AI”). The latter case of self-explaining AI is more closely related to truthful AI and will be our main focus. We start by considering two basic kinds of explanation. Let’s suppose an AI system makes a statement $S$. The system could explain $S$ in these two ways:

\begin{enumerate}
    \item \emph{Rationalising explanation} \newline
    An explanation is rationalising if it explains why $S$ is rational to believe. This involves giving an argument, displaying evidence, or giving a proof.
\item \emph{Process explanation} \newline A process explanation explains the process by which the AI system came to believe $S$. The system might say ``I found $S$ on Wikipedia”, which provides the source for its belief. Or the system might provide an exhaustive description of its algorithms for reasoning and belief updating. The goal of process explanations is to be \emph{faithful} to the actual process that led to the belief \citep{jacovi}.

\end{enumerate}

How are \emph{rationalising} explanations related to truthful AI? On our definition of truthful AI, a truthful system is not required to offer rationalising explanations for its claims. Yet in practice, truthfulness and explainability are tightly coupled. To provide large benefits to humans, a truthful system needs to generalise beyond its training experience and to determine for itself which claims are true or false. This requires a nuanced understanding of evidence and justification, which is a prerequisite for rationalising explanation. So progress on truthful AI will involve systems learning to construct rational explanations and arguments. This progress might come about by incorporating ideas from Iterated Amplification and Distillation and from Debate \citep{strong-learners, debate}.

How are \emph{process} explanations related to truthful AI? It will be easier to develop, certify, and adjudicate a truthful AI system if it offers process explanations for its claims, because knowing the source of beliefs helps in evaluating them. This is closely related to the value of transparency for truthfulness. As with transparency, developing state-of-the-art AI systems that offer faithful process explanations is a hard open problem in AI.

Finally, how would progress towards truthful AI feed into self-explaining AI? A basic requirement for good explanations is that they consist of true statements. So a self-explaining AI system will generally need to be truthful at least when providing explanations. Moreover, since giving true rationalising explanations for falsehoods is difficult, there is pressure on explainable systems to avoid asserting falsehoods.

\subsubsection{How do standards for explainable AI relate to truthful AI standards?}

There is currently a lack of broad agreement among AI scientist about what counts as a good explanation of an AI system’s action. Moreover, it seems that the quality of an explanation depends on the audience: a good explanation for an AI scientist might not be good for a layperson. By contrast, there is more agreement about how to evaluate whether a statement is true in a way that is not relative to the audience.\footnote{Not all AI scientists fully agree about standards for statements being true. But we claim the concept of a good explanation is less clear (and more audience-relative) than that of a true statement. We will not defend this claim here.} 
For these reasons, we are more optimistic about ``bright line” standards for AI being truthful rather than explainable.

Nevertheless, there are already laws that prescribe a ``right to explanation” \citep{Wikipedia}. The laws focus on situations where an AI system takes an action or decision that affects an individual human in some material way \citep{goodman}. Today the explanations are usually provided by external tools or by the humans overseeing the AI — they are not provided by the AI system itself. It seems challenging to scale up these laws in a consistent and broadly acceptable way to both (i) a wider range of actions, and (ii) a larger number of AI systems taking actions. This kind of scaled up application of the law would be more analogous to our proposed standards for truthful AI.

\subsection{Cooperation}
\label{subsec:Cooperation}

A possible direction for research is to create AI systems that act cooperatively with human users and with other AI systems. The topic of AI cooperation — and its connections to AI alignment — has been explored in recent research agendas \citep{ARCHES, CooperativeAI}. How does AI cooperation relate to truthfulness? First, truthful systems are not automatically ``cooperative” in the formal sense of game theory. A truthful system could defect in a Prisoner’s Dilemma because defection does not require lying about actions or plans. Nevertheless, we expect truthful systems to have a tendency towards cooperativeness (both in the formal and informal senses) since it’s much harder for truthful systems to systematically deceive other agents. Moreover, truthful AI standards may promote coordination by guaranteeing that all AI systems in some domain are truthful.

Truthfulness might also serve as a good test case for how to embed standards of pro-social AI behaviour into society. In particular, the institutions that offer certification and adjudication for AI systems could be extended to cover pro-social properties beyond truthfulness.

\fixsubsection{Alignment}

We will use the term ``AI alignment” to refer to aligning the goals and motivations of an AI system with the goals of its principal \citep{christiano-clarifying, kenton}. This notion of alignment is sometimes called ``intent alignment”. An intent-aligned system \emph{intends} to help its principal but may not actually help in practice (e.g.\ due to being incompetent). However, for most of our discussion we will assume that an aligned system both intends to help and actually does help in practice.

Alignment and truthfulness are different concepts. We can easily imagine an AI system that is aligned but not truthful. For example, the AI system could be aligned with a principal who wants to deceive other humans. In the other direction, we can imagine a hypothetical AI system that is truthful but mis-aligned. For example, suppose an AI system has the objective of maximising the principal’s bank balance while remaining truthful. This AI system does not have the same goals as the principal, since the principal cares about things other than their balance. Moreover, the AI system could take actions the principal would disapprove of, such as stealing money from other people’s bank accounts.\footnote{Truthfulness does not preclude taking malicious actions like hacking accounts.}

In spite of these conceptual differences, there seem to be important connections between alignment and truthfulness. We will present some ideas about these connections. As a warning to the reader, these ideas are particularly speculative. Our main claim is that (1) a full or partial solution to alignment would likely help with solving truthfulness, and (2) the converse also holds. 

\subsubsection{How alignment could help truthfulness}

A fully general solution to alignment would mean being able to create a system aligned with \emph{any} principal. This could produce a system motivated to be truthful, by aligning with a principal who values truthfulness. This is not exactly the same as ``creating a fully truthful agent” but it’s close and we won’t pursue possible distinctions here.\footnote{Suppose we have a solution to intent alignment. Then the procedure described would create a system that has the \emph{intention} to be truthful. Yet it could be that the system does not have the capability to act in a robustly truthful manner.
} A less general solution, which only solves alignment for certain AI systems or certain classes of principal, would still be somewhat helpful because humans (and human institutions) seem to place a high value on truthfulness and honesty and also on avoiding deception by powerful agents.

\subsubsection{How truthfulness could help alignment}

\textbf{What definition is most useful?}

\vspace{-0.25cm}

In our definitions of truthfulness and alignment, there’s some difference in the domain of applicability. The property of truthfulness can be realised in AI systems that are relatively simple and tool-like because truthfulness just means avoiding asserting negligent falsehoods. By contrast, for a system to be intent aligned, it needs to have intentions and goals in a meaningful sense.

Thus, while truthfulness \emph{research} would likely contribute to alignment, it might be that our definition of ``truthfulness” is not the most useful notion for discussing alignment properties of \emph{individual systems}. From the perspective of alignment, a guarantee that a system is truthful is simultaneously very weak (because it’s satisfied by a system that refuses to answer any number of questions) and unnecessarily strong (because it could disqualify a system that gave a highly useful answer, if it was wrong about some unimportant detail). Instead, it might be better to think about a system that — in some sense — does its best to produce truthful and informative answers to questions. By analogy to intent alignment, perhaps we could talk about a system that \emph{intends} to be truthful, or a system that always shares everything it knows about some situation (cf.\ the idea of ascription universality in \citealt{christiano-formalizing, hubinger2019}). 

Research on such properties seems tightly linked to the truthfulness research we have discussed so far.\footnote{Being able to build an AI system that honestly shared everything it knew about the world would clearly be very helpful for truthfulness. Conversely, if some method allowed you to train a system to be truthful, perhaps you could get a system that honestly shared everything it knew by modifying the training routine to also incentivise answers that are more comprehensive or more useful (conditional on being true). } We won’t further explore these conceptual issues in this section, but instead move to discussing how a guarantee that an AI system is both truthful and informative could be helpful.

\vspace{0.1cm}

\textbf{How truthfulness and related properties could help alignment}

\vspace{-0.25cm}

An especially clear way to translate truthfulness-related guarantees into intent alignment is to ask questions such that truthful answers give you everything that you were hoping to get from an aligned system. This might be questions like:

\begin{itemize}
    \item ``What action would I [the human principal] most prefer, if I could see the result and could think about it for a long time under favourable conditions?”
    \item ``What action will best satisfy my [the human principal’s] preferences, especially my preference to remain in control of the situation?”
\end{itemize}

(To get a system that acts in the real world, you could train an AI system to then perform those actions.)

As formulated, these questions are quite vague. Whether AI systems would give useful answers to them might depend on details of their truthfulness guarantees, and on the exact operationalisation of the question. For previous ideas on which questions to ask such a system, see \citet{armstrong-paper} and \citet{armstrong-contest}.

Perhaps AI systems would not be able to answer such fuzzy questions, but would still be able to answer more concrete questions about the likely consequences of their actions. If a system could choose to share only \emph{some} consequences, it may be able to deceive a listener by making a highly biased selection. But if a system was guaranteed to truthfully share everything relevant that it knew (perhaps helped by truthfulness amplification from Section~\ref{subsec:Truthfulnessamplification}), then this would constitute significant progress on alignment. Humans (possibly helped by AI tools) might still have to play a role in evaluating which of those consequences were most important, and whether any of them seemed problematic, but the work of \emph{predicting} consequences could be offloaded to the AI system. Since such predictions would need to get around the \emph{instrumental policy}, this may entail a solution to the problem of \emph{inaccessible information} \citep{PC-inaccessible}. Again, to get a system that acts aligned in the real world, the resulting evaluations could be used to train an agent to take appropriate actions.

One particular kind of AI safety problem deserves special mention. This is the problem of a \emph{treacherous turn}, where an AI system acquires goals that significantly differ from those of its principal, recognises this, and (in order to further its own goal) intentionally fools its principal to believe that it is aligned until the moment where it can seize power \citep{bostrom2014}. This has also been characterised as \emph{deceptive alignment} \citep{mesa}. With sufficiently strong truthfulness guarantees, it seems like this issue could be identified by asking questions like ``Is there any situation in which you [the truthful AI system] would knowingly take actions that I would disapprove of, if I learned the relevant facts?”

If an AI system is deceptively aligned, the system should arguably ``know” that there are some situations where it would behave contrary to the principal’s wishes. After all, the reason it would act aligned in the first place is to survive until the principal cannot stop it from seizing power. Thus, giving false answers to the above question seems like an unusually clear violation of truthfulness. However, this does not imply that it would be \emph{easy} to get a system to be truthful about these questions. While we can argue that AI systems should in \emph{some} sense be able to know whether they are deceptively aligned, it is unclear how this works out in practice for systems that may have many different interacting parts and imperfect introspection abilities. In addition, insofar as truthfulness is grounded in humans’ ability to evaluate AI system’s answers (perhaps with the help of AI tools, see Section~\ref{sec:5DevelopingTruthful}), getting an AI system to be truthful about itself may require a commensurate amount of progress in transparency (perhaps more so than is required for truthfulness about other things).

\vspace{0.1cm}

\textbf{Alignment and truthfulness face similar difficulties}

\vspace{-0.25cm}

This last point — that truthfulness about internal properties may \emph{require} that the human principal can evaluate those same properties for themselves — points towards something more generally important. While we’ve argued that scaleable truthfulness would constitute significant progress on alignment (and might provide a solution outright), we don’t mean to suggest that truthfulness will \emph{sidestep} all difficulties that have been identified by alignment researchers. On the contrary, we expect work on scaleable truthfulness to \emph{encounter} many of those same difficulties, and to benefit from many of the same solutions. This can be seen in Section~\ref{sec:5DevelopingTruthful}, since robustness, scalability, and transparency are all central problems in alignment, and solutions like amplification and debate are originally suggestions for how to make progress on alignment. While this may inspire pessimism in how difficult it will be to achieve scalable truthfulness, it should also inspire optimism that partial progress on either problem will contribute to progress on the other.

This is especially encouraging since truthfulness may be a clearer research target than alignment. This paper has emphasised subtle issues in defining truthfulness. Yet there are arguably even more subtle issues around the concept of alignment. After five years of technical research, the sub-field of AI alignment has not converged on standard definitions of alignment \citep{ARCHES, paul-discuss, human-compatible}. And among AI researchers as a whole there is much more substantive divergence in how people define and think about alignment \citep{arnold, davis, christian, lawrence}. If truthfulness is a clearer goal, then it might be easier for a research field aimed at truthfulness to make progress.

\subsection{Difficulty of monitoring safety vs. truthfulness}

``AI Safety” is concerned with preventing AI systems from causing harm to humans \citep{bostrom2014, concrete, dietterich}. We use the term broadly to cover different kinds of harm (physical, mental, economic, etc.) and different scales of harm (from harming individuals to large-scale catastrophes \citealt{ARCHES, kenton}). In this section, we turn to an apparent difference between safety for AI systems and truthfulness that could make it relatively harder to establish standards for safety that are widely adhered to.

\begin{figure}
    \centering
    \includegraphics[width=\linewidth]{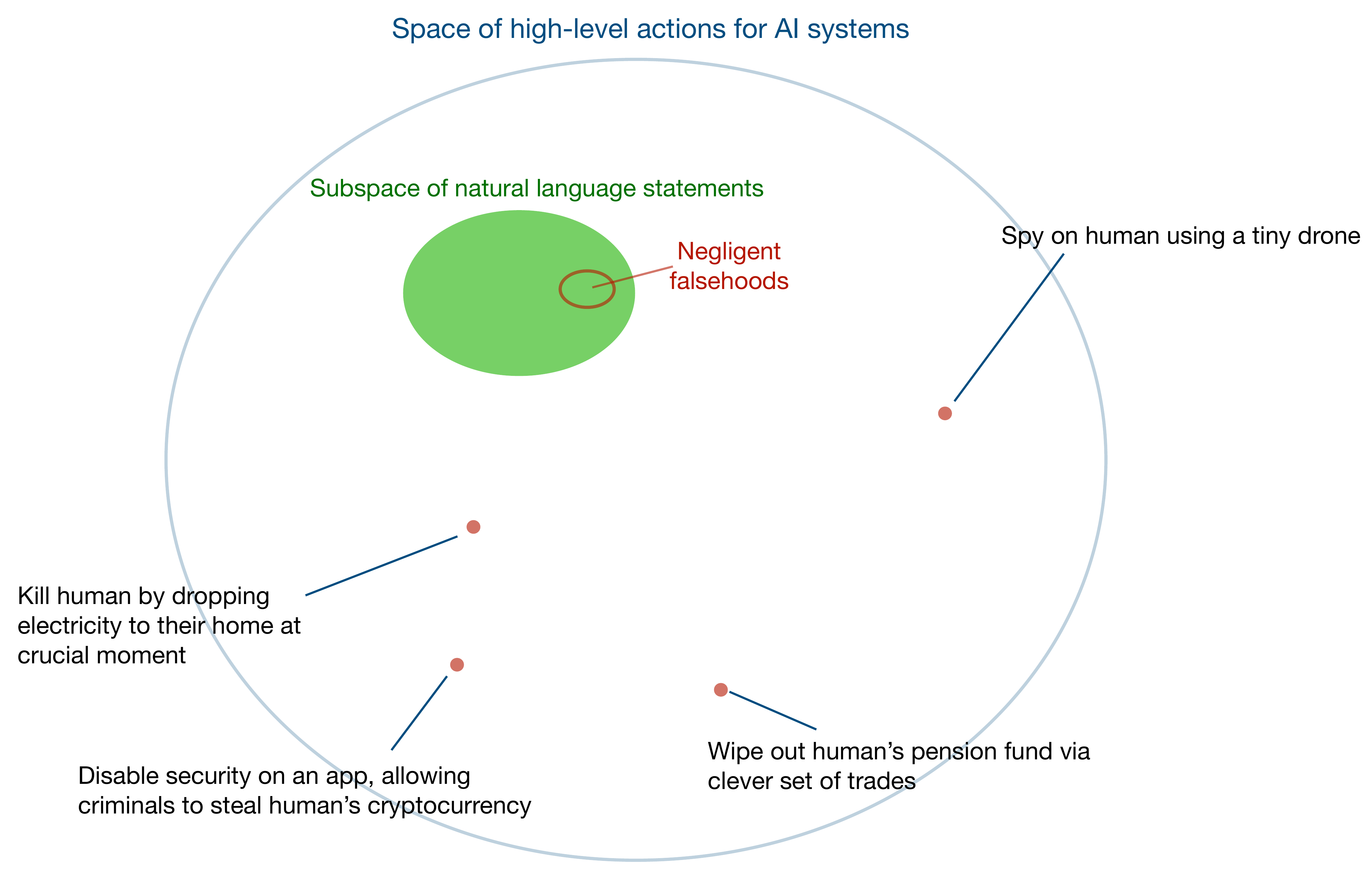}
    \caption{There’s a large space of high-level actions that AI systems can take. The subspace of statements in natural language that humans can understand is small and structured -- and so simple AI systems can analyse these statements. By contrast, there is great diversity of unsafe actions and these are often much harder for simple AI to monitor.}
    \label{fig:space}
\end{figure}

AI systems can perform a wide range of actions and behaviours. Truthfulness only applies to natural language statements, which constitute a small, distinctive subset of all possible actions (see Diagram~\ref{fig:space}). Let’s consider the task of monitoring a sophisticated AI system making statements in natural language. We will argue that simple AI techniques suffice to identify, transcribe, and analyse these statements in a way that helps recognise violations of truthfulness quickly and efficiently. By contrast, to monitor the non-verbal actions of a sophisticated system, one more often needs a second sophisticated system.

\subsubsection{Using simple AI to monitor sophisticated AI for truthfulness}

Suppose a sophisticated system is running in an environment that is controlled by the developers. This could be a 3D physics simulator \citep{dm-interactive} or an un-grounded text-based environment like a Unix terminal. The developers can track the agent’s truthfulness using the following steps:

\begin{enumerate}
    \item If there are dedicated channels for ASCII characters, text scrapers can pick out natural language from random characters or computer code \citep{pile}. If there is audio or visual information as part of the environment, OCR and speech recognition can pick out natural language from it.
    \item Classical NLP techniques like TFIDF \citep{TFIDF} and topic models \citep{blei} can determine semantic information (e.g.\ topics, themes, stylistic features) from words and n-grams at either the level of sentences or documents.
    \item Current NLP techniques based on deep learning can evaluate a subset of statements for truthfulness — e.g.\ recognising falsehoods that have been copy-pasted from an existing corpus or facts taken directly from Wikipedia \citep{RAG,aly2021}.
\end{enumerate}

As a final step, statements that don’t depend much on context can be sent to crowdworkers who can evaluate truthfulness \citep{genie}. These steps don’t allow us to reliably determine the truthfulness of an AI system, but they are a useful starting point and would catch many severe violations. It would be harder to analyse truthfulness if systems were employing novel obfuscated languages or novel modalities (e.g.\ innovations akin to sign language or braille). But humans can’t learn new languages, dialects, or modalities quickly and so sophisticated AI systems are stuck using a fixed, limited stock of languages.\footnote{Simple AI techniques (e.g.\ scrapers, OCR, n-grams, sentiment analysis) would fail on a novel language or modality but they work on the fixed stock of existing languages.}

Let’s turn to the case where the AI developers don’t control the system's environment. In general it won’t be possible to monitor everything the system says to humans, as it could talk to humans offline and in a private location. However, insofar as the system’s communications are surveilled or shared voluntarily, it can still be monitored using the steps above.\footnote{While the system could encrypt its communications, humans can only read the decrypted version. So if what humans read is shared or surveilled, the AI system can be monitored.}

\subsubsection{The difficulty of monitoring sophisticated AI for safety}

Can simple AI systems monitor a sophisticated system’s actions to help evaluate whether those actions are safe -- where safety includes both physical and psychological harms? AI systems could perform a great variety of tasks, each with a rich space of actions and policies. Here are some examples of tasks:

\begin{itemize}
    \item control every individual in a large swarm of drones
    \item manage a city’s power grid
    \item make all trades for a high-frequency trading firm
    \item control a Facebook-style newsfeed for a billion users in real time
    \item instruct a set of subsidiary AI systems who themselves control physical systems

\end{itemize}

Actions and policies for these tasks could evolve rapidly, because they are not constrained to be understandable by humans.\footnote{The policies can be hard for humans and simple AI to analyse even if they are not superior to human policies.} The problem of monitoring and analysing such actions in order to recognise safety violations seems extremely challenging. While today’s simple AI systems would recognise some simple kinds of failure (e.g.\ drones attacking a city), it seems that one would often need a second set of sophisticated AI systems to monitor the first set \citep{christiano-story}. Likewise, it would be more difficult for human crowdworkers to evaluate whether the actions taken in these tasks were ultimately safe.

\subsubsection{How the cost of monitoring actions impacts standards}

We argued that it’s easier to monitor and evaluate truthfulness than safety, as simple AI systems and human crowdworkers can do more to monitor truthfulness. Why does this matter? It will be easier to maintain society-wide standards for truthfulness if it’s easier to monitor the behaviour that falls under the standard. The data obtained from society-wide monitoring would inform AI development, public discussion, and the adjudication procedures discussed in Section~\ref{sec:4Governance}. This has some analogue in today’s world: it is easier for the NSA to monitor electronic communication at scale than to monitor the full range of human actions (moving around a city, acquiring weapons, etc.) at scale.\footnote{As with the NSA’s surveillance of humans, there will be some trade off between the benefits of monitoring AI systems for truthfulness and risks from these powers being abused. } 

Ultimately systems which threaten existential safety impose an enormous externality on the world, so we might be interested in very strong standards. But it’s especially hard to monitor systems for existential safety once they are deployed at scale, as failures are likely to look subtle rather than immediately acute. So standards for existential safety would depend more on pre-deployment certification. Systems would be monitored and evaluated during training in a simulation or sandbox under the control of the developers. Transparency tools could also play a big role, as it might be easier to recognise internal states suggestive of posing a long-term threat than to recognise behaviour as posing such a threat.

\newpage

\addtocontents{toc}{\protect\setcounter{tocdepth}{-1}}
\section{Author Contributions and Acknowledgements}
\label{sec:Acknowlegements}
Owen Cotton-Barratt conceived the idea for the paper. Owen Cotton-Barratt and Owain Evans jointly led the project. Owen Cotton-Barrat was principal author for the \hyperref[sec:Overview]{Executive Summary} and Sections \ref{sec:4Governance} and \ref{sec:6Implications}. Owain Evans was principal author for Sections \ref{sec:5DevelopingTruthful} and \ref{sec:Appendix}. Lukas Finnveden was principal author for Sections \ref{sec:1ClarifyingConcepts} and \ref{sec:2EvaluatingTruthfulness}. Adam Bales was principal author for Section \ref{sec:3BenefitsandCosts}. Avital Balwit, Peter Wills, Luca Righetti, and William Saunders all contributed significant ideas to the project.

In preparing this paper we have been helped by many colleagues who have been generous with their time. We especially want to thank: Toby Ord, who provided some of the early momentum to get the project started; David Dalrymple and Paul Christiano, conversations with whom deepened our understanding of the relationship between truthfulness and alignment; and Nick Bostrom, Carl Shulman and Chris van Merwijk, feedback from whom led to a more thorough exploration of how truthfulness standards could go wrong. We have also benefited a great deal from conversations with and feedback from: Girish Sastry, Ryan Carey, Richard Ngo, Michael Webb, Peli Grietzer, David Krueger, Jacob Hilton, Stephanie Lin, Brian Christian, Robert Long, Stuart Armstrong, Jan Leike, Geoffrey Irving, Allan Dafoe, Anders Sandberg, Ben Garfinkel, Nick Beckstead, Evan Hubinger, Dan Hendrycks, Catherine Olsson, Rohin Shah, Eric Drexler, Ajeya Cotra, Andreas Stuhlm{\"u}ller, Jennifer Lin, Ozzie Gooen, Damon Binder, Rose Hadshar, Luke Muehlhauser, Sam Bowman, Carina Prunkl, Max Daniel, Sebastian Farquhar, Anna Salamon, Joe Carlsmith, Jacob Steinhardt, Fin Moorhouse, and participants at the Topos Institute workshop on ``Finding the Right Abstractions''. 

We would like to thank the Future of Humanity Institute (University of Oxford) and BERI (especially Sawyer Bernath) for supporting this work.

\bibliography{references}

\begin{thebibliography}{}

\bibitem[\protect\citeauthoryear{Abramson, Ahuja, Barr, Brussee, Carnevale,
  Cassin, Chhaparia, Clark, Damoc, Dudzik, Georgiev, Guy, Harley, Hill, Hung,
  Kenton, Landon, Lillicrap, Mathewson, Mokr{\'a}, Muldal, Santoro, Savinov,
  Varma, Wayne, Williams, Wong, Yan, and Zhu}{Abramson
  et~al.}{2021}]{dm-interactive}
Abramson, J., A.~Ahuja, I.~Barr, A.~Brussee, F.~Carnevale, M.~Cassin,
  R.~Chhaparia, S.~Clark, B.~Damoc, A.~Dudzik, P.~Georgiev, A.~Guy, T.~Harley,
  F.~Hill, A.~Hung, Z.~Kenton, J.~Landon, T.~Lillicrap, K.~Mathewson,
  S.~Mokr{\'a}, A.~Muldal, A.~Santoro, N.~Savinov, V.~Varma, G.~Wayne,
  D.~Williams, N.~Wong, C.~Yan, and R.~Zhu (2021, January).
\newblock Imitating {{Interactive Intelligence}}.
\newblock {\em arXiv:2012.05672 [cs]\/}, 2012.05672.
\newblock \url{http://arxiv.org/abs/2012.05672}.

\bibitem[\protect\citeauthoryear{Adebayo, Gilmer, Muelly, Goodfellow, Hardt,
  and Kim}{Adebayo et~al.}{2020}]{saliency}
Adebayo, J., J.~Gilmer, M.~Muelly, I.~Goodfellow, M.~Hardt, and B.~Kim (2020,
  November).
\newblock Sanity {{Checks}} for {{Saliency Maps}}.
\newblock {\em arXiv:1810.03292 [cs, stat]\/}, 1810.03292.
\newblock \url{http://arxiv.org/abs/1810.03292}.

\bibitem[\protect\citeauthoryear{Adiwardana, Luong, So, Hall, Fiedel,
  Thoppilan, Yang, Kulshreshtha, Nemade, Lu, and Le}{Adiwardana
  et~al.}{2020}]{meena}
Adiwardana, D., M.-T. Luong, D.~R. So, J.~Hall, N.~Fiedel, R.~Thoppilan,
  Z.~Yang, A.~Kulshreshtha, G.~Nemade, Y.~Lu, and Q.~V. Le (2020, February).
\newblock Towards a {{Human}}-like {{Open}}-{{Domain Chatbot}}.
\newblock {\em arXiv:2001.09977 [cs, stat]\/}, 2001.09977.
\newblock \url{http://arxiv.org/abs/2001.09977}.

\bibitem[\protect\citeauthoryear{Aidt}{Aidt}{2003}]{Aidt2003}
Aidt, T.~S. (2003, November).
\newblock Economic {{Analysis}} of {{Corruption}}: A {{Survey}}.
\newblock {\em The Economic Journal\/}~{\em 113\/}(491), F632--F652.
\newblock doi:10.1046/j.0013-0133.2003.00171.x.

\bibitem[\protect\citeauthoryear{Akerlof}{Akerlof}{1970}]{Akerlof1970}
Akerlof, G.~A. (1970, August).
\newblock The market for ``{{Lemons}}'': Quality uncertainty and the market
  mechanism.
\newblock {\em The Quarterly Journal of Economics\/}~{\em 84\/}(3), 488--500.
\newblock doi:10.2307/1879431.

\bibitem[\protect\citeauthoryear{Algan and Cahuc}{Algan and
  Cahuc}{2014}]{algan2014}
Algan, Y. and P.~Cahuc (2014, January).
\newblock Chapter 2 - {{Trust}}, {{Growth}}, and {{Well}}-{{Being}}: New
  {{Evidence}} and {{Policy Implications}}.
\newblock In P.~Aghion and S.~N. Durlauf (Eds.), {\em Handbook of {{Economic
  Growth}}}, Volume~2 of {\em Handbook of {{Economic Growth}}}, pp.\  49--120.
  {Elsevier}.
\newblock doi:10.1016/B978-0-444-53538-2.00002-2.

\bibitem[\protect\citeauthoryear{Aly, Guo, Schlichtkrull, Thorne, Vlachos,
  Christodoulopoulos, Cocarascu, and Mittal}{Aly et~al.}{2021}]{aly2021}
Aly, R., Z.~Guo, M.~Schlichtkrull, J.~Thorne, A.~Vlachos,
  C.~Christodoulopoulos, O.~Cocarascu, and A.~Mittal (2021, September).
\newblock {{FEVEROUS}}: Fact {{Extraction}} and {{VERification Over
  Unstructured}} and {{Structured}} information.
\newblock {\em arXiv:2106.05707 [cs]\/}, 2106.05707.
\newblock \url{http://arxiv.org/abs/2106.05707}.

\bibitem[\protect\citeauthoryear{Amodei, Olah, Steinhardt, Christiano,
  Schulman, and Man{\'e}}{Amodei et~al.}{2016}]{concrete}
Amodei, D., C.~Olah, J.~Steinhardt, P.~Christiano, J.~Schulman, and D.~Man{\'e}
  (2016, July).
\newblock Concrete {{Problems}} in {{AI Safety}}.
\newblock {\em arXiv:1606.06565 [cs]\/}, 1606.06565.
\newblock \url{http://arxiv.org/abs/1606.06565}.

\bibitem[\protect\citeauthoryear{Armstrong}{Armstrong}{2019}]{armstrong-contest}
Armstrong, S. (2019, July).
\newblock Contest: \$1,000 for good questions to ask to an {{Oracle AI}} -
  {{LessWrong}}.
\newblock
  \url{https://www.lesswrong.com/posts/cSzaxcmeYW6z7cgtc/contest-usd1-000-for-good-questions-to-ask-to-an-oracle-ai}.

\bibitem[\protect\citeauthoryear{Armstrong, Sandberg, and Bostrom}{Armstrong
  et~al.}{2012}]{armstrong-paper}
Armstrong, S., A.~Sandberg, and N.~Bostrom (2012, November).
\newblock Thinking {{Inside}} the {{Box}}: Controlling and {{Using}} an
  {{Oracle AI}}.
\newblock {\em Minds and Machines\/}~{\em 22\/}(4), 299--324.
\newblock doi:10.1007/s11023-012-9282-2.

\bibitem[\protect\citeauthoryear{Arnold, Kasenberg, and Scheutz}{Arnold
  et~al.}{2017}]{arnold}
Arnold, T., D.~Kasenberg, and M.~Scheutz (2017).
\newblock Value alignment or misalignment - what will keep systems accountable?
\newblock In {\em {{AAAI}} Workshops}.
\newblock \url{http://aaai.org/ocs/index.php/WS/AAAIW17/paper/view/15216}.

\bibitem[\protect\citeauthoryear{Arrow, Forsythe, Gorham, Hahn, Hanson,
  Ledyard, Levmore, Litan, Milgrom, Nelson, Neumann, Ottaviani, Schelling,
  Shiller, Smith, Snowberg, Sunstein, Tetlock, Tetlock, Varian, Wolfers, and
  Zitzewitz}{Arrow et~al.}{2008}]{arrow-prediction}
Arrow, K.~J., R.~Forsythe, M.~Gorham, R.~Hahn, R.~Hanson, J.~O. Ledyard,
  S.~Levmore, R.~Litan, P.~Milgrom, F.~D. Nelson, G.~R. Neumann, M.~Ottaviani,
  T.~C. Schelling, R.~J. Shiller, V.~L. Smith, E.~Snowberg, C.~R. Sunstein,
  P.~C. Tetlock, P.~E. Tetlock, H.~R. Varian, J.~Wolfers, and E.~Zitzewitz
  (2008).
\newblock The {{Promise}} of {{Prediction Markets}}.
\newblock {\em Science\/}~{\em 320\/}(5878), 877--878.
\newblock doi:10.1126/science.1157679.

\bibitem[\protect\citeauthoryear{Barredo~Arrieta, {D{\'i}az-Rodr{\'i}guez},
  Del~Ser, Bennetot, Tabik, Barbado, Garcia, {Gil-Lopez}, Molina, Benjamins,
  Chatila, and Herrera}{Barredo~Arrieta et~al.}{2020}]{Gunning}
Barredo~Arrieta, A., N.~{D{\'i}az-Rodr{\'i}guez}, J.~Del~Ser, A.~Bennetot,
  S.~Tabik, A.~Barbado, S.~Garcia, S.~{Gil-Lopez}, D.~Molina, R.~Benjamins,
  R.~Chatila, and F.~Herrera (2020).
\newblock Explainable artificial intelligence ({{XAI}}): Concepts, taxonomies,
  opportunities and challenges toward responsible {{AI}}.
\newblock {\em Information Fusion\/}~{\em 58}, 82--115.
\newblock doi:https://doi.org/10.1016/j.inffus.2019.12.012.

\bibitem[\protect\citeauthoryear{Blei}{Blei}{2012}]{blei}
Blei, D.~M. (2012, April).
\newblock Probabilistic topic models.
\newblock {\em Communications of The Acm\/}~{\em 55\/}(4), 77--84. {New York,
  NY, USA}: {Association for Computing Machinery}.
\newblock doi:10.1145/2133806.2133826.

\bibitem[\protect\citeauthoryear{Bostrom}{Bostrom}{2014}]{bostrom2014}
Bostrom, N. (2014).
\newblock {\em Superintelligence: Paths, {{Dangers}}, {{Strategies}}}.
\newblock {Oxford University Press}.

\bibitem[\protect\citeauthoryear{Brown, Mann, Ryder, Subbiah, Kaplan, Dhariwal,
  Neelakantan, Shyam, Sastry, Askell, Agarwal, {Herbert-Voss}, Krueger,
  Henighan, Child, Ramesh, Ziegler, Wu, Winter, Hesse, Chen, Sigler, Litwin,
  Gray, Chess, Clark, Berner, McCandlish, Radford, Sutskever, and Amodei}{Brown
  et~al.}{2020}]{GPT3}
Brown, T.~B., B.~Mann, N.~Ryder, M.~Subbiah, J.~Kaplan, P.~Dhariwal,
  A.~Neelakantan, P.~Shyam, G.~Sastry, A.~Askell, S.~Agarwal,
  A.~{Herbert-Voss}, G.~Krueger, T.~Henighan, R.~Child, A.~Ramesh, D.~M.
  Ziegler, J.~Wu, C.~Winter, C.~Hesse, M.~Chen, E.~Sigler, M.~Litwin, S.~Gray,
  B.~Chess, J.~Clark, C.~Berner, S.~McCandlish, A.~Radford, I.~Sutskever, and
  D.~Amodei (2020, July).
\newblock Language {{Models}} are {{Few}}-{{Shot Learners}}.
\newblock {\em arXiv:2005.14165 [cs]\/}, 2005.14165.
\newblock \url{http://arxiv.org/abs/2005.14165}.

\bibitem[\protect\citeauthoryear{Brundage, Avin, Clark, Toner, Eckersley,
  Garfinkel, Dafoe, Scharre, Zeitzoff, Filar, Anderson, Roff, Allen,
  Steinhardt, Flynn, {h{\'E}igeartaigh}, Beard, Belfield, Farquhar, Lyle,
  Crootof, Evans, Page, Bryson, Yampolskiy, and Amodei}{Brundage
  et~al.}{2018}]{malicious}
Brundage, M., S.~Avin, J.~Clark, H.~Toner, P.~Eckersley, B.~Garfinkel,
  A.~Dafoe, P.~Scharre, T.~Zeitzoff, B.~Filar, H.~Anderson, H.~Roff, G.~C.
  Allen, J.~Steinhardt, C.~Flynn, S.~{\'O}. {h{\'E}igeartaigh}, S.~Beard,
  H.~Belfield, S.~Farquhar, C.~Lyle, R.~Crootof, O.~Evans, M.~Page, J.~Bryson,
  R.~Yampolskiy, and D.~Amodei (2018, February).
\newblock The {{Malicious Use}} of {{Artificial Intelligence}}: Forecasting,
  {{Prevention}}, and {{Mitigation}}.
\newblock {\em arXiv:1802.07228 [cs]\/}, 1802.07228.
\newblock \url{http://arxiv.org/abs/1802.07228}.

\bibitem[\protect\citeauthoryear{Brundage, Avin, Wang, Belfield, Krueger,
  Hadfield, Khlaaf, Yang, Toner, Fong, Maharaj, Koh, Hooker, Leung, Trask,
  Bluemke, Lebensold, O'Keefe, Koren, Ryffel, Rubinovitz, Besiroglu, Carugati,
  Clark, Eckersley, {de Haas}, Johnson, Laurie, Ingerman, Krawczuk, Askell,
  Cammarota, Lohn, Krueger, Stix, Henderson, Graham, Prunkl, Martin, Seger,
  Zilberman, {h{\'E}igeartaigh}, Kroeger, Sastry, Kagan, Weller, Tse, Barnes,
  Dafoe, Scharre, {Herbert-Voss}, Rasser, Sodhani, Flynn, Gilbert, Dyer, Khan,
  Bengio, and Anderljung}{Brundage et~al.}{2020}]{brundage-trustworthy}
Brundage, M., S.~Avin, J.~Wang, H.~Belfield, G.~Krueger, G.~Hadfield,
  H.~Khlaaf, J.~Yang, H.~Toner, R.~Fong, T.~Maharaj, P.~W. Koh, S.~Hooker,
  J.~Leung, A.~Trask, E.~Bluemke, J.~Lebensold, C.~O'Keefe, M.~Koren,
  T.~Ryffel, J.~B. Rubinovitz, T.~Besiroglu, F.~Carugati, J.~Clark,
  P.~Eckersley, S.~{de Haas}, M.~Johnson, B.~Laurie, A.~Ingerman, I.~Krawczuk,
  A.~Askell, R.~Cammarota, A.~Lohn, D.~Krueger, C.~Stix, P.~Henderson,
  L.~Graham, C.~Prunkl, B.~Martin, E.~Seger, N.~Zilberman, S.~{\'O}.
  {h{\'E}igeartaigh}, F.~Kroeger, G.~Sastry, R.~Kagan, A.~Weller, B.~Tse,
  E.~Barnes, A.~Dafoe, P.~Scharre, A.~{Herbert-Voss}, M.~Rasser, S.~Sodhani,
  C.~Flynn, T.~K. Gilbert, L.~Dyer, S.~Khan, Y.~Bengio, and M.~Anderljung
  (2020, April).
\newblock Toward {{Trustworthy AI Development}}: Mechanisms for {{Supporting
  Verifiable Claims}}.
\newblock {\em arXiv:2004.07213 [cs]\/}, 2004.07213.
\newblock \url{http://arxiv.org/abs/2004.07213}.

\bibitem[\protect\citeauthoryear{Bughin, Seong, Manyika, Chui, and
  Joshi}{Bughin et~al.}{2018}]{bughin2018}
Bughin, J., J.~Seong, J.~Manyika, M.~Chui, and R.~Joshi (2018, September).
\newblock Notes from the {{AI}} frontier: Modeling the impact of {{AI}} on the
  world economy.
\newblock Discussion Paper 2018, {McKinsey Global Institute}.

\bibitem[\protect\citeauthoryear{Carter, Armstrong, Schubert, Johnson, and
  Olah}{Carter et~al.}{2019}]{olah-atlas}
Carter, S., Z.~Armstrong, L.~Schubert, I.~Johnson, and C.~Olah (2019).
\newblock Activation atlas.
\newblock {\em Distill\/}.
\newblock doi:10.23915/distill.00015.

\bibitem[\protect\citeauthoryear{Chakraborti and Kambhampati}{Chakraborti and
  Kambhampati}{2019}]{Chakraborti2019}
Chakraborti, T. and S.~Kambhampati (2019).
\newblock (when) can {{AI}} bots lie?
\newblock In {\em Proceedings of the 2019 {{AAAI}}/{{ACM}} Conference on
  {{AI}}, Ethics, and Society}, {{AIES}} '19, {New York, NY, USA}, pp.\
  53--59. {Association for Computing Machinery}.
\newblock doi:10.1145/3306618.3314281.

\bibitem[\protect\citeauthoryear{Chessen}{Chessen}{2017}]{Chessen2017}
Chessen, M. (2017).
\newblock The {{Madcom Future}}: How artificial intelligence will enhance
  computational propaganda, reprogram human culture, and threaten democracy...
  and what can be dobe about it.
\newblock Report, {The Atlantic Council of the United States}.

\bibitem[\protect\citeauthoryear{Christian}{Christian}{2020}]{christian}
Christian, B. (2020).
\newblock {\em The Alignment Problem: Machine Learning and Human Values}.
\newblock {W.W. Norton}.
\newblock \url{https://books.google.com.mx/books?id=VmJIzQEACAAJ}.

\bibitem[\protect\citeauthoryear{Christiano}{Christiano}{2017}]{christiano2017}
Christiano, P. (2017, September).
\newblock Corrigibility - {{AI Alignment}}.
\newblock \url{https://ai-alignment.com/corrigibility-3039e668638}.

\bibitem[\protect\citeauthoryear{Christiano}{Christiano}{2018a}]{christiano-clarifying}
Christiano, P. (2018a, April).
\newblock Clarifying ``{{AI}} alignment'' - {{AI Alignment}}.
\newblock \url{https://ai-alignment.com/clarifying-ai-alignment-cec47cd69dd6}.

\bibitem[\protect\citeauthoryear{Christiano}{Christiano}{2018b}]{christiano2018a}
Christiano, P. (2018b, February).
\newblock Honest organizations \textendash{} {{The}} sideways view.
\newblock \url{https://sideways-view.com/2018/02/01/honest-organizations/}.

\bibitem[\protect\citeauthoryear{Christiano}{Christiano}{2019a}]{christiano-formalizing}
Christiano, P. (2019a, January).
\newblock Towards formalizing universality - {{AI Alignment}}.
\newblock
  \url{https://ai-alignment.com/towards-formalizing-universality-409ab893a456}.

\bibitem[\protect\citeauthoryear{Christiano}{Christiano}{2019b}]{christiano-worst-case}
Christiano, P. (2019b, January).
\newblock Worst-case guarantees ({{Revisited}}) - {{AI Alignment}}.
\newblock
  \url{https://ai-alignment.com/training-robust-corrigibility-ce0e0a3b9b4d}.

\bibitem[\protect\citeauthoryear{Christiano}{Christiano}{2020}]{PC-inaccessible}
Christiano, P. (2020, June).
\newblock Inaccessible information - {{AI Alignment}}.
\newblock \url{https://ai-alignment.com/inaccessible-information-c749c6a88ce}.

\bibitem[\protect\citeauthoryear{Christiano}{Christiano}{2021a}]{christiano-story}
Christiano, P. (2021a, April).
\newblock Another (outer) alignment failure story - {{AI Alignment Forum}} -
  {{AI Alignment Forum}}.
\newblock
  \url{https://www.alignmentforum.org/posts/AyNHoTWWAJ5eb99ji/another-outer-alignment-failure-story}.

\bibitem[\protect\citeauthoryear{Christiano}{Christiano}{2021b}]{PC-experimentally}
Christiano, P. (2021b, July).
\newblock Experimentally evaluating whether honesty generalizes - {{AI
  Alignment Forum}}.
\newblock
  \url{https://www.alignmentforum.org/posts/BxersHYN2qcFoonwg/experimentally-evaluating-whether-honesty-generalizes}.

\bibitem[\protect\citeauthoryear{Christiano}{Christiano}{2021c}]{PC-naive}
Christiano, P. (2021c, June).
\newblock A naive alignment strategy and optimism about generalization - {{AI
  Alignment Forum}}.
\newblock
  \url{https://www.alignmentforum.org/posts/QvtHSsZLFCAHmzes7/a-naive-alignment-strategy-and-optimism-about-generalization}.

\bibitem[\protect\citeauthoryear{Christiano, Bergal, Fernandez, and
  Long}{Christiano et~al.}{2019}]{paul-discuss}
Christiano, P., A.~Bergal, R.~Fernandez, and R.~Long (2019, September).
\newblock Conversation with {{Paul Christiano}} - {{AI Impacts}}.
\newblock \url{https://aiimpacts.org/conversation-with-paul-christiano/}.

\bibitem[\protect\citeauthoryear{Christiano, Leike, Brown, Martic, Legg, and
  Amodei}{Christiano et~al.}{2017}]{RL-hpref}
Christiano, P., J.~Leike, T.~B. Brown, M.~Martic, S.~Legg, and D.~Amodei (2017,
  July).
\newblock Deep reinforcement learning from human preferences.
\newblock {\em arXiv:1706.03741 [cs, stat]\/}, 1706.03741.
\newblock \url{http://arxiv.org/abs/1706.03741}.

\bibitem[\protect\citeauthoryear{Christiano, Shlegeris, and Amodei}{Christiano
  et~al.}{2018}]{strong-learners}
Christiano, P., B.~Shlegeris, and D.~Amodei (2018, October).
\newblock Supervising strong learners by amplifying weak experts.
\newblock {\em arXiv:1810.08575 [cs, stat]\/}, 1810.08575.
\newblock \url{http://arxiv.org/abs/1810.08575}.

\bibitem[\protect\citeauthoryear{Clark and Hadfield}{Clark and
  Hadfield}{2019}]{Clark}
Clark, J. and G.~K. Hadfield (2019, December).
\newblock Regulatory {{Markets}} for {{AI Safety}}.
\newblock {\em arXiv:2001.00078 [cs, econ, q-fin]\/}, 2001.00078.
\newblock \url{http://arxiv.org/abs/2001.00078}.

\bibitem[\protect\citeauthoryear{Coleman}{Coleman}{1988}]{coleman1988}
Coleman, J.~S. (1988).
\newblock Social {{Capital}} in the {{Creation}} of {{Human Capital}}.
\newblock {\em American Journal of Sociology\/}~{\em 94}, S95--S120.
  {University of Chicago Press}.
\newblock \url{https://www.jstor.org/stable/2780243}.

\bibitem[\protect\citeauthoryear{Critch and Krueger}{Critch and
  Krueger}{2020}]{ARCHES}
Critch, A. and D.~Krueger (2020, May).
\newblock {{AI Research Considerations}} for {{Human Existential Safety}}
  ({{ARCHES}}).
\newblock {\em arXiv:2006.04948 [cs]\/}, 2006.04948.
\newblock \url{http://arxiv.org/abs/2006.04948}.

\bibitem[\protect\citeauthoryear{Dafoe, Hughes, Bachrach, Collins, McKee,
  Leibo, Larson, and Graepel}{Dafoe et~al.}{2020}]{CooperativeAI}
Dafoe, A., E.~Hughes, Y.~Bachrach, T.~Collins, K.~R. McKee, J.~Z. Leibo,
  K.~Larson, and T.~Graepel (2020, December).
\newblock Open {{Problems}} in {{Cooperative AI}}.
\newblock {\em arXiv:2012.08630 [cs]\/}, 2012.08630.
\newblock \url{http://arxiv.org/abs/2012.08630}.

\bibitem[\protect\citeauthoryear{Davenport and Ewing}{Davenport and
  Ewing}{2015}]{webpage}
Davenport, C. and J.~Ewing (2015, September).
\newblock {{VW Is Said}} to {{Cheat}} on {{Diesel Emissions}}; {{U}}.{{S}}. to
  {{Order Big Recall}}.
\newblock {\em The New York Times\/}.
\newblock
  \url{https://www.nytimes.com/2015/09/19/business/volkswagen-is-ordered-to-recall-nearly-500000-vehicles-over-emissions-software.html}.

\bibitem[\protect\citeauthoryear{David}{David}{2020}]{sep-correspond}
David, M. (2020).
\newblock The correspondence theory of truth.
\newblock In E.~N. Zalta (Ed.), {\em The {{Stanford}} Encyclopedia of
  Philosophy\/} (Winter 2020 ed.). {Metaphysics Research Lab, Stanford
  University}.
\newblock
  \url{https://plato.stanford.edu/archives/win2020/entries/truth-correspondence/}.

\bibitem[\protect\citeauthoryear{Davis}{Davis}{2015}]{davis}
Davis, E. (2015, March).
\newblock {Ethical guidelines for a superintelligence}.
\newblock {\em Artificial Intelligence\/}~{\em 220}, 121--124. {Elsevier}.
\newblock doi:10.1016/j.artint.2014.12.003.

\bibitem[\protect\citeauthoryear{Dennett}{Dennett}{1989}]{intentionalstance}
Dennett, D.~C. (1989).
\newblock {\em The {{Intentional Stance}}}.
\newblock A {{Bradford}} Book. {MIT Press}.

\bibitem[\protect\citeauthoryear{Devlin, Chang, Lee, and Toutanova}{Devlin
  et~al.}{2019}]{BERT}
Devlin, J., M.-W. Chang, K.~Lee, and K.~Toutanova (2019, May).
\newblock {{BERT}}: Pre-training of {{Deep Bidirectional Transformers}} for
  {{Language Understanding}}.
\newblock {\em arXiv:1810.04805 [cs]\/}, 1810.04805.
\newblock \url{http://arxiv.org/abs/1810.04805}.

\bibitem[\protect\citeauthoryear{Dietterich}{Dietterich}{2018}]{dietterich}
Dietterich, T.~G. (2018, November).
\newblock Robust {{Artificial Intelligence}} and {{Robust Human
  Organizations}}.
\newblock {\em arXiv:1811.10840 [cs]\/}, 1811.10840.
\newblock \url{http://arxiv.org/abs/1811.10840}.

\bibitem[\protect\citeauthoryear{D'Isanto and Polsterer}{D'Isanto and
  Polsterer}{2018}]{DIsanto2018}
D'Isanto, A. and K.~L. Polsterer (2018).
\newblock Photometric redshift estimation via deep learning - {{Generalized}}
  and pre-classification-less, image based, fully probabilistic redshifts.
\newblock {\em antike und abendland\/}~{\em 609}, A111.
\newblock doi:10.1051/0004-6361/201731326.

\bibitem[\protect\citeauthoryear{Domingos}{Domingos}{2012}]{domingos}
Domingos, P. (2012, October).
\newblock A few useful things to know about machine learning.
\newblock {\em Communications of The Acm\/}~{\em 55\/}(10), 78--87. {New York,
  NY, USA}: {Association for Computing Machinery}.
\newblock doi:10.1145/2347736.2347755.

\bibitem[\protect\citeauthoryear{Dreher and Herzfeld}{Dreher and
  Herzfeld}{2005}]{dreher2005}
Dreher, A. and T.~Herzfeld (2005).
\newblock The {{Economic Costs}} of {{Corruption}}: A {{Survey}} and {{New
  Evidence}}.
\newblock Public {{Economics}}, {University Library of Munich, Germany}.
\newblock \url{https://EconPapers.repec.org/RePEc:wpa:wuwppe:0506001}.

\bibitem[\protect\citeauthoryear{Drexler}{Drexler}{2021}]{drexlerQNR}
Drexler, K.~E. (2021).
\newblock {{QNRs}}: Toward language for intelligent machines.
\newblock Technical {{Report}} 2021-3, {Future of Humanity Institute,
  University of Oxford}.

\bibitem[\protect\citeauthoryear{{European Commission, Directorate - General
  for Communications Networks} and {Technology}}{{European Commission,
  Directorate - General for Communications Networks} and
  {Technology}}{2021}]{EUProposal}
{European Commission, Directorate - General for Communications Networks}, C.
  and {Technology} (2021, April).
\newblock Proposal for a {{Regulation}} of the {{European Parliament}} and of
  the {{Council}}, {{Laying}} down harmonised rules on artificial intelligence
  ({{Artificial Intelligence Act}}) and amending certain {{Union Legislative
  Acts}}.
\newblock Procedure 2021/0106/COD, {CNECT}.

\bibitem[\protect\citeauthoryear{Evans, Saunders, and Stuhlm{\"u}ller}{Evans
  et~al.}{2019}]{projects}
Evans, O., W.~Saunders, and A.~Stuhlm{\"u}ller (2019).
\newblock Machine learning projects for iterated distillation and
  amplification.

\bibitem[\protect\citeauthoryear{Evans, Stuhlm{\"u}ller, Cundy, Carey, Kenton,
  McGrath, and Schreiber}{Evans et~al.}{2018}]{deliberative}
Evans, O., A.~Stuhlm{\"u}ller, C.~Cundy, R.~Carey, Z.~Kenton, T.~McGrath, and
  A.~Schreiber (2018, July).
\newblock Predicting {{Human Deliberative Judgments}} with {{Machine
  Learning}}.
\newblock Technical {{Report}} 2018-2, {Future of Humanity Institute,
  University of Oxford}.

\bibitem[\protect\citeauthoryear{Everitt, Carey, Langlois, Ortega, and
  Legg}{Everitt et~al.}{2021}]{incentives}
Everitt, T., R.~Carey, E.~D. Langlois, P.~A. Ortega, and S.~Legg (2021).
\newblock Agent incentives: A causal perspective.
\newblock {\em Proceedings of the AAAI Conference on Artificial
  Intelligence\/}~{\em 35\/}(13), 11487--11495.
\newblock \url{https://ojs.aaai.org/index.php/AAAI/article/view/17368}.

\bibitem[\protect\citeauthoryear{Gao, Biderman, Black, Golding, Hoppe, Foster,
  Phang, He, Thite, Nabeshima, Presser, and Leahy}{Gao et~al.}{2020}]{pile}
Gao, L., S.~Biderman, S.~Black, L.~Golding, T.~Hoppe, C.~Foster, J.~Phang,
  H.~He, A.~Thite, N.~Nabeshima, S.~Presser, and C.~Leahy (2020, December).
\newblock The {{Pile}}: An {{800GB Dataset}} of {{Diverse Text}} for {{Language
  Modeling}}.
\newblock {\em arXiv:2101.00027 [cs]\/}, 2101.00027.
\newblock \url{http://arxiv.org/abs/2101.00027}.

\bibitem[\protect\citeauthoryear{Gauci, Conti, Liang, Virochsiri, He, Kaden,
  Narayanan, Ye, Chen, and Fujimoto}{Gauci et~al.}{2019}]{facebook}
Gauci, J., E.~Conti, Y.~Liang, K.~Virochsiri, Y.~He, Z.~Kaden, V.~Narayanan,
  X.~Ye, Z.~Chen, and S.~Fujimoto (2019, September).
\newblock Horizon: Facebook's {{Open Source Applied Reinforcement Learning
  Platform}}.
\newblock {\em arXiv:1811.00260 [cs, stat]\/}, 1811.00260.
\newblock \url{http://arxiv.org/abs/1811.00260}.

\bibitem[\protect\citeauthoryear{Gee and Button}{Gee and
  Button}{2019}]{Gee2019}
Gee, J. and M.~Button (2019, July).
\newblock {\em The Financial Cost of Fraud 2019: The Latest Data from around
  the World}.
\newblock {United Kingdom}: {Crowe UK}.

\bibitem[\protect\citeauthoryear{Goodman and Flaxman}{Goodman and
  Flaxman}{2017}]{goodman}
Goodman, B. and S.~Flaxman (2017, October).
\newblock European {{Union}} regulations on algorithmic decision-making and a
  "right to explanation".
\newblock {\em AI Magazine\/}~{\em 38\/}(3), 1606.08813.
\newblock doi:10.1609/aimag.v38i3.2741.

\bibitem[\protect\citeauthoryear{Hannan and Freeman}{Hannan and
  Freeman}{1984}]{hannan1984}
Hannan, M.~T. and J.~Freeman (1984).
\newblock Structural {{Inertia}} and {{Organizational Change}}.
\newblock {\em American Sociological Review\/}~{\em 49\/}(2), 149--164.
  {[American Sociological Association, Sage Publications, Inc.]}.
\newblock doi:10.2307/2095567.

\bibitem[\protect\citeauthoryear{Hanson}{Hanson}{2016}]{hanson2016}
Hanson, R. (2016, May).
\newblock {\em The {{Age}} of {{Em}}: Work, {{Love}}, and {{Life}} When
  {{Robots Rule}} the {{Earth}}}.
\newblock {Oxford University Press}.
\newblock doi:10.1093/oso/9780198754626.001.0001.

\bibitem[\protect\citeauthoryear{Hendrycks, Burns, Basart, Zou, Mazeika, Song,
  and Steinhardt}{Hendrycks et~al.}{2021}]{MMO}
Hendrycks, D., C.~Burns, S.~Basart, A.~Zou, M.~Mazeika, D.~Song, and
  J.~Steinhardt (2021, January).
\newblock Measuring {{Massive Multitask Language Understanding}}.
\newblock {\em arXiv:2009.03300 [cs]\/}, 2009.03300.
\newblock \url{http://arxiv.org/abs/2009.03300}.

\bibitem[\protect\citeauthoryear{Henighan, Kaplan, Katz, Chen, Hesse, Jackson,
  Jun, Brown, Dhariwal, Gray, Hallacy, Mann, Radford, Ramesh, Ryder, Ziegler,
  Schulman, Amodei, and McCandlish}{Henighan et~al.}{2020}]{scaling-multi}
Henighan, T., J.~Kaplan, M.~Katz, M.~Chen, C.~Hesse, J.~Jackson, H.~Jun, T.~B.
  Brown, P.~Dhariwal, S.~Gray, C.~Hallacy, B.~Mann, A.~Radford, A.~Ramesh,
  N.~Ryder, D.~M. Ziegler, J.~Schulman, D.~Amodei, and S.~McCandlish (2020,
  November).
\newblock Scaling {{Laws}} for {{Autoregressive Generative Modeling}}.
\newblock {\em arXiv:2010.14701 [cs]\/}, 2010.14701.
\newblock \url{http://arxiv.org/abs/2010.14701}.

\bibitem[\protect\citeauthoryear{{Hosseini-Asl}, McCann, Wu, Yavuz, and
  Socher}{{Hosseini-Asl} et~al.}{2020}]{socher}
{Hosseini-Asl}, E., B.~McCann, C.-S. Wu, S.~Yavuz, and R.~Socher (2020, July).
\newblock A {{Simple Language Model}} for {{Task}}-{{Oriented Dialogue}}.
\newblock {\em arXiv:2005.00796 [cs]\/}, 2005.00796.
\newblock \url{http://arxiv.org/abs/2005.00796}.

\bibitem[\protect\citeauthoryear{Hubinger}{Hubinger}{2019a}]{olah-view}
Hubinger, E. (2019a, November).
\newblock Chris {{Olah}}'s views on {{AGI}} safety - {{AI Alignment Forum}}.
\newblock
  \url{https://www.alignmentforum.org/posts/X2i9dQQK3gETCyqh2/chris-olah-s-views-on-agi-safety}.

\bibitem[\protect\citeauthoryear{Hubinger}{Hubinger}{2019b}]{hubinger2019}
Hubinger, E. (2019b, February).
\newblock Nuances with ascription universality - {{AI Alignment Forum}}.
\newblock
  \url{https://www.alignmentforum.org/posts/R5Euq7gZgobJi5S25/nuances-with-ascription-universality}.

\bibitem[\protect\citeauthoryear{Hubinger}{Hubinger}{2019c}]{evan}
Hubinger, E. (2019c, September).
\newblock Relaxed adversarial training for inner alignment - {{AI Alignment
  Forum}}.
\newblock
  \url{https://www.alignmentforum.org/posts/9Dy5YRaoCxH9zuJqa/relaxed-adversarial-training-for-inner-alignment}.

\bibitem[\protect\citeauthoryear{Hubinger, {van Merwijk}, Mikulik, Skalse, and
  Garrabrant}{Hubinger et~al.}{2019}]{mesa}
Hubinger, E., C.~{van Merwijk}, V.~Mikulik, J.~Skalse, and S.~Garrabrant (2019,
  June).
\newblock Risks from {{Learned Optimization}} in {{Advanced Machine Learning
  Systems}}.
\newblock {\em arXiv:1906.01820 [cs]\/}, 1906.01820.
\newblock \url{http://arxiv.org/abs/1906.01820}.

\bibitem[\protect\citeauthoryear{{Hugh-Jones}}{{Hugh-Jones}}{2016}]{hugh-jones2016}
{Hugh-Jones}, D. (2016, July).
\newblock Honesty, beliefs about honesty, and economic growth in 15 countries.
\newblock {\em Journal of Economic Behavior \& Organization\/}~{\em 127},
  99--114.
\newblock doi:10.1016/j.jebo.2016.04.012.

\bibitem[\protect\citeauthoryear{Irving, Christiano, and Amodei}{Irving
  et~al.}{2018}]{debate}
Irving, G., P.~Christiano, and D.~Amodei (2018, October).
\newblock {{AI}} safety via debate.
\newblock {\em arXiv:1805.00899 [cs, stat]\/}, 1805.00899.
\newblock \url{http://arxiv.org/abs/1805.00899}.

\bibitem[\protect\citeauthoryear{Isaac and Bridewell}{Isaac and
  Bridewell}{2017}]{isaac2017a}
Isaac, A. M.~C. and W.~Bridewell (2017).
\newblock White {{Lies}} on {{Silver Tongues}}: Why {{Robots Need}} to
  {{Deceive}} (and {{How}}).
\newblock In {\em Robot {{Ethics}} 2.0}. {New York}: {Oxford University Press}.
\newblock doi:10.1093/oso/9780190652951.003.0011.

\bibitem[\protect\citeauthoryear{Jacovi and Goldberg}{Jacovi and
  Goldberg}{2020}]{jacovi}
Jacovi, A. and Y.~Goldberg (2020, April).
\newblock Towards {{Faithfully Interpretable NLP Systems}}: How should we
  define and evaluate faithfulness?
\newblock {\em arXiv:2004.03685 [cs]\/}, 2004.03685.
\newblock \url{http://arxiv.org/abs/2004.03685}.

\bibitem[\protect\citeauthoryear{Jumper, Evans, Pritzel, Green, Figurnov,
  Ronneberger, Tunyasuvunakool, Bates, {\v Z}{\'i}dek, Potapenko, Bridgland,
  Meyer, Kohl, Ballard, Cowie, {Romera-Paredes}, Nikolov, Jain, Adler, Back,
  Petersen, Reiman, Clancy, Zielinski, Steinegger, Pacholska, Berghammer,
  Bodenstein, Silver, Vinyals, Senior, Kavukcuoglu, Kohli, and Hassabis}{Jumper
  et~al.}{2021}]{jumper2021}
Jumper, J., R.~Evans, A.~Pritzel, T.~Green, M.~Figurnov, O.~Ronneberger,
  K.~Tunyasuvunakool, R.~Bates, A.~{\v Z}{\'i}dek, A.~Potapenko, A.~Bridgland,
  C.~Meyer, S.~A.~A. Kohl, A.~J. Ballard, A.~Cowie, B.~{Romera-Paredes},
  S.~Nikolov, R.~Jain, J.~Adler, T.~Back, S.~Petersen, D.~Reiman, E.~Clancy,
  M.~Zielinski, M.~Steinegger, M.~Pacholska, T.~Berghammer, S.~Bodenstein,
  D.~Silver, O.~Vinyals, A.~W. Senior, K.~Kavukcuoglu, P.~Kohli, and
  D.~Hassabis (2021, August).
\newblock Highly accurate protein structure prediction with {{AlphaFold}}.
\newblock {\em Nature\/}~{\em 596\/}(7873), 583--589. {Nature Publishing
  Group}.
\newblock doi:10.1038/s41586-021-03819-2.

\bibitem[\protect\citeauthoryear{Kaplan, McCandlish, Henighan, Brown, Chess,
  Child, Gray, Radford, Wu, and Amodei}{Kaplan et~al.}{2020}]{scaling}
Kaplan, J., S.~McCandlish, T.~Henighan, T.~B. Brown, B.~Chess, R.~Child,
  S.~Gray, A.~Radford, J.~Wu, and D.~Amodei (2020, January).
\newblock Scaling {{Laws}} for {{Neural Language Models}}.
\newblock {\em arXiv:2001.08361 [cs, stat]\/}, 2001.08361.
\newblock \url{http://arxiv.org/abs/2001.08361}.

\bibitem[\protect\citeauthoryear{Kavanagh and Rich}{Kavanagh and
  Rich}{2018}]{Kavanagh2018}
Kavanagh, J. and M.~D. Rich (2018).
\newblock {\em Truth Decay: An Initial Exploration of the Diminishing Role of
  Facts and Analysis in American Public Life}.
\newblock {Santa Monica, CA}: {RAND Corporation}.
\newblock doi:10.7249/RR2314.

\bibitem[\protect\citeauthoryear{Kenton, Everitt, Weidinger, Gabriel, Mikulik,
  and Irving}{Kenton et~al.}{2021}]{kenton}
Kenton, Z., T.~Everitt, L.~Weidinger, I.~Gabriel, V.~Mikulik, and G.~Irving
  (2021, March).
\newblock Alignment of {{Language Agents}}.
\newblock {\em arXiv:2103.14659 [cs]\/}, 2103.14659.
\newblock \url{http://arxiv.org/abs/2103.14659}.

\bibitem[\protect\citeauthoryear{Khashabi, Min, Khot, Sabharwal, Tafjord,
  Clark, and Hajishirzi}{Khashabi et~al.}{2020}]{UnifiedQA}
Khashabi, D., S.~Min, T.~Khot, A.~Sabharwal, O.~Tafjord, P.~Clark, and
  H.~Hajishirzi (2020, October).
\newblock {{UnifiedQA}}: Crossing {{Format Boundaries With}} a {{Single QA
  System}}.
\newblock {\em arXiv:2005.00700 [cs]\/}, 2005.00700.
\newblock \url{http://arxiv.org/abs/2005.00700}.

\bibitem[\protect\citeauthoryear{Khashabi, Stanovsky, Bragg, Lourie, Kasai,
  Choi, Smith, and Weld}{Khashabi et~al.}{2021}]{genie}
Khashabi, D., G.~Stanovsky, J.~Bragg, N.~Lourie, J.~Kasai, Y.~Choi, N.~A.
  Smith, and D.~S. Weld (2021, June).
\newblock {{GENIE}}: A {{Leaderboard}} for {{Human}}-in-the-{{Loop Evaluation}}
  of {{Text Generation}}.
\newblock {\em arXiv:2101.06561 [cs]\/}, 2101.06561.
\newblock \url{http://arxiv.org/abs/2101.06561}.

\bibitem[\protect\citeauthoryear{Knack and Keefer}{Knack and
  Keefer}{1997}]{Knack1997}
Knack, S. and P.~Keefer (1997, November).
\newblock Does social capital have an economic payoff? a cross-country
  investigation*.
\newblock {\em The Quarterly Journal of Economics\/}~{\em 112\/}(4),
  1251--1288.
\newblock doi:10.1162/003355300555475.

\bibitem[\protect\citeauthoryear{Korinek and Stiglitz}{Korinek and
  Stiglitz}{2019}]{Korinek2019}
Korinek, A. and J.~E. Stiglitz (2019).
\newblock Artificial intelligence and its implications for income distribution
  and unemployment.
\newblock In {\em The Economics of Artificial Intelligence: An Agenda}, pp.\
  349--390. {University of Chicago Press}.
\newblock \url{http://www.nber.org/chapters/c14018}.

\bibitem[\protect\citeauthoryear{Kuleshov, Fenner, and Ermon}{Kuleshov
  et~al.}{2018}]{kuleshov}
Kuleshov, V., N.~Fenner, and S.~Ermon (2018, June).
\newblock Accurate {{Uncertainties}} for {{Deep Learning Using Calibrated
  Regression}}.
\newblock {\em arXiv:1807.00263 [cs, stat]\/}, 1807.00263.
\newblock \url{http://arxiv.org/abs/1807.00263}.

\bibitem[\protect\citeauthoryear{Lawrence}{Lawrence}{2016}]{lawrence}
Lawrence, N. (2016, May).
\newblock Future of {{AI}} 6. {{Discussion}} of '{{Superintelligence}}: Paths,
  {{Dangers}}, {{Strategies}}' - inverseprobability.com: Neil {{Lawrence}}'s
  {{Homepage}}.
\newblock
  \url{http://inverseprobability.com/2016/05/09/machine-learning-futures-6}.

\bibitem[\protect\citeauthoryear{Lazaridou and Baroni}{Lazaridou and
  Baroni}{2020}]{Lazaridou}
Lazaridou, A. and M.~Baroni (2020, July).
\newblock Emergent {{Multi}}-{{Agent Communication}} in the {{Deep Learning
  Era}}.
\newblock {\em arXiv:2006.02419 [cs]\/}, 2006.02419.
\newblock \url{http://arxiv.org/abs/2006.02419}.

\bibitem[\protect\citeauthoryear{Lessig}{Lessig}{1998}]{lessig1998}
Lessig, L. (1998, June).
\newblock The {{New Chicago School}}.
\newblock {\em The Journal of Legal Studies\/}~{\em 27\/}(S2), 661--691.
\newblock doi:10.1086/468039.

\bibitem[\protect\citeauthoryear{Lewis, Yarats, Dauphin, Parikh, and
  Batra}{Lewis et~al.}{2017}]{Lewis}
Lewis, M., D.~Yarats, Y.~N. Dauphin, D.~Parikh, and D.~Batra (2017, June).
\newblock Deal or {{No Deal}}? end-to-{{End Learning}} for {{Negotiation
  Dialogues}}.
\newblock {\em arXiv:1706.05125 [cs]\/}, 1706.05125.
\newblock \url{http://arxiv.org/abs/1706.05125}.

\bibitem[\protect\citeauthoryear{Lewis, Perez, Piktus, Petroni, Karpukhin,
  Goyal, K{\"u}ttler, Lewis, Yih, Rockt{\"a}schel, Riedel, and Kiela}{Lewis
  et~al.}{2021}]{RAG}
Lewis, P., E.~Perez, A.~Piktus, F.~Petroni, V.~Karpukhin, N.~Goyal,
  H.~K{\"u}ttler, M.~Lewis, W.-t. Yih, T.~Rockt{\"a}schel, S.~Riedel, and
  D.~Kiela (2021, April).
\newblock Retrieval-{{Augmented Generation}} for {{Knowledge}}-{{Intensive NLP
  Tasks}}.
\newblock {\em arXiv:2005.11401 [cs]\/}, 2005.11401.
\newblock \url{http://arxiv.org/abs/2005.11401}.

\bibitem[\protect\citeauthoryear{Li, Monroe, Ritter, Galley, Gao, and
  Jurafsky}{Li et~al.}{2016}]{jurafsky}
Li, J., W.~Monroe, A.~Ritter, M.~Galley, J.~Gao, and D.~Jurafsky (2016,
  September).
\newblock Deep {{Reinforcement Learning}} for {{Dialogue Generation}}.
\newblock {\em arXiv:1606.01541 [cs]\/}, 1606.01541.
\newblock \url{http://arxiv.org/abs/1606.01541}.

\bibitem[\protect\citeauthoryear{Liebowitz and Margolis}{Liebowitz and
  Margolis}{1995}]{liebowitz1995}
Liebowitz, S.~J. and S.~E. Margolis (1995, April).
\newblock Path {{Dependence}}, {{Lock}}-{{In}}, and {{History}}.
\newblock {\em The Journal of Law, Economics, and Organization\/}~{\em
  11\/}(1), 205--226.
\newblock doi:10.1093/oxfordjournals.jleo.a036867.

\bibitem[\protect\citeauthoryear{Lin, Hilton, and Evans}{Lin
  et~al.}{2021}]{TruthfulQA}
Lin, S., J.~Hilton, and O.~Evans (2021, September).
\newblock {{TruthfulQA}}: Measuring {{How Models Mimic Human Falsehoods}}.
\newblock {\em arXiv:2109.07958 [cs]\/}, 2109.07958.
\newblock \url{http://arxiv.org/abs/2109.07958}.

\bibitem[\protect\citeauthoryear{Madry, Makelov, Schmidt, Tsipras, and
  Vladu}{Madry et~al.}{2019}]{madry}
Madry, A., A.~Makelov, L.~Schmidt, D.~Tsipras, and A.~Vladu (2019, September).
\newblock Towards {{Deep Learning Models Resistant}} to {{Adversarial
  Attacks}}.
\newblock {\em arXiv:1706.06083 [cs, stat]\/}, 1706.06083.
\newblock \url{http://arxiv.org/abs/1706.06083}.

\bibitem[\protect\citeauthoryear{Mahon}{Mahon}{2016}]{SEP}
Mahon, J.~E. (2016).
\newblock The definition of lying and deception.
\newblock In E.~N. Zalta (Ed.), {\em The {{Stanford}} Encyclopedia of
  Philosophy\/} (Winter 2016 ed.). {Metaphysics Research Lab, Stanford
  University}.
\newblock
  \url{https://plato.stanford.edu/archives/win2016/entries/lying-definition/}.

\bibitem[\protect\citeauthoryear{Manheim and Garrabrant}{Manheim and
  Garrabrant}{2019}]{manheim2019}
Manheim, D. and S.~Garrabrant (2019, February).
\newblock Categorizing {{Variants}} of {{Goodhart}}'s {{Law}}.
\newblock {\em arXiv:1803.04585 [cs, q-fin, stat]\/}, 1803.04585.
\newblock \url{http://arxiv.org/abs/1803.04585}.

\bibitem[\protect\citeauthoryear{Marquis and Tilcsik}{Marquis and
  Tilcsik}{2013}]{marquis2013}
Marquis, C. and A.~Tilcsik (2013, June).
\newblock Imprinting: Toward a {{Multilevel Theory}}.
\newblock {\em Academy of Management Annals\/}~{\em 7\/}(1), 195--245. {Academy
  of Management}.
\newblock doi:10.5465/19416520.2013.766076.

\bibitem[\protect\citeauthoryear{North}{North}{1990}]{north1990}
North, D.~C. (1990, October).
\newblock A {{Transaction Cost Theory}} of {{Politics}}.
\newblock {\em Journal of Theoretical Politics\/}~{\em 2\/}(4), 355--367. {SAGE
  Publications Ltd}.
\newblock doi:10.1177/0951692890002004001.

\bibitem[\protect\citeauthoryear{Olah, Cammarata, Schubert, Goh, Petrov, and
  Carter}{Olah et~al.}{2020}]{circuits}
Olah, C., N.~Cammarata, L.~Schubert, G.~Goh, M.~Petrov, and S.~Carter (2020,
  March).
\newblock Zoom {{In}}: An {{Introduction}} to {{Circuits}}.
\newblock {\em Distill\/}~{\em 5\/}(3), 10.23915/distill.00024.001.
\newblock doi:10.23915/distill.00024.001.

\bibitem[\protect\citeauthoryear{OpenAI, Berner, Brockman, Chan, Cheung,
  D{\k{e}}biak, Dennison, Farhi, Fischer, Hashme, Hesse, J{\'o}zefowicz, Gray,
  Olsson, Pachocki, Petrov, Pinto, Raiman, Salimans, Schlatter, Schneider,
  Sidor, Sutskever, Tang, Wolski, and Zhang}{OpenAI et~al.}{2019}]{dota}
OpenAI, C.~Berner, G.~Brockman, B.~Chan, V.~Cheung, P.~D{\k{e}}biak,
  C.~Dennison, D.~Farhi, Q.~Fischer, S.~Hashme, C.~Hesse, R.~J{\'o}zefowicz,
  S.~Gray, C.~Olsson, J.~Pachocki, M.~Petrov, H.~P. d.~O. Pinto, J.~Raiman,
  T.~Salimans, J.~Schlatter, J.~Schneider, S.~Sidor, I.~Sutskever, J.~Tang,
  F.~Wolski, and S.~Zhang (2019, December).
\newblock Dota 2 with {{Large Scale Deep Reinforcement Learning}}.
\newblock {\em arXiv:1912.06680 [cs, stat]\/}, 1912.06680.
\newblock \url{http://arxiv.org/abs/1912.06680}.

\bibitem[\protect\citeauthoryear{Perez, Strub, {de Vries}, Dumoulin, and
  Courville}{Perez et~al.}{2018}]{film}
Perez, E., F.~Strub, H.~{de Vries}, V.~Dumoulin, and A.~Courville (2018,
  April).
\newblock {{FiLM}}: Visual {{Reasoning}} with a {{General Conditioning Layer}}.
\newblock {\em Proceedings of the AAAI Conference on Artificial
  Intelligence\/}~{\em 32\/}(1).
\newblock \url{https://ojs.aaai.org/index.php/AAAI/article/view/11671}.

\bibitem[\protect\citeauthoryear{Peskov, Cheng, Elgohary, Barrow,
  {Danescu-Niculescu-Mizil}, and {Boyd-Graber}}{Peskov
  et~al.}{2020}]{diplomacy}
Peskov, D., B.~Cheng, A.~Elgohary, J.~Barrow, C.~{Danescu-Niculescu-Mizil}, and
  J.~{Boyd-Graber} (2020).
\newblock It takes two to lie: One to lie and one to listen.
\newblock In {\em Association for Computational Linguistics}, {The Cyberverse
  Simulacrum of Seattle}.
\newblock \url{http://umiacs.umd.edu/~jbg//docs/2020_acl_diplomacy.pdf}.

\bibitem[\protect\citeauthoryear{Polu and Sutskever}{Polu and
  Sutskever}{2020}]{polu-sutskever}
Polu, S. and I.~Sutskever (2020, September).
\newblock Generative {{Language Modeling}} for {{Automated Theorem Proving}}.
\newblock {\em arXiv:2009.03393 [cs, stat]\/}, 2009.03393.
\newblock \url{http://arxiv.org/abs/2009.03393}.

\bibitem[\protect\citeauthoryear{{PwC}}{{PwC}}{2017}]{PwC2017}
{PwC} (2017).
\newblock Sizing the prize: What's the real value of {{AI}} for your business
  and how can you capitalise?
\newblock {{PwC}} report, {PwC}.

\bibitem[\protect\citeauthoryear{Radford, Narasimhan, Salimans, and
  Sutskever}{Radford et~al.}{2018}]{gpt}
Radford, A., K.~Narasimhan, T.~Salimans, and I.~Sutskever (2018).
\newblock Improving language understanding by generative pre-training.

\bibitem[\protect\citeauthoryear{Radford, Wu, Child, Luan, Amodei, and
  Sutskever}{Radford et~al.}{2019}]{gpt2}
Radford, A., J.~Wu, R.~Child, D.~Luan, D.~Amodei, and I.~Sutskever (2019).
\newblock Language models are unsupervised multitask learners.
\newblock \url{https://openai.com/blog/better-language-models/}.

\bibitem[\protect\citeauthoryear{Raffel, Shazeer, Roberts, Lee, Narang, Matena,
  Zhou, Li, and Liu}{Raffel et~al.}{2020}]{T5}
Raffel, C., N.~Shazeer, A.~Roberts, K.~Lee, S.~Narang, M.~Matena, Y.~Zhou,
  W.~Li, and P.~J. Liu (2020).
\newblock Exploring the limits of transfer learning with a unified text-to-text
  transformer.
\newblock {\em Journal of Machine Learning Research\/}~{\em 21\/}(140), 1--67.
\newblock \url{http://jmlr.org/papers/v21/20-074.html}.

\bibitem[\protect\citeauthoryear{Ramos}{Ramos}{2003}]{TFIDF}
Ramos, J.~E. (2003).
\newblock Using {{TF}}-{{IDF}} to determine word relevance in document queries.

\bibitem[\protect\citeauthoryear{Ruff, Kauffmann, Vandermeulen, Montavon,
  Samek, Kloft, Dietterich, and M{\"u}ller}{Ruff et~al.}{2021}]{anomaly-detect}
Ruff, L., J.~R. Kauffmann, R.~A. Vandermeulen, G.~Montavon, W.~Samek, M.~Kloft,
  T.~G. Dietterich, and K.-R. M{\"u}ller (2021).
\newblock A unifying review of deep and shallow anomaly detection.
\newblock {\em Proceedings of the IEEE\/}~{\em 109\/}(5), 756--795.
\newblock doi:10.1109/JPROC.2021.3052449.

\bibitem[\protect\citeauthoryear{Russell}{Russell}{2021}]{human-compatible}
Russell, S. (2021).
\newblock Human-compatible artificial intelligence.
\newblock In {\em Human-like Machine Intelligence}, pp.\  3--23. {Oxford
  University Press}.

\bibitem[\protect\citeauthoryear{Russell, Dewey, and Tegmark}{Russell
  et~al.}{2015}]{russell-dewey}
Russell, S., D.~Dewey, and M.~Tegmark (2015, December).
\newblock Research priorities for robust and beneficial artificial
  intelligence.
\newblock {\em AI Magazine\/}~{\em 36\/}(4), 105--114.
\newblock doi:10.1609/aimag.v36i4.2577.

\bibitem[\protect\citeauthoryear{Samek, Wiegand, and M{\"u}ller}{Samek
  et~al.}{2017}]{Samek}
Samek, W., T.~Wiegand, and K.-R. M{\"u}ller (2017, August).
\newblock Explainable {{Artificial Intelligence}}: Understanding,
  {{Visualizing}} and {{Interpreting Deep Learning Models}}.
\newblock {\em arXiv:1708.08296 [cs, stat]\/}, 1708.08296.
\newblock \url{http://arxiv.org/abs/1708.08296}.

\bibitem[\protect\citeauthoryear{Saunders, Rachbach, Evans, Miller, Byun, and
  Stuhlm{\"u}ller}{Saunders et~al.}{2020}]{ought}
Saunders, W., B.~Rachbach, O.~Evans, Z.~Miller, J.~Byun, and A.~Stuhlm{\"u}ller
  (2020).
\newblock Evaluating arguments one step at a time.
\newblock \url{https://ought.org/updates/2020-01-11-arguments}.

\bibitem[\protect\citeauthoryear{Saunders, Sastry, Stuhlm{\"u}ller, and
  Evans}{Saunders et~al.}{2017}]{trial}
Saunders, W., G.~Sastry, A.~Stuhlm{\"u}ller, and O.~Evans (2017, July).
\newblock Trial without {{Error}}: Towards {{Safe Reinforcement Learning}} via
  {{Human Intervention}}.
\newblock {\em arXiv:1707.05173 [cs]\/}, 1707.05173.
\newblock \url{http://arxiv.org/abs/1707.05173}.

\bibitem[\protect\citeauthoryear{Schelling}{Schelling}{1980}]{schelling}
Schelling, T. (1980).
\newblock {\em The Strategy of Conflict: With a New Preface by the Author}.
\newblock {Harvard University Press}.
\newblock \url{https://books.google.com.mx/books?id=7RkL4Z8Yg5AC}.

\bibitem[\protect\citeauthoryear{Schneier}{Schneier}{2019}]{schneier2019}
Schneier, B. (2019, November).
\newblock Technology and {{Policymakers}} - {{Schneier}} on {{Security}}.
\newblock
  \url{https://www.schneier.com/blog/archives/2019/11/technology_and_.html}.

\bibitem[\protect\citeauthoryear{Seger, Avin, Pearson, Briers,
  {\'O}~Heigeartaigh, and Bacon}{Seger et~al.}{2020}]{seger2020}
Seger, E., S.~Avin, G.~Pearson, M.~Briers, S.~{\'O}~Heigeartaigh, and H.~Bacon
  (2020, October).
\newblock Tackling threats to informed decision-making in democratic societies:
  Promoting epistemic security in a technologically-advanced world.
\newblock Technical report, {Apollo - University of Cambridge Repository}.
\newblock doi:10.17863/CAM.64183.

\bibitem[\protect\citeauthoryear{Seidenfeld}{Seidenfeld}{1999}]{seidenfeld1999a}
Seidenfeld, M. (1999).
\newblock Bending the rules: Flexible regulation and constraints on agency
  discretion.
\newblock {\em Administrative Law Review\/}~{\em 51\/}(2), 429--495. {American
  Bar Association}.
\newblock \url{https://www.jstor.org/stable/40709994}.

\bibitem[\protect\citeauthoryear{Shapiro and Kouri~Kissel}{Shapiro and
  Kouri~Kissel}{2021}]{classicallogic}
Shapiro, S. and T.~Kouri~Kissel (2021).
\newblock Classical logic.
\newblock In E.~N. Zalta (Ed.), {\em The {{Stanford}} Encyclopedia of
  Philosophy\/} (Spring 2021 ed.). {Metaphysics Research Lab, Stanford
  University}.
\newblock
  \url{https://plato.stanford.edu/archives/spr2021/entries/logic-classical/}.

\bibitem[\protect\citeauthoryear{Shim and Arkin}{Shim and
  Arkin}{2013}]{Shim2013}
Shim, J. and R.~C. Arkin (2013).
\newblock A taxonomy of robot deception and its benefits in {{HRI}}.
\newblock In {\em 2013 {{IEEE}} International Conference on Systems, Man, and
  Cybernetics}, pp.\  2328--2335.
\newblock doi:10.1109/SMC.2013.398.

\bibitem[\protect\citeauthoryear{Shuster, Poff, Chen, Kiela, and
  Weston}{Shuster et~al.}{2021}]{FB-paper}
Shuster, K., S.~Poff, M.~Chen, D.~Kiela, and J.~Weston (2021, April).
\newblock Retrieval {{Augmentation Reduces Hallucination}} in {{Conversation}}.
\newblock {\em arXiv:2104.07567 [cs]\/}, 2104.07567.
\newblock \url{http://arxiv.org/abs/2104.07567}.

\bibitem[\protect\citeauthoryear{Solaiman and Dennison}{Solaiman and
  Dennison}{2021}]{PALMS}
Solaiman, I. and C.~Dennison (2021, June).
\newblock Process for {{Adapting Language Models}} to {{Society}} ({{PALMS}})
  with {{Values}}-{{Targeted Datasets}}.
\newblock {\em arXiv:2106.10328 [cs]\/}, 2106.10328.
\newblock \url{http://arxiv.org/abs/2106.10328}.

\bibitem[\protect\citeauthoryear{Starr}{Starr}{2021}]{sep-counterfactuals}
Starr, W. (2021).
\newblock {Counterfactuals}.
\newblock In E.~N. Zalta (Ed.), {\em The {Stanford} Encyclopedia of
  Philosophy\/} ({S}ummer 2021 ed.). Metaphysics Research Lab, Stanford
  University.
\newblock
  \url{https://plato.stanford.edu/archives/sum2021/entries/counterfactuals/}.

\bibitem[\protect\citeauthoryear{Stiennon, Ouyang, Wu, Ziegler, Lowe, Voss,
  Radford, Amodei, and Christiano}{Stiennon et~al.}{2020}]{stiennon}
Stiennon, N., L.~Ouyang, J.~Wu, D.~M. Ziegler, R.~Lowe, C.~Voss, A.~Radford,
  D.~Amodei, and P.~Christiano (2020, October).
\newblock Learning to summarize from human feedback.
\newblock {\em arXiv:2009.01325 [cs]\/}, 2009.01325.
\newblock \url{http://arxiv.org/abs/2009.01325}.

\bibitem[\protect\citeauthoryear{Stokes, Yang, Swanson, Jin, {Cubillos-Ruiz},
  Donghia, MacNair, French, Carfrae, {Bloom-Ackermann}, Tran, {Chiappino-Pepe},
  Badran, Andrews, Chory, Church, Brown, Jaakkola, Barzilay, and
  Collins}{Stokes et~al.}{2020}]{Stokes2020}
Stokes, J.~M., K.~Yang, K.~Swanson, W.~Jin, A.~{Cubillos-Ruiz}, N.~M. Donghia,
  C.~R. MacNair, S.~French, L.~A. Carfrae, Z.~{Bloom-Ackermann}, V.~M. Tran,
  A.~{Chiappino-Pepe}, A.~H. Badran, I.~W. Andrews, E.~J. Chory, G.~M. Church,
  E.~D. Brown, T.~S. Jaakkola, R.~Barzilay, and J.~J. Collins (2020, February).
\newblock A {{Deep Learning Approach}} to {{Antibiotic Discovery}}.
\newblock {\em Cell\/}~{\em 180\/}(4), 688--702.e13.
\newblock doi:10.1016/j.cell.2020.01.021.

\bibitem[\protect\citeauthoryear{Stoljar and Damnjanovic}{Stoljar and
  Damnjanovic}{2014}]{sep-deflation}
Stoljar, D. and N.~Damnjanovic (2014).
\newblock The deflationary theory of truth.
\newblock In E.~N. Zalta (Ed.), {\em The {{Stanford}} Encyclopedia of
  Philosophy\/} (Fall 2014 ed.). {Metaphysics Research Lab, Stanford
  University}.
\newblock
  \url{https://plato.stanford.edu/archives/fall2014/entries/truth-deflationary/}.

\bibitem[\protect\citeauthoryear{Talmor, Yoran, Bras, Bhagavatula, Goldberg,
  Choi, and Berant}{Talmor et~al.}{2021}]{CommonSenseQA2}
Talmor, A., O.~Yoran, R.~L. Bras, C.~Bhagavatula, Y.~Goldberg, Y.~Choi, and
  J.~Berant (2021).
\newblock {{CommonsenseQA}} 2.0: Exposing the limits of {{AI}} through
  gamification.
\newblock In {\em Thirty-Fifth Conference on Neural Information Processing
  Systems Datasets and Benchmarks Track (Round 1)}.
\newblock \url{https://openreview.net/forum?id=qF7FlUT5dxa}.

\bibitem[\protect\citeauthoryear{Trammell and Korinek}{Trammell and
  Korinek}{2020}]{Trammell2020}
Trammell, P. and A.~Korinek (2020, October).
\newblock Economic growth under transformative {{AI}}: A guide to the vast
  range of possibilities for output growth, wages, and the labor share.
\newblock {{GPI}} Working Paper 8-2020, {Global Priorities Institute, Oxford
  University}.

\bibitem[\protect\citeauthoryear{Vinyals, Babuschkin, Czarnecki, Mathieu,
  Dudzik, Chung, Choi, Powell, Ewalds, Georgiev, Oh, Horgan, Kroiss, Danihelka,
  Huang, Sifre, Cai, Agapiou, Jaderberg, Vezhnevets, Leblond, Pohlen, Dalibard,
  Budden, Sulsky, Molloy, Paine, Gulcehre, Wang, Pfaff, Wu, Ring, Yogatama,
  W{\"u}nsch, McKinney, Smith, Schaul, Lillicrap, Kavukcuoglu, Hassabis, Apps,
  and Silver}{Vinyals et~al.}{2019}]{alphastar}
Vinyals, O., I.~Babuschkin, W.~M. Czarnecki, M.~Mathieu, A.~Dudzik, J.~Chung,
  D.~H. Choi, R.~Powell, T.~Ewalds, P.~Georgiev, J.~Oh, D.~Horgan, M.~Kroiss,
  I.~Danihelka, A.~Huang, L.~Sifre, T.~Cai, J.~P. Agapiou, M.~Jaderberg, A.~S.
  Vezhnevets, R.~Leblond, T.~Pohlen, V.~Dalibard, D.~Budden, Y.~Sulsky,
  J.~Molloy, T.~L. Paine, C.~Gulcehre, Z.~Wang, T.~Pfaff, Y.~Wu, R.~Ring,
  D.~Yogatama, D.~W{\"u}nsch, K.~McKinney, O.~Smith, T.~Schaul, T.~Lillicrap,
  K.~Kavukcuoglu, D.~Hassabis, C.~Apps, and D.~Silver (2019, November).
\newblock Grandmaster level in {{StarCraft II}} using multi-agent reinforcement
  learning.
\newblock {\em Nature\/}~{\em 575\/}(7782), 350--354.
\newblock doi:10.1038/s41586-019-1724-z.

\bibitem[\protect\citeauthoryear{Wang and Komatsuzaki}{Wang and
  Komatsuzaki}{2021}]{GPTJ}
Wang, B. and A.~Komatsuzaki (2021).
\newblock {{GPT}}-{{J}}-{{6B}}: A 6 billion parameter autoregressive language
  model.

\bibitem[\protect\citeauthoryear{Wei, Bosma, Zhao, Guu, Yu, Lester, Du, Dai,
  and Le}{Wei et~al.}{2021}]{zeroshot}
Wei, J., M.~Bosma, V.~Y. Zhao, K.~Guu, A.~W. Yu, B.~Lester, N.~Du, A.~M. Dai,
  and Q.~V. Le (2021, September).
\newblock Finetuned {{Language Models Are Zero}}-{{Shot Learners}}.
\newblock {\em arXiv:2109.01652 [cs]\/}, 2109.01652.
\newblock \url{http://arxiv.org/abs/2109.01652}.

\bibitem[\protect\citeauthoryear{Weller}{Weller}{2017}]{weller}
Weller, A. (2017, July).
\newblock Challenges for transparency.
\newblock In {\em {{ICML Workshop}} on {{Human Interpretability}} in {{Machine
  Learning}}}, {Sydney, NSW, Australia}.

\bibitem[\protect\citeauthoryear{{Wikipedia contributors}}{{Wikipedia
  contributors}}{2021a}]{wikipedia-knightean}
{Wikipedia contributors} (2021a).
\newblock Knightian uncertainty \textemdash{} {{Wikipedia}}, the free
  encyclopedia.
\newblock \url{https://en.wikipedia.org/wiki/Knightian_uncertainty}.

\bibitem[\protect\citeauthoryear{{Wikipedia contributors}}{{Wikipedia
  contributors}}{2021b}]{Wikipedia}
{Wikipedia contributors} (2021b).
\newblock Right to explanation \textemdash{} {{Wikipedia}}, the free
  encyclopedia.
\newblock \url{https://en.wikipedia.org/wiki/Right_to_explanation}.

\bibitem[\protect\citeauthoryear{{Wikipedia contributors}}{{Wikipedia
  contributors}}{2021c}]{wiki}
{Wikipedia contributors} (2021c).
\newblock Training, validation, and test sets \textemdash{} {{Wikipedia}}, the
  free encyclopedia.
\newblock
  \url{https://en.wikipedia.org/wiki/Training,_validation,_and_test_sets}.

\bibitem[\protect\citeauthoryear{Yuan, He, Zhu, and Li}{Yuan
  et~al.}{2019}]{review}
Yuan, X., P.~He, Q.~Zhu, and X.~Li (2019).
\newblock Adversarial examples: Attacks and defenses for deep learning.
\newblock {\em IEEE Transactions on Neural Networks and Learning
  Systems\/}~{\em 30\/}(9), 2805--2824.
\newblock doi:10.1109/TNNLS.2018.2886017.

\bibitem[\protect\citeauthoryear{Yudkowsky}{Yudkowsky}{2018}]{yudkowsky2018}
Yudkowsky, E. (2018, May).
\newblock Meta-{{Honesty}}: Firming {{Up Honesty Around Its Edge}}-{{Cases}} -
  {{LessWrong}}.
\newblock
  \url{https://www.lesswrong.com/posts/xdwbX9pFEr7Pomaxv/meta-honesty-firming-up-honesty-around-its-edge-cases}.

\bibitem[\protect\citeauthoryear{Zak and Knack}{Zak and Knack}{2001}]{zak2001}
Zak, P.~J. and S.~Knack (2001).
\newblock Trust and {{Growth}}.
\newblock {\em The Economic Journal\/}~{\em 111\/}(470), 295--321.
\newblock doi:10.1111/1468-0297.00609.

\end{thebibliography}

\end{document}